\documentclass{aa}
\usepackage{graphicx}
\usepackage{times}
\usepackage{astron}
\usepackage{psfig}
\def\tp{{\it type}}
\def\Tp{{\it Type}}
\begin{document}

\title{The COMBO-17 Survey:\\ Evolution of the Galaxy Luminosity Function\\
 from 25,000 Galaxies with $0.2<z<1.2$ }

\author{C. Wolf\inst{1,2} \and K. Meisenheimer\inst{1} \and H.-W. Rix\inst{1}
\and A. Borch\inst{1} \and S. Dye\inst{1,3} \and M. Kleinheinrich\inst{1,4} }

\institute{ Max-Planck-Institut f\"ur Astronomie, K\"onigstuhl 17,
            D-69117 Heidelberg, Germany 
       \and Department of Physics, Denys Wilkinson Bldg.,
            University of Oxford, Keble Road, Oxford, OX1 3RH, U.K. 
       \and Astrophysics Group, Blackett Lab,
            Imperial College, Prince Consort Road, London, U.K.  
       \and IAEF, Universit\"at Bonn, 
            Auf dem H\"ugel 71, D-53121 Bonn, Germany }

\date{Received / Accepted }

\abstract{
We present a detailed empirical assessment of how the galaxy luminosity function 
and stellar luminosity density evolves over the last half of the universe's age 
($0.2<z<1.2$) for galaxies of different spectral energy distributions (SED). The 
results are based on $\sim 25000$ galaxies ($R\la24$) with redshift measurements 
($\sigma_z\approx 0.03$) and SEDs across $\lambda_\mathrm{obs}\approx 350\ldots 
930$~nm. The redshifts and SEDs were derived from medium-band photometry in 17 
filters, observed as part of the COMBO-17 survey (``Classifying Objects by 
Medium-Band Observations in 17 Filters'') over three disjoint fields with a total 
area of 0.78 square degrees. Luminosity functions (LF), binned in redshift and 
SED-type, are presented in the restframe passbands of the SDSS r-band, the 
Johnson B-band and a synthetic UV continuum band at 280~nm. \\
We find that the luminosity function depends strongly on SED-type at all 
redshifts covered. The shape of the LF, i.e. the faint-end power-law slope, does 
depend on SED type, but not on redshift. However, the redshift evolution of the 
characteristic luminosity $M^*$ and density $\phi^*$ depends strongly on SED-type: 
(1) Early-type galaxies, defined as redder than a present-day reference Sa 
spectrum, become drastically more abundant towards low redshift, by a factor of 10 
in the number density $\phi^*$ from z=1.1 to now, and by a factor of 4 in their 
contribution to the co-moving r-band luminosity density, $j_\mathrm{r}$. 
(2) Galaxies resembling present-day Sa- to Sbc-colours show a co-moving number 
density and contribution to $j_\mathrm{r}$ that does not vary much with redshift. 
(3) Galaxies with blue spectra reflecting strong star formation decrease towards 
low redshift both in luminosity and density, and by a factor of 4 in their 
$j_\mathrm{r}$ contribution.
Summed over all SED types and galaxy luminosities, the comoving luminosity density
decreases towards low redshift, between z=1.1 and now, by a small amount in restframe
r and B, but by a factor of $\sim6$ in restframe 280~nm. At $z=1.1$, galaxies 
redder than Sbc's, contribute 40\% to the total $j_\mathrm{r}$, which increases to 
75\% by z=0. For $\lambda_\mathrm{rest}=$280~nm, this increase is from 12\% to 25\% 
over the same redshift interval. \\
Comparison of the three independent sight-lines shows that our results are not
significantly affected by large-scale structure. Our lowest redshift bin at $z=
[0.2,0.4]$ largely agrees with the recent assessment of the present-day galaxy 
population by SDSS and 2dFGRS and deviates only by an excess of ``faint blue 
galaxies'' at $z\sim 0.3$ compared to very local samples. Overall our findings 
provide a set of new and much more precise constraints to model the waning of 
overall star formation activity, the demise of star-bursts and the strong emergence 
of ``old'' galaxies, with hardly any young population, over the last 6-8 Gigayears.
}
\titlerunning{The galaxy luminosity function from COMBO-17}
\authorrunning{Wolf et al.}
\maketitle

\section{Introduction}

The formation and subsequent evolution of galaxies are determined both by the
overall gravitational growth of structure and by the physics of gas cooling, star 
formation and feed-back which determine the successive conversion of gas into stars. 
Hierarchical structure formation within cold dark matter scenarios and their various 
extension to address star formation, now provide a comprehensive, but parameterized, 
framework for galaxy formation (e.g. Cole et al. 2000).

The onset of galaxy formation seems to take place at such 
high redshifts that it has so far escaped direct observation. Instead, most 
observations of galaxies have so far concentrated on obtaining large samples at
lower, more easily accessible redshifts.

At the high redshift end, pioneering galaxy survey work has meanwhile
reached redshifts $z\sim 5$ (e.g. Ouchi etal. 2002). In the local universe,
two large surveys are currently characterizing in detail the luminosity function 
of the galaxy population, based on large samples with $>10^5$ 
objects, the 2dF Galaxy Redshift Survey (2dFGRS, e.g. Madgwick et al. 2002) 
and the Sloan Digital Sky Survey (SDSS, e.g. Blanton et al. 2001). 

During the past ten years several studies have aimed to map out the evolution of
the luminosity function and the total luminosity density from the local universe 
to a redshift of $\sim 1$ \cite{Lilly95,Mad98,Lin99,Fri01}. But samples sizes
well in excess of a few thousand objects are only becoming available now
or in the near future, e.g. with 
the 17-colour survey COMBO-17 presented here and with the large spectroscopic
campaigns DEEP \cite{Koo01,Im02} and VIRMOS \cite{LeF01}. The scientific inferences
from existing, faint surveys out to $z \ga 1$ have been limited mainly by 
their sample sizes, aggravated by the strong influence of large-scale structure when
observing small co-moving volumes. The present survey, and other ongoing initiatives,
aim at improving the measurement of the luminosity function by smoothing over 
structure and increasing the volume. 

The COMBO-17 project (``Classifying Objects by Medium-Band Observations in 17 
Filters'') was designed to provide a sample of $\sim$50,000 galaxies and 
$\la$1,000 quasars with rather precise photometric redshifts based on 17 colours. 
In practice, such a filter set provides a redshift accuracy of $\sigma_\mathrm{z,gal} 
\approx 0.03$, $\sigma_\mathrm{z,QSO} \la 0.1$, smoothing the true redshift distribution 
of the sample only slightly and allowing the derivation of luminosity functions.

The foremost data analysis goal of the COMBO-17 approach is to convert the photometric
observations into a very-low-resolution {\it spectrum} that allows simultaneously a 
reliable spectral classification of stars, galaxies of different types and QSOs as well 
as an accurate redshift (or SED) estimation for the latter two. The full survey 
catalogue should contain about 75,000 objects with classifications and redshifts 
on 1.5~$\sq\degr$ of area. This {\it fuzzy spectroscopy} consciously compromises 
on redshift accuracy ($\sigma_z\approx 0.03$) in order to 
obtain very large samples of galaxies with a reasonable observational effort.
While both characteristics are well suited for the analysis of an evolving
population, they understandably do not permit dynamical or chemical studies
which require quite detailed spectroscopic information.

While the photometric redshift technique has already been applied to galaxy 
samples about 40 years ago \cite{Baum63,But83}, we have optimized the
technique by increasing the number of filters and narrowing their bandwidth to
obtain better spectral resolution and more spectral bins. Therefore, COMBO-17
also provides identifications and reasonably accurate redshifts for quasars (see 
also Koo 1999 for a nice overview on photometric redshifts and SubbaRao et al. 
1996 on applying the technique in the context of luminosity functions).  

The goal of the present paper is the use of COMBO-17 redshifts and SEDs for 
25,000 galaxies over 0.78~$\sq\degr$ to draw up a detailed, empirical picture of
how the population of galaxies evolved over the last half of the universe's age.

Our paper is organized as follows: in Sect.~2 we present the observations that
have led to the current sample of $\sim 25$,000 galaxies. Our techniques for 
obtaining their redshifts, SED classification, luminosities and completeness 
are described in Sect.~3. The resulting sample properties are discussed in Sect.~4.
In Sect.~5 we derive our ``quasi-local'' luminosity function, 
drawn from the redshift interval of $z=[0.2,0.4]$ and compare it with the results
from more local samples obtained by the 2dF Galaxy Redshift Survey and the Sloan 
Digital Sky Survey. Finally, we show the evolution of the luminosity function and 
the luminosity density out to $z<1.2$.

\section{Observations - the COMBO-17 survey}

\begin{figure*}
\centerline{\hbox{
\psfig{figure=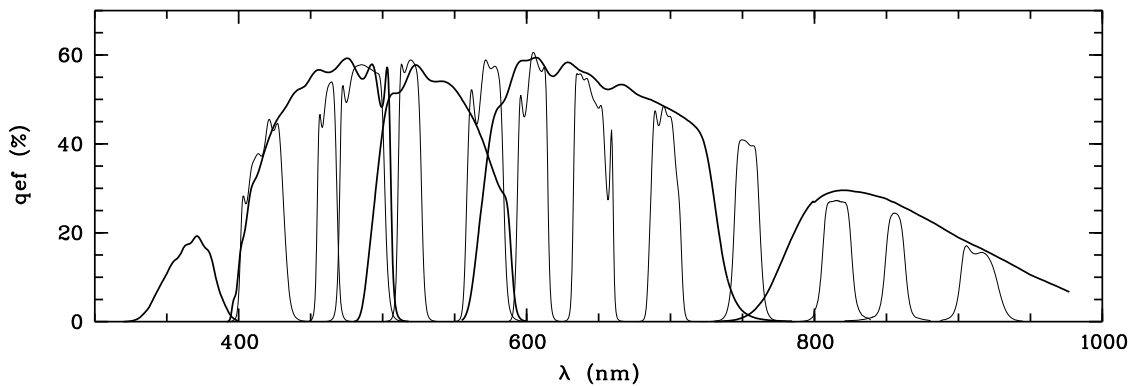,angle=270,clip=t,width=15cm}}}
\caption[ ]{COMBO-17 filter set: Total system efficiencies are shown in the 
COMBO-17 passbands, including two telecope mirrors, WFI instrument, CCD 
detector and average La Silla atmosphere. Combining all observations provides a 
low-resolution spectrum for all objects in the field. Photometric calibrations
of such ``multi''-colour datasets are best achieved with spectrophotometric
standards inside the target fields. \label{qeff}}
\end{figure*}

The COMBO-17 survey has produced multi-colour data in 17 optical filters on 
1~$\sq\degr$ of sky at high galactic latitudes, including to date the Chandra
Deep Field South (CDFS) and the field of the supercluster Abell 901/902. The 
filter set (Fig.~\ref{qeff} and Tab.~\ref{filterset}) contains five broad-band 
filters (UBVRI) and 12 medium-band filters stretching from 400 to 930~nm in 
wavelength coverage. 

All observations presented here were obtained with the Wide Field Imager (WFI, 
Baade et al., 1998, 1999) at the MPG/ESO 2.2-m telescope on La Silla, Chile. 
They encompass a total exposure time of $\sim$160~ksec per field including a
$\sim$20~ksec exposure in the R-band with seeing below 0\farcs8. The WFI provides
a field of view of $34\arcmin \times 33\arcmin$ on a CCD mosaic consisting
of eight 2k $\times$ 4k CCDs with $\sim67$ million pixels providing a scale of 
$0\farcs238$/pixel. The observations started in the commissioning phase of the 
WFI in January 1999 and are continuing as the area is extended to cover more fields.

The instrument design and the survey concept have been matched to the
requirements of deep extragalactic surveys. The morphological and 
spectral dataset of COMBO-17 is primarily intended for studies of
(i) gravitational lensing and (ii) evolution of galaxies and QSOs. Indeed,
the optics of the instrument have been designed with the prime application
of lensing in mind, while the filter set was tailored for the task of
object classification and redshift estimation.  

Observations and data analysis have been completed for three fields (see 
Table\,\ref{fields}) covering an area of 0.78~$\sq\degr$ and providing a 
catalogue of $\sim$200,000 objects found by SExtractor \cite{BA96} on deep, 
high-resolution R-band images with 5$\sigma$ point source limits of $R\approx 26$. 

These deep R-band images provide very sensitive surface brightness limits and 
allow to establish the total object photometry using the SExtractor measurement 
MAG-AUTO. Except for L-stars and quasars at $z>5$, they provide the highest 
signal-to-noise ratio for object detection and position measurement among all 
data available in the survey.

The spectral shapes of the objects in the R-band selected catalogue were 
measured with a different approach. Photometry was obtained in 17 different 
passbands by projecting the object coordinates into the frames of reference 
of each single exposure and measuring the object fluxes at the given locations.
In order to optimize the signal-to-noise ratio, we measure the spectral shape
in the high surface brightness regions of the objects and ignore potential low
surface brightness features at large distance from the center.
 
Since seeing variations among the different bands would introduce artificial 
colour offsets by changing observing conditions typical for ground-based 
observations, we need a non-standard photometry approach to measure spectral
shapes accurately. In fact, we need to measure the same central fraction of 
an object in every band as it would appear in equal seeing. To this end, we
employ a seeing-adaptive, weighted aperture photometry as performed by the 
package MPIAPHOT \cite{RM91,Mei02}.  

MPIAPHOT measures the central surface brightness of objects after convolving 
their appearance outside the atmosphere to an effective PSF of $1\farcs5$ 
diameter. In detail, the procedure measures the observed stellar PSF on each 
individual frame and chooses the necessary Gaussian smoothing for reaching a
common effective PSF of $1\farcs5$ uniformly on all frames in all bands. For
most objects this measurement is similar to a flux measurement in an aperture 
of $\sim2\arcsec$ diameter in $1\farcs5$ seeing.

The photometric calibration is based on a system of faint standard stars in 
the COMBO-17 fields, which we established by spectrophotometric calibration
with respect to spectrophotometric standard stars in photometric nights. Our
standards were selected from the Hamburg/ESO survey database \cite{Wis00} of 
digital objective prism spectra (see Wolf et al., 2001b, for procedure).
By having standard stars within each survey exposure, we were independent 
from photometric conditions for imaging. 

In summary, all luminosities used in the paper are based on SExtractor MAG-AUTO
measurements on the deep R-band stack, while all redshift and SED fits are 
based on seeing-adjusted aperture measurements (with MPIAPHOT) across all 
bandpasses. For more details of the data reduction, we like to refer the reader 
to a forthcoming technical survey paper (Wolf et al. {\it in prep.}). In this
paper, all magnitudes are cited with reference to Vega as a zero point.

\begin{table}
\caption{The COMBO-17 filter set: Exposure times and 10$\sigma$ magnitude limits 
reached for point sources, averaged over all three fields. The $R$-band 
observations were selected to be taken under the best seeing conditions (FWHM 
$0\farcs55\ldots0\farcs8$). \label{filterset}}
\begin{tabular}{llrc}
\multicolumn{2}{l}{$\lambda_\mathrm{\mathrm{cen}}$/fwhm (nm)}  & $t_\mathrm{\mathrm{exp}}$/sec & 
 $m_\mathrm{\mathrm{lim},10\sigma}$ \\ 
\noalign{\smallskip} \hline \noalign{\smallskip} 
364/38 &  $U$ &  20000 & 23.7 \\ 
456/99 &  $B$ &  14000 & 25.5 \\ 
540/89 &  $V$ &   6000 & 24.4  \\
652/162 & $R$ &  20000 & 25.2 \\ 
850/150 & $I$ &   7500 & 23.0 \\
\noalign{\smallskip} 
420/30  &      &  8000 & 24.0 \\ 
462/14  &      & 10000 & 24.0 \\ 
485/31  &      &  5000 & 23.8 \\ 
518/16  &      &  6000 & 23.6 \\ 
571/25  &      &  4000 & 23.4 \\ 
604/21  &      &  5000 & 23.4 \\ 
646/27  &      &  4500 & 22.7 \\
696/20  &      &  6000 & 22.8 \\ 
753/18  &      &  8000 & 22.5 \\ 
815/20  &      & 20000 & 22.8 \\ 
856/14  &      & 15000 & 21.8 \\ 
914/27  &      & 15000 & 22.0 \\
\noalign{\smallskip} \hline
\end{tabular}
\end{table}

\begin{table}
\caption{Positions and galactic reddening (Schlegel et al. 1998) for the 
three COMBO-17 fields analysed. All observations were obtained 
at the Wide Field Imager at the MPG/ESO 2.2\,m-telescope at La Silla. 
\label{fields} }
\begin{tabular}{llllll}
Field   & $\alpha_\mathrm{\mathrm{J2000}}$ & $\delta_\mathrm{\mathrm{J2000}}$ & 
        $l_\mathrm{\mathrm{gal}}$ & $b_\mathrm{\mathrm{gal}}$ & $E_\mathrm{B-V}$ \\ 
\noalign{\smallskip} \hline \noalign{\smallskip} 
CDFS    & $03^{\mathrm{h}} 32^{\mathrm{m}} 25^{\mathrm{s}}$ & $-27\degr 48' 50''$ &
        $223\fdg 6$ & $-54\fdg 5$ & 0.01 \\ 
A 901   & $09^{\mathrm{h}} 56^{\mathrm{m}} 17^{\mathrm{s}}$ & $-10\degr 01' 25''$ & 
        $248\fdg 0$ & $+33\fdg 6$ & 0.06 \\ 
S 11    & $11^{\mathrm{h}} 42^{\mathrm{m}} 58^{\mathrm{s}}$ & $-01\degr 42' 50''$ & 
        $270\fdg 5$ & $+56\fdg 8$ & 0.02 \\
\noalign{\smallskip} \hline
\end{tabular}
\end{table}

\section{The galaxy catalogue}

The galaxy catalogue is extracted from the full survey catalogue purely on 
the basis of spectral information. There are no morphological criteria used 
to differentiate between stars, galaxies and quasars. Indeed, many faint 
galaxies appear compact in typical ground-based seeing, 
while binary stars can produce objects with stellar spectra but 
extended appearance. Therefore, the abundance of photometric information 
provides a safer separation between the object classes if analysed 
with a classification technique as presented in the following section.

\subsection{Classification and redshift estimation}

The photometric measurements from 17 filters provide a {\it low-resolution 
spectrum} for each object to be analysed by a statistical technique for 
classification and redshift estimation based on spectral templates \cite{WMR01}. 
This approach has already been applied to the Calar Alto Deep Imaging Survey 
(CADIS) and to a mass determination of the galaxy cluster Abell 1689 
\cite{Wolf01a,Dye00}. 

Since these initial analyses, we have improved the galaxy templates in the 
restframe UV region where the original Kinney templates E, S0, Sa and Sb show 
fairly noisy patches in the wavelength interval of $\lambda = 290\ldots 
330$~nm. These patches have been replaced by spectra obtained with the stellar
population synthesis code PEGASE \cite{FRV97} 
by first matching these to the Kinney templates. For the purpose of 
efficiency, we have also changed the redshift axis in the grid by making it 
equidistant on a $log(1+z)$ axis rather than on a linear $z$ axis. We have not 
yet incorporated trustworthy template information bluewards of the Lyman-alpha 
line, and still restrict the redshift range such that the existing templates 
always cover our entire filter set. This constrains the investigated redshifts 
to $z<1.55$ for now and leads to a deliberate exclusion of higher redshift
objects from the catalogue. While they are not within the scope of this paper, 
some high-redshift galaxies, e.g. Lyman-break objects at $z\sim 3\ldots 4$, 
could be mistaken by the redshift estimation to reside at low-redshift and 
contaminate the sample to a very small degree at the faintest end.

The quasar library has also been improved by deriving it from the more modern 
SDSS QSO template spectrum \cite{Bud01} rather than from the Francis et al. (1991)
emission line contour. This will make a difference mainly for a better detection 
of low-redshift quasars and hardly affect the rich class of galaxies in a 
statistical sense. Likewise the stellar library has been improved by omitting 
stars of spectral type O, B and A from the Pickles (1998) atlas, which we do 
not expect to see in the Galactic halo anyway, while template 
spectra of white dwarfs, subdwarfs and blue horizontal branch stars have been 
included. For all details we like to refer the reader to our forthcoming paper 
on the accuracy of the classification and redshift estimation \cite{Wolf02}.

The galaxy catalogue of concern in this paper is based on the observed average 
templates from Kinney et al. (1996), except for the modifications mentioned
above. These ten templates cover typical local 
galaxy SEDs from elliptical galaxies to starbursts (see Wolf, Meisenheimer \& 
R\"oser, 2001, for details). Altogether, they probably encompass the widest 
range of average ages possible for stellar populations in galaxies, but they 
explicitly do not contain templates for deeply dust-enshrouded starbursts as 
they may be part of the ERO galaxy population believed to reside at $z=1\ldots2$. 
If galaxies with restframe SEDs similar to that of dust-enshrouded EROs were 
contained in this catalogue, they would not be identified as such, but rather 
with the best-fitting SED type among the ten Kinney templates, i.e. as likely 
old populations typical of local ellipticals. However, from an analysis of the
CADIS galaxy sample we know that EROs are not sufficiently abundant to change 
the conclusions of our study \cite{Tho99}.

Our classification and redshift estimation provides full probabilities for 
every object class based on all photometric measurements, rather than merely 
searching for a single best-fitting template. We therefore assign to an object 
the class which yields the largest sum over all probabilities for each spectral 
fit within that class. This prevents unreliable assignments of any class which 
may have a single spurious highly probable template, but otherwise fits very 
poorly over all other templates. While such a simple $\chi^2$-minimisation is 
a powerful technique when plenty of spectral information is available (e.g. 500 
channels), the analysis of low-dimensional colour space benefits significantly 
from a wholesome probability-based approach which handles ambiguities better 
(which is even more relevant in broad-band surveys with only five filters).

While discriminative power is an important concern in every classification 
problem, completeness is no less important to avoid missing relevant fractions
of a population. Therefore, it is reassuring, that less than 1\% of the objects 
observed in COMBO-17 appear to have peculiar spectra which do not resemble any 
template, but instead are outliers at more than a 3~$\sigma$ level
of significance. Among these are interesting individual objects, but 
also a few blended objects and stars with uncorrected short-term variability 
skewing the observed spectra beyond our control.

\subsection{Galaxy spectral types}

\begin{table}
\caption{Galaxy spectral types, their corresponding template range in the
sequence of Kinney et al. (1996), and COMBO-17 sample sizes.  
\label{typedef} }
\begin{tabular}{clcrr}
TYPE & Template range   & SED type	& objects at	& objects at \\
     & (Kinney et al.)  & range  	& $z=$0.2-0.4	& $z=$0.2-1.2 \\ 
\noalign{\smallskip} \hline \noalign{\smallskip} 
1    & E--Sa		&  0-30 	&  344	&  1365 \\ 
2    & Sa--Sbc		& 30-55 	&  986	&  5489 \\ 
3    & Sbc--SB6		& 55-75 	& 1398	&  7520 \\
4    & SB6--SB1		& 75-99 	& 2946	& 11057 \\
\noalign{\smallskip} \hline \noalign{\smallskip} 
all  & E--SB1		&  0-99		& 5674	& 25431 \\
\noalign{\smallskip} \hline
\end{tabular}
\end{table}

\begin{figure}
\centerline{\hbox{
\psfig{figure=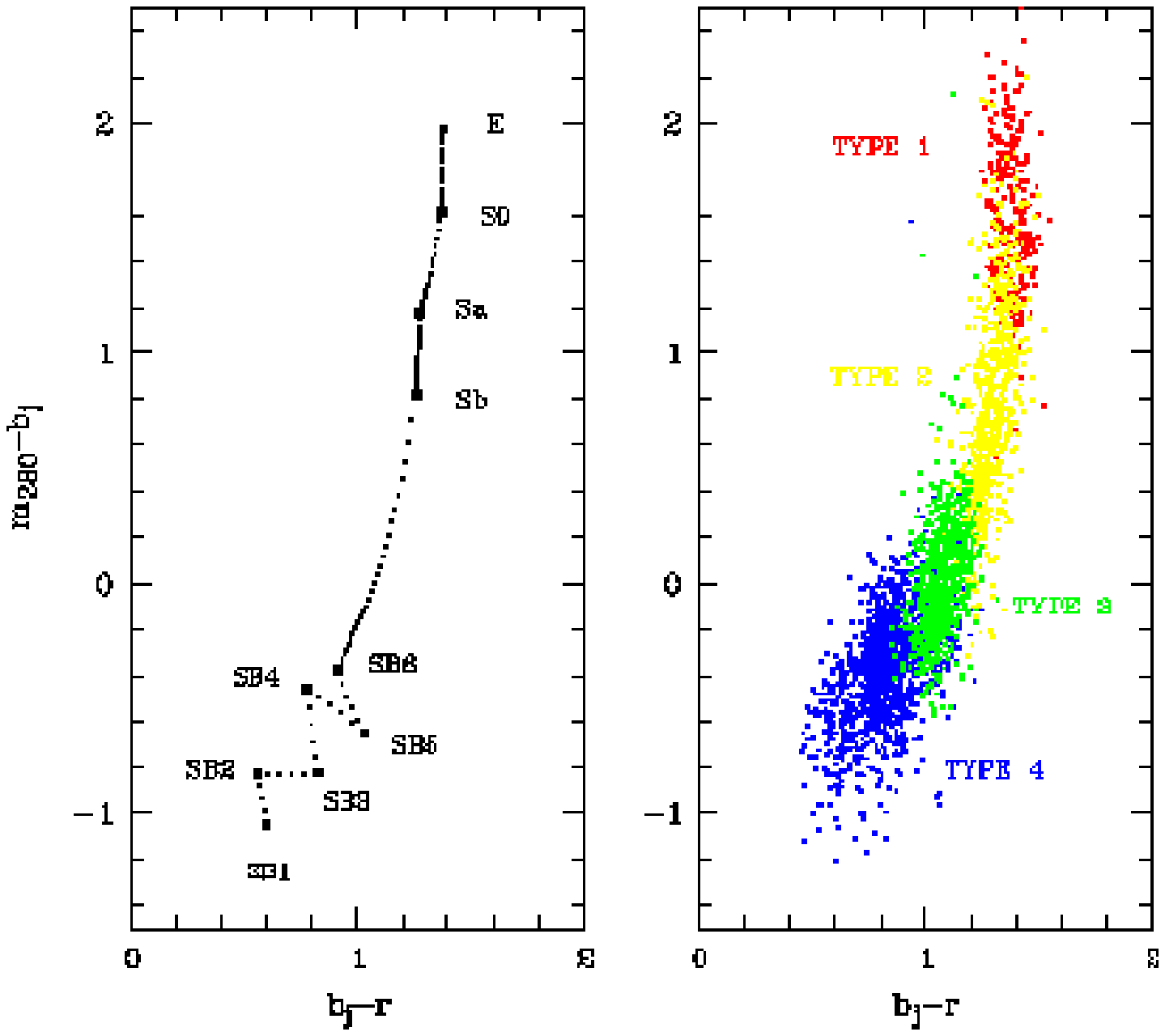,angle=270,clip=t,width=8.7cm}}}
\caption[ ]{Restframe colour of spectral templates and type definition: \\ 
{\it Left-hand side:} Restframe colours calculated for the sequence of 
galaxy templates along the entire SED parameter range. Shown are the restframe 
colours $(m_\mathrm{280}-B)$ vs. $(B-r)$. The ten original 
Kinney templates are large labelled dots. The template sequence was mapped onto 
a grid of 100 templates by linear interpolation in flux space. \\
{\it Right-hand side:} Restframe colours measured for the galaxy sample at $R<23$ 
and $z=[0.2,0.4]$ as obtained from the 17-filter spectra (see Sect.~\ref{restlum}). 
The four different SED types used in this paper are rendered in different colours. 
Since the SED types are determined from 17 passbands, they do not correspond to 
precisely defined intervals of any single colour axis. 
\label{restcol_UBR}}
\end{figure}

The just described photometric data analysis of COMBO-17 provides luminosity and 
restframe spectral type as the main observables of galaxies. As a consequence, the 
evolution of their luminosity function can be investigated in sub-samples split by 
restframe spectral types. Note that, in contrast, types of morphology or explicit 
star formation history are not subject of the discussion presented here.

There is no unique definition of SED types along a spectral parameter axis, since 
galaxies cover a continuum of parameter values. This fact not only applies to our 
own parameter definition which is based on the grid of templates from Kinney et al. 
(1996), but is found in every definition of spectral type, whether it is based on 
restframe continuum colour as in the SDSS analysis or derived from a principal 
component analysis (PCA) of the spectra themselves as in the 2dFGRS, an equivalent 
width of H$\alpha$-emission, or a mean age of a stellar population. We illustrate 
our choices and definitions in Fig.~\ref{restcol_UBR} and Tab.~\ref{typedef}.

Having obtained very-low-resolution spectra from 17 passbands in COMBO-17 it is 
only natural to use all available information for the determination of spectral 
types, which is similar to the approach taken by the 2dFGRS. In fact, our 
classification procedure is based on a single-parameter set of templates and 
provides SED parameter estimates just as it provides redshift estimates.
\begin{figure}

\centerline{\hbox{
\psfig{figure=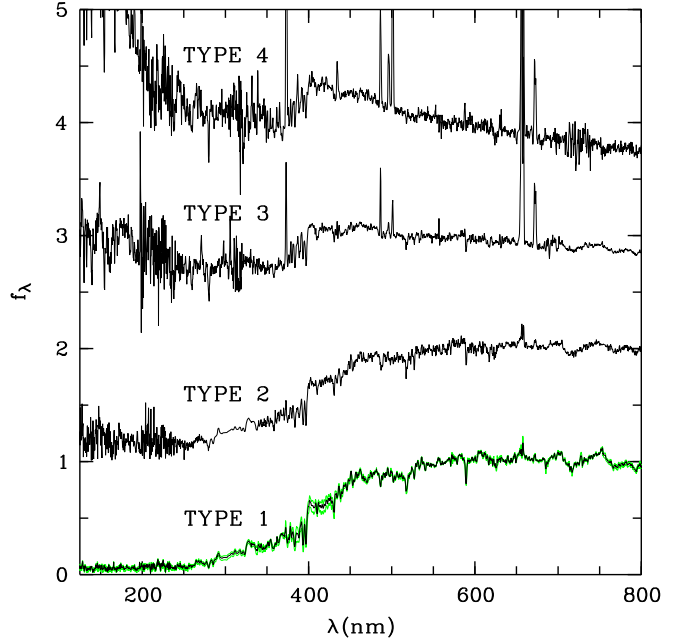,angle=270,clip=t,width=8.7cm}}}
\caption[ ]{Average templates for galaxy types: The average restframe templates of
the four SED types at $0.2<z<1.2$ are plotted with an offset of one flux unit each.
The average template in \tp 1 becomes slightly redder from $z\sim1.1$ to $z\sim0.3$,
but we see no change in the other types. The high and low redshift versions of the
\tp 1 template are plotted in grey, and the mean over all redshifts is shown in black.
\label{SEDtemps}}
\end{figure}

\begin{figure*}
\centerline{\hbox{
\psfig{figure=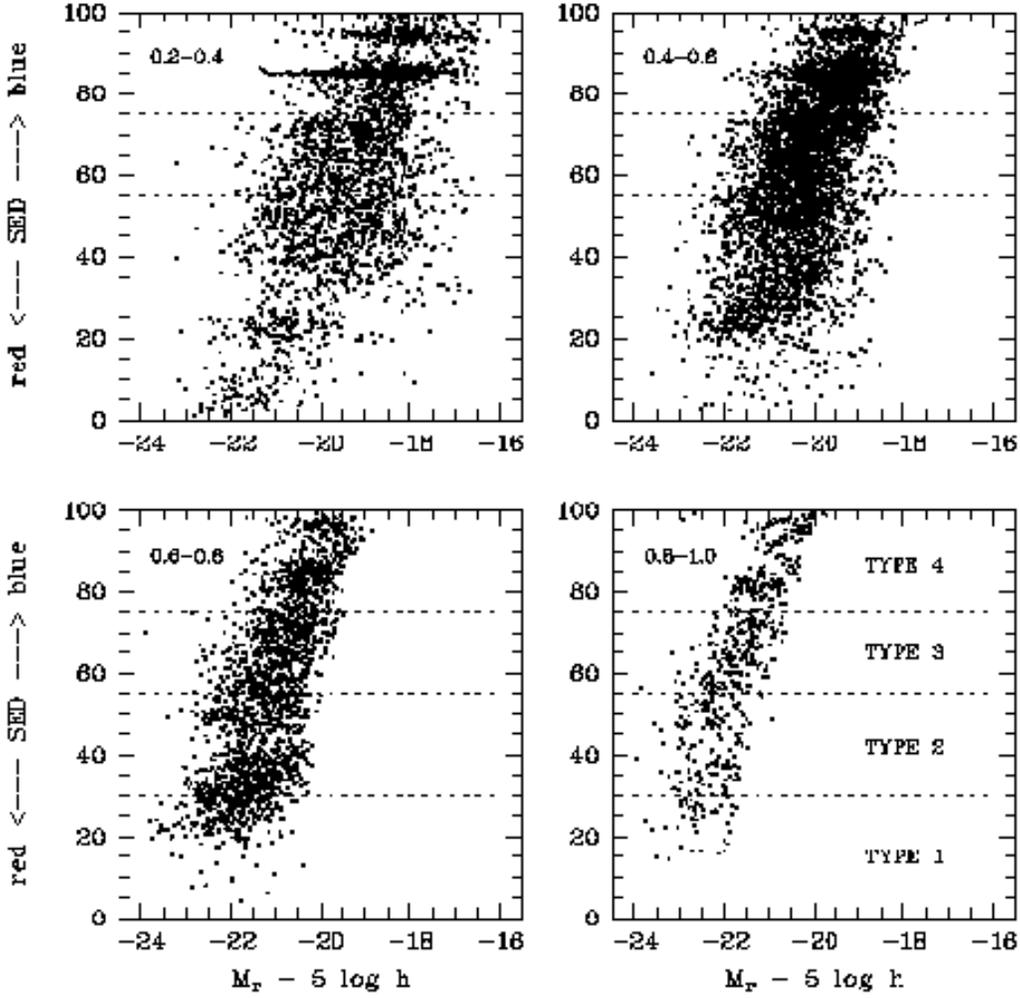,angle=270,clip=t,width=14cm}}}
\caption[ ]{The galaxy sample (here $R<23$) split by redshift interval: Shown is 
the spectral type parameter SED over the luminosity $M_\mathrm{r}$ in the restframe 
SDSS r-band for $(\Omega_m,\Omega_\Lambda) = (0.3,0.7)$. The intervals occupied 
by the four spectral types used in this paper are delineated by dashed lines. 
At all redshifts a clear trend between luminosity and SED is demonstrated showing 
the well-known fact that more luminous galaxies are redder on average. The dense
horizontal concentration around the SED 85 originates from the fact, that the
template library is not entirely monotonic (see relative position of templates
SB4 and SB5 in Fig.~\ref{restcol_UBR}).  
\label{sampleredshift}}
\end{figure*}

The range of SED parameter values just maps the Kinney et al. (1996) templates
onto a sequence of values from 0 (E galaxy template) to 99 (SB1 starburst template),
as in Wolf, Meisenheimer \& R\"oser (2001). The distribution of SED values in the
sample shows no obvious structures suggesting particular bins (see 
Fig.~\ref{sampleredshift}). For the purpose of this paper, we decided to use a set 
of four SED types, derived by eye inspection of the evolutionary patterns observed, 
aiming at producing a clear picture of differential behaviour between the types 
(see Table~\ref{typedef} for the relation). Fig.~\ref{SEDtemps} shows the mean
templates representing the four types of galaxies at $0.2<z<1.2$.

The relationship between these types and the restframe colour indices formed among 
the passpands in our luminosity definition can be seen in Fig.~\ref{restcol_UBR}. 
It shows that our template sequence (left panel) covers the colour distribution of 
the observed galaxies (right panel). It also illustrates that the SED classification 
(based on 17 filters) reflects a single-parameter sequence, but does not correspond 
to precisely defined limits on a single colour axis, due to (a) the presence of 
noise, (b) the not entirely monotonic behaviour of the template sequence, and (c) 
the fact that the SED parameters of individual objects are determined from all 17 
passbands and not from a single colour index.

In comparison, the 2dFGRS analysis by Madgwick et al. (2002) measures a spectral
type from a principal component analysis of all present spectra. An individual 
galaxy is then assigned an $\eta$ value, corresponding to the main parameter 
arising from the PCA. Much like in COMBO-17, information from the {\it whole} 
spectrum was mapped onto a single parameter defining a sequence of spectral types, 
which could presumably be mapped onto our SED axis. The sequence was then split into 
four bins with the first one covering a prominent peak of no-emission-line galaxies 
in the $\eta$-histogram. However, we did not attempt to reproduce their types since
we lack the necessary colour data to perform such a task.

On the contrary, the type definition in the SDSS analysis by Blanton et al. (2001) 
was based on a single colour index. The spread in reconstructed restframe 
$(g-r)$-colour was split into five bins of equal width on the
magnitude scale. This is roughly equivalent to drawing equidistant vertical 
lines into Fig.~\ref{restcol_UBR} for separating the subsamples along their
restframe $(B-r)$-colour. The figure demonstrates that the colour index 
on the vertical axis, $(m_\mathrm{280}-B)$, provides a larger spread and more 
sensitivity to the mean age of the stellar population by enclosing the 
4000~\AA-break. Especially in the presence of noise and scatter around the mean 
colour relation of the sample, type limits defined on the $(B-r)$-axis lead 
to considerable smoothing over types defined along the sequence of the relation.
It will therefore prove difficult to compare directly our results split by type 
with either 2dFGRS or SDSS data. 

Fig.~\ref{restcol_UBR} also highlights a problem with a monotonic interpretation 
of the SED axis among the starburst templates. It seems that the templates SB4 
and SB5 reside at roughly the same location {\it along} the main axis of the 
distribution, which perhaps means physically that their stellar population could 
have a similar mean age. They differ indeed by their location {\it across} the 
main axis, maybe suggesting a second-order feature in the distribution of stellar
ages. As a matter of speculation, the brighter B-band flux combined 
with a fainter $m_\mathrm{280}$ flux might correspond to a post-starburst galaxy 
with less current star formation as in SB5, but a stronger A star population 
boosting the B-band. However, for the purpose of this paper, we will 
not further speculate about the detailed star formation histories of the templates. 
We just note, that the starburst range of the templates do not form a perfectly 
monotonic, one-dimensional sequence!

\begin{figure}
\centerline{\hbox{
\psfig{figure=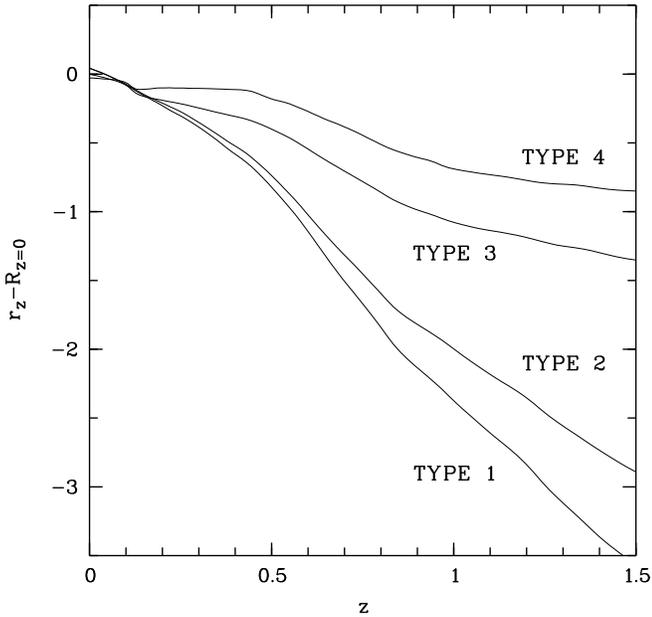,angle=270,clip=t,width=8.7cm}}}
\caption[ ]{Average K-correction for galaxy types: The mean magnitude K-correction
with redshift is shown as a difference between observed frame WFI-R-band and 
restframe SDSS r-band for the average galaxies in the four SED types.
\label{Kcorr}}
\end{figure}

Furthermore, part of the present sample of galaxies is shown split by redshift 
layer in Fig.~\ref{sampleredshift}, depicting the continuous SED type parameter 
over luminosity. In this diagram, the intervals corresponding to \tp 1 to \tp 
4 used in the following analysis are indicated. A dense horizontal feature can
be seen around SED values of 85 (=SB4), which is probably caused by the not quite
monotonic distribution of the starburst templates discussed above, which makes a
proper statistical SED parameter estimate very difficult, given that it is based
on a probability distribution over a locally non-monotonic parameter.

\subsection{Restframe luminosities and sample definition}\label{restlum}

Instead of using generic K-corrections, the restframe luminosity of all galaxies 
are individually measured from their 17-filter spectrum. For each galaxy, three 
restframe passbands are considered, (i) the SDSS r-band, (ii) the Johnson B-band 
and (iii) a synthetic UV continuum band centered at $\lambda_\mathrm{rest} = 
280$~nm with 40~nm FWHM and a rectangular transmission function. This is achieved 
by precisely matching the redshifted template corresponding to the galaxy SED 
classification into the observed multi-colour photometry and integrate its spectrum 
over redshifted versions of the three restframe passbands to derive the flux to be 
observed in them. At $0.2<z<1.2$, this approach allows a reliable measurement of the
luminosity in the B- and 280-band, but does require an extrapolation 
for the r-band at $z\ga0.5$, where it is redshifted beyond our longest wavelength 
filter. The mean K-correction of the average galaxies in each SED type is shown 
in Fig.~\ref{Kcorr} as a function of redshift. Throughout the paper, we use $H_0 = 
h~\times$ 100~km/(s~Mpc) in combination with $(\Omega_m,\Omega_\Lambda)=(0.3,0.7)$.

The sample used for all analyses in this paper is defined by limits in aperture
magnitude, in redshift and in luminosity. It is still affected by incompleteness 
within these limits as discussed in the following section. Objects are selected
to have an aperture magnitude of $17<R<24$, because in the saturation range of
the individual frames as well as in the noisy magnitude regime we can not reliably 
measure spectral shapes and redshifts. They are further selected to have a redshift 
of $0.2<z<1.2$: At low redshift the solid angle covered by our survey is too small 
to obtain useful samples; and at higher redshift we currently have no spectroscopic 
information on our redshift accuracy, not even from the 
CADIS observations of $\sim100$ galaxies. The resulting catalogue contains more
than 25,000 galaxies of which $\sim50$ appear to have luminosities of $M_\mathrm{r}<-24$,
corresponding to 0.2\% of the sample. We assume that these objects are mostly
unreliable measurements, due to (i) Seyfert 1 galaxies contaminating the sample
with potentially entirely wrong redshift estimations, (ii) catastrophic mistakes
in the redshift assignments leading to completely wrong luminosities, (iii) any
photometric artifacts of unknown origin. We therefore restrict the sample to
$M_\mathrm{r}>-24$. For a discussion of the average reliability of redshift 
measurements we like to refer the reader to Sect.~3.5.

\subsection{Completeness correction}\label{completeness}

\begin{figure*}
\vbox{
\centerline{\hbox{
\psfig{figure=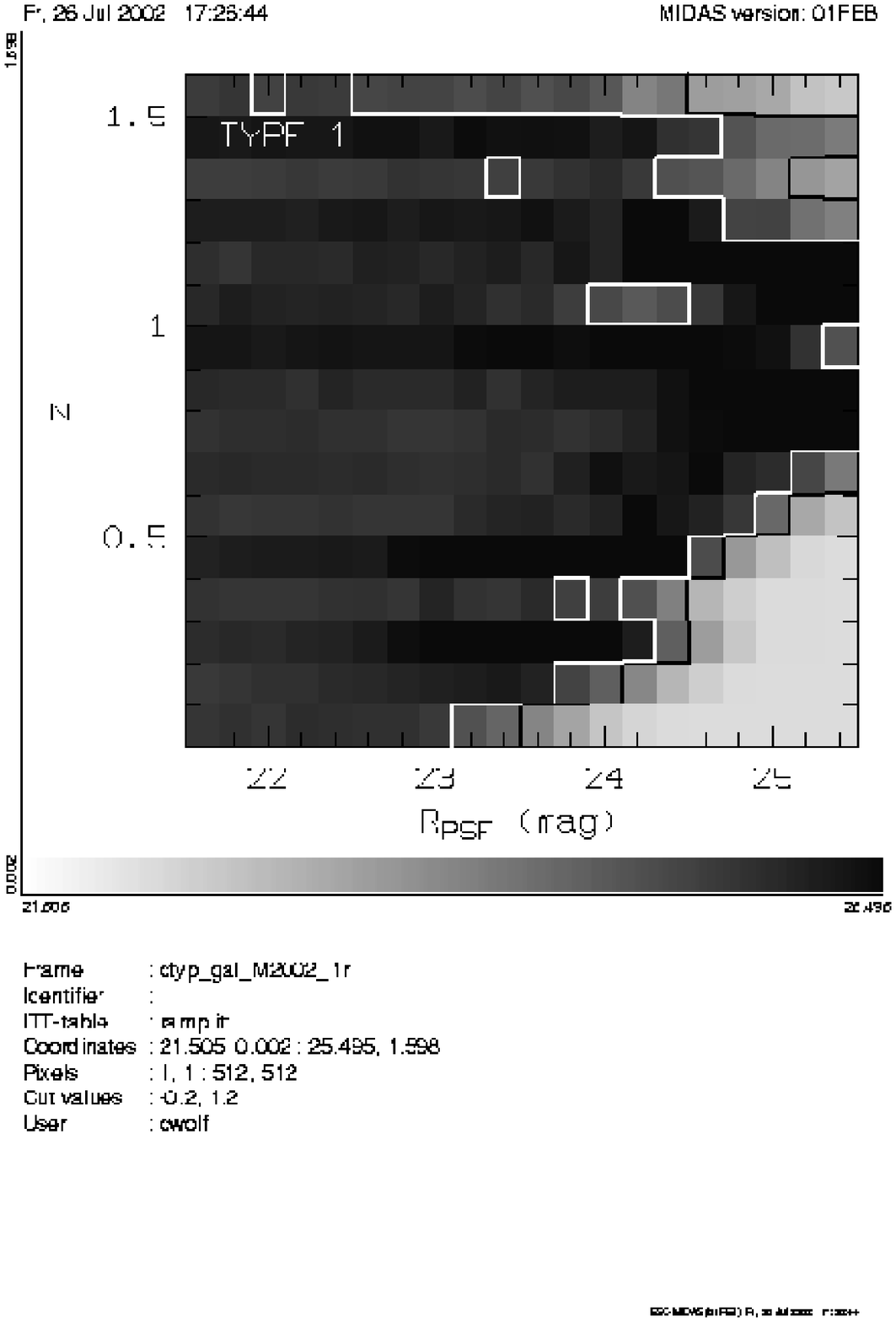,angle=0,clip=t,width=8cm}
\psfig{figure=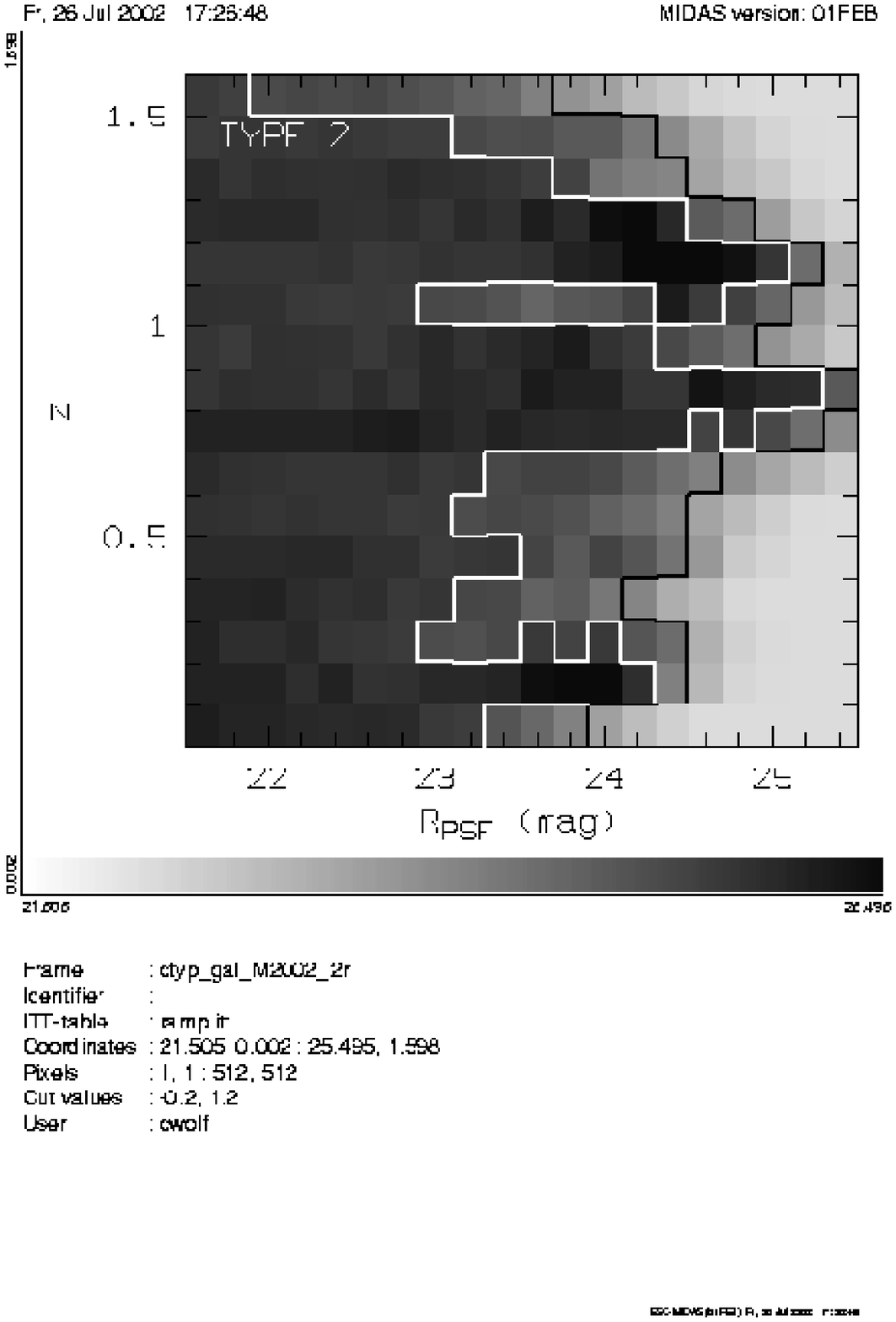,angle=0,clip=t,width=8cm}}}
\centerline{\hbox{
\psfig{figure=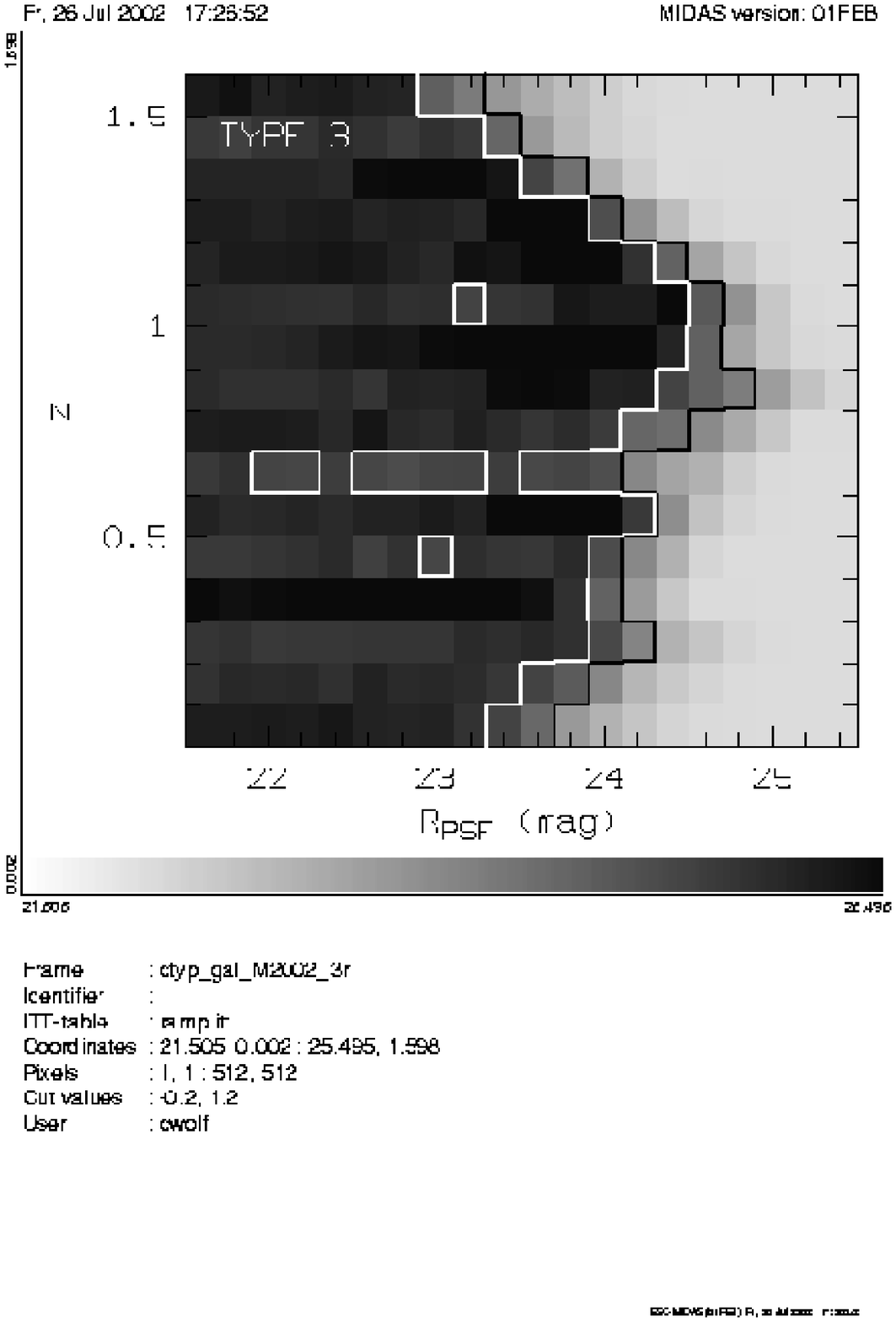,angle=0,clip=t,width=8cm}
\psfig{figure=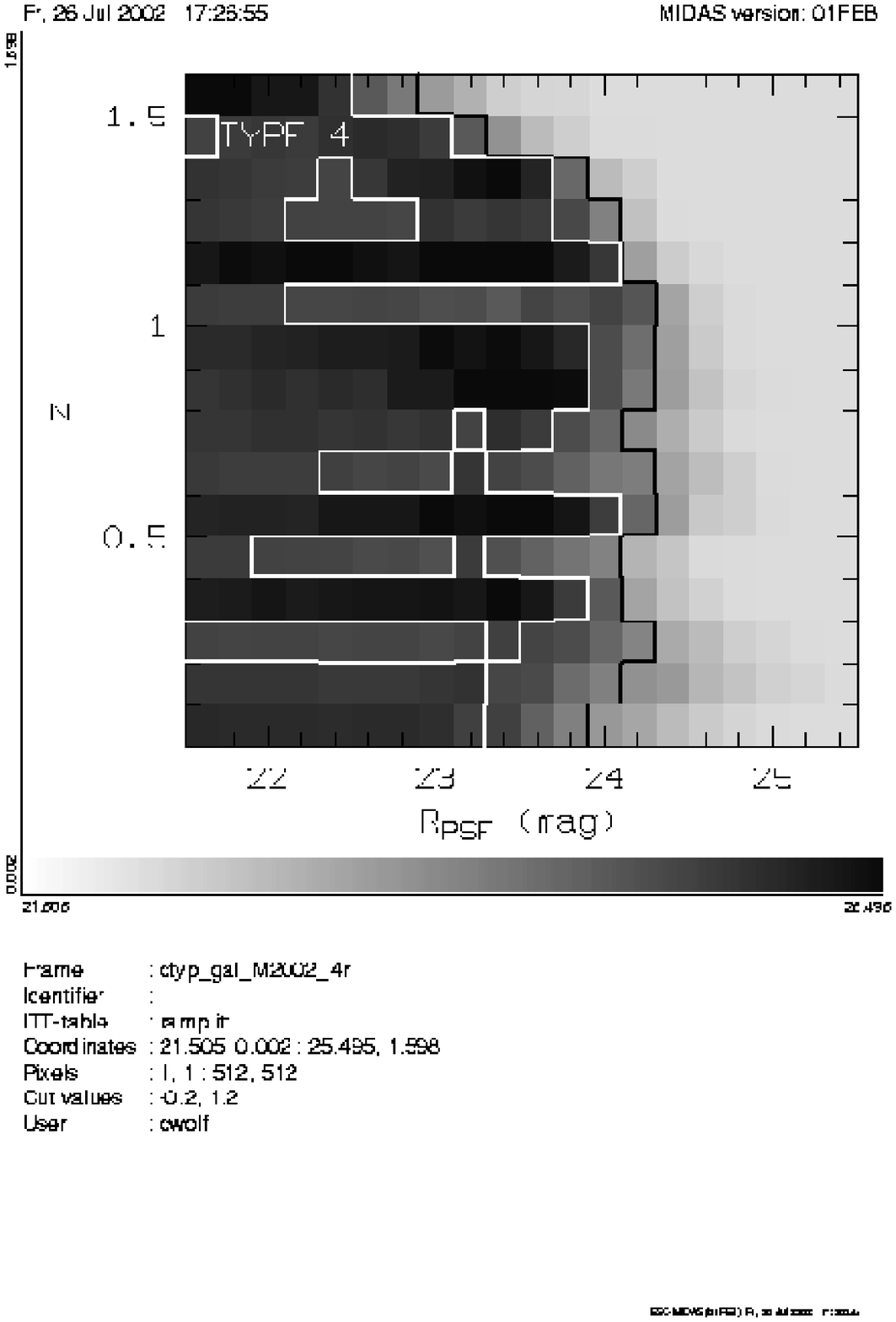,angle=0,clip=t,width=8cm}}}
}
\caption[ ]{Completeness maps for unresolved galaxies: Grey-scale and contour
maps demonstrating how the fraction of galaxies having successful redshift 
measurements depends on magnitude, redshift and spectral type. The greyscale 
shows completeness levels from 0\% (light grey) to 120\% (black). The contour
lines are drawn for 90\% (white) and 50\% completeness (black).
This completeness map corresponds to $C(\mu_\mathrm{central},z,SED)$ as in Eq. (1), 
but plotted over the R-band magnitude of unresolved objects. Values above 100\% 
occur when redshift aliasing creates local overdensities in the observed, 
estimated z/SED-distribution based on a flat underlying simulated distribution. 
These maps are based on Monte-Carlo simulations of the survey. See Sect.~3.4 for
a detailed discussion of the completeness correction.
\label{ctyp_gal}}
\end{figure*}

\begin{figure*}
\vbox{
\centerline{\hbox{
\psfig{figure=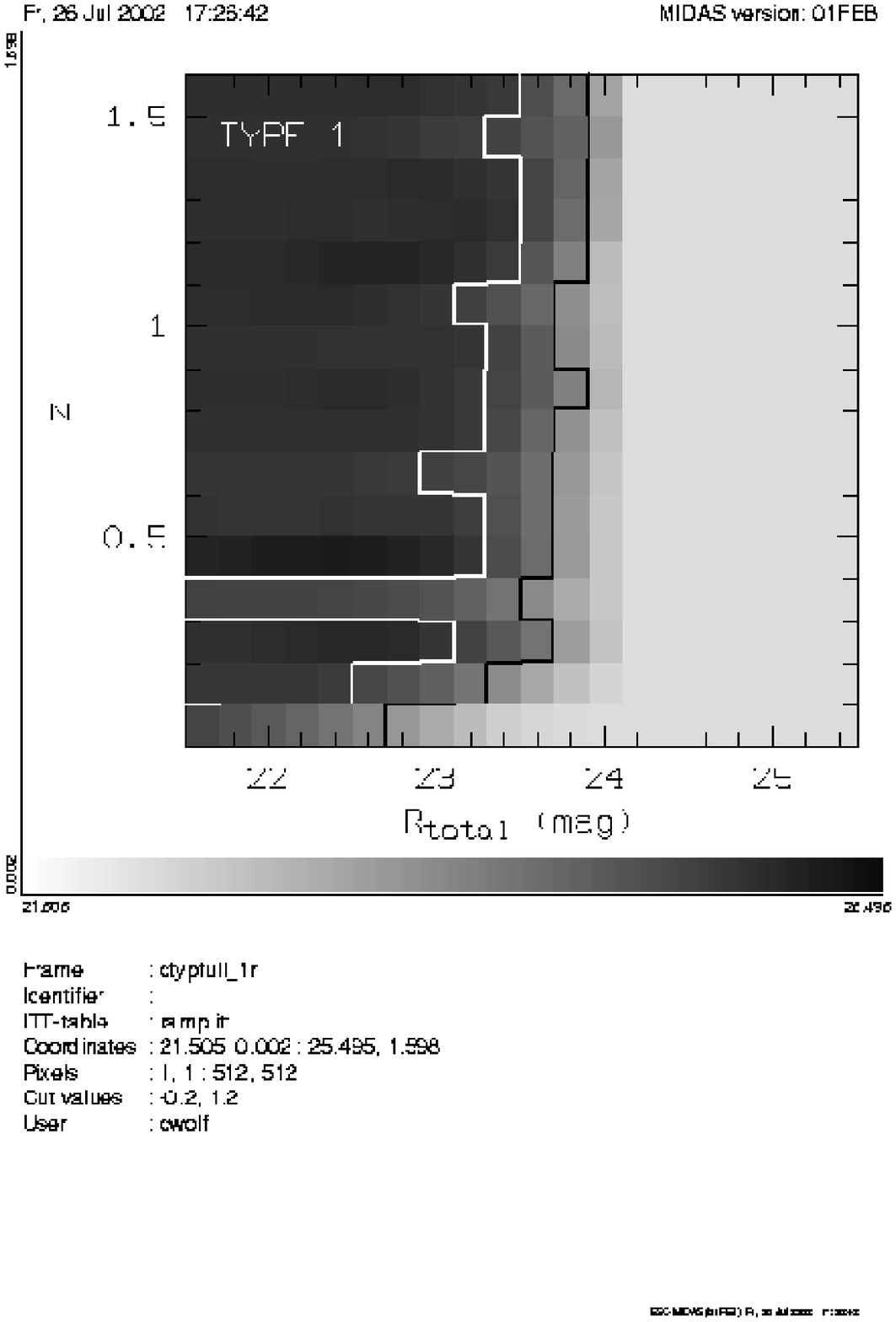,angle=0,clip=t,width=8cm}
\psfig{figure=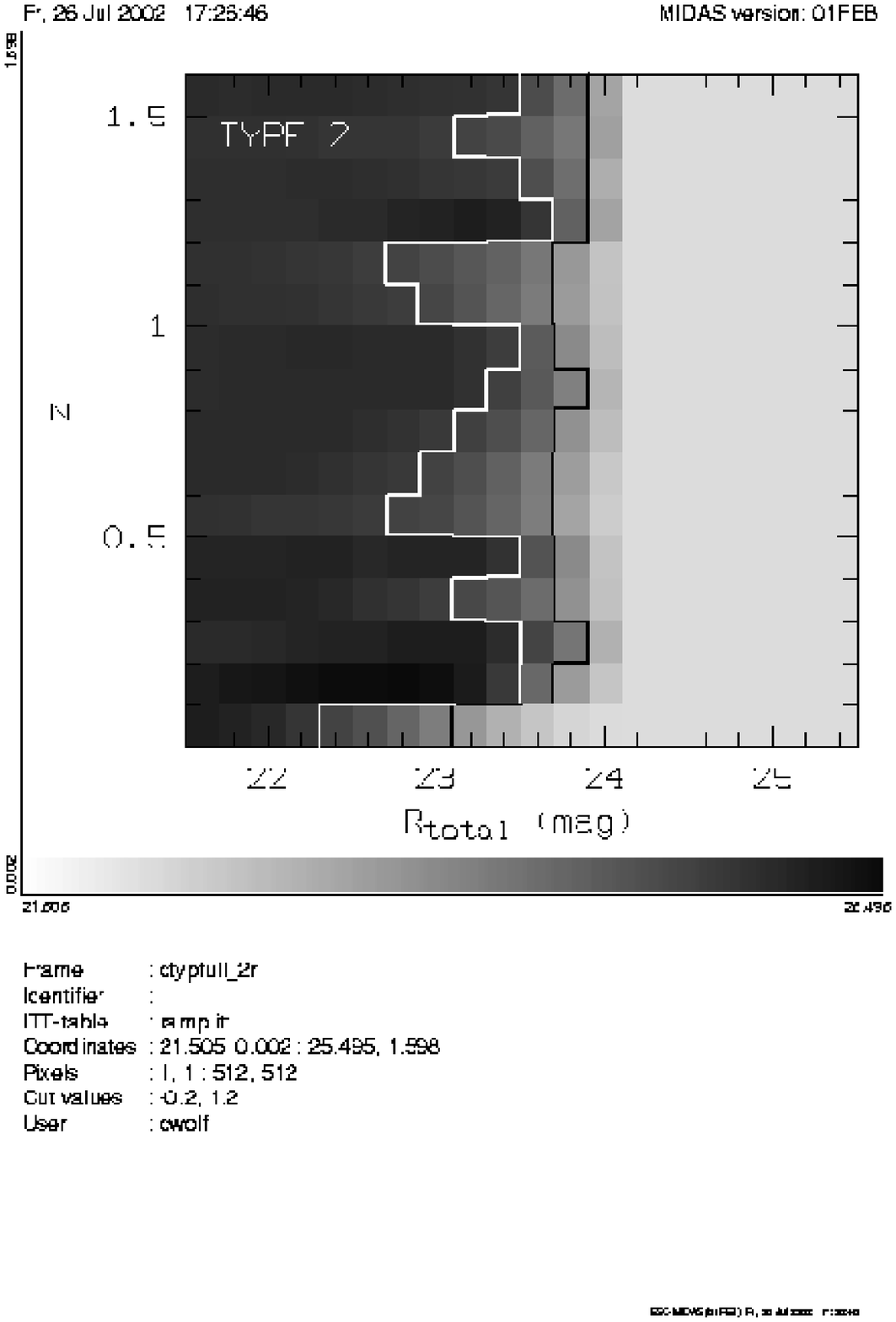,angle=0,clip=t,width=8cm}}}
\centerline{\hbox{
\psfig{figure=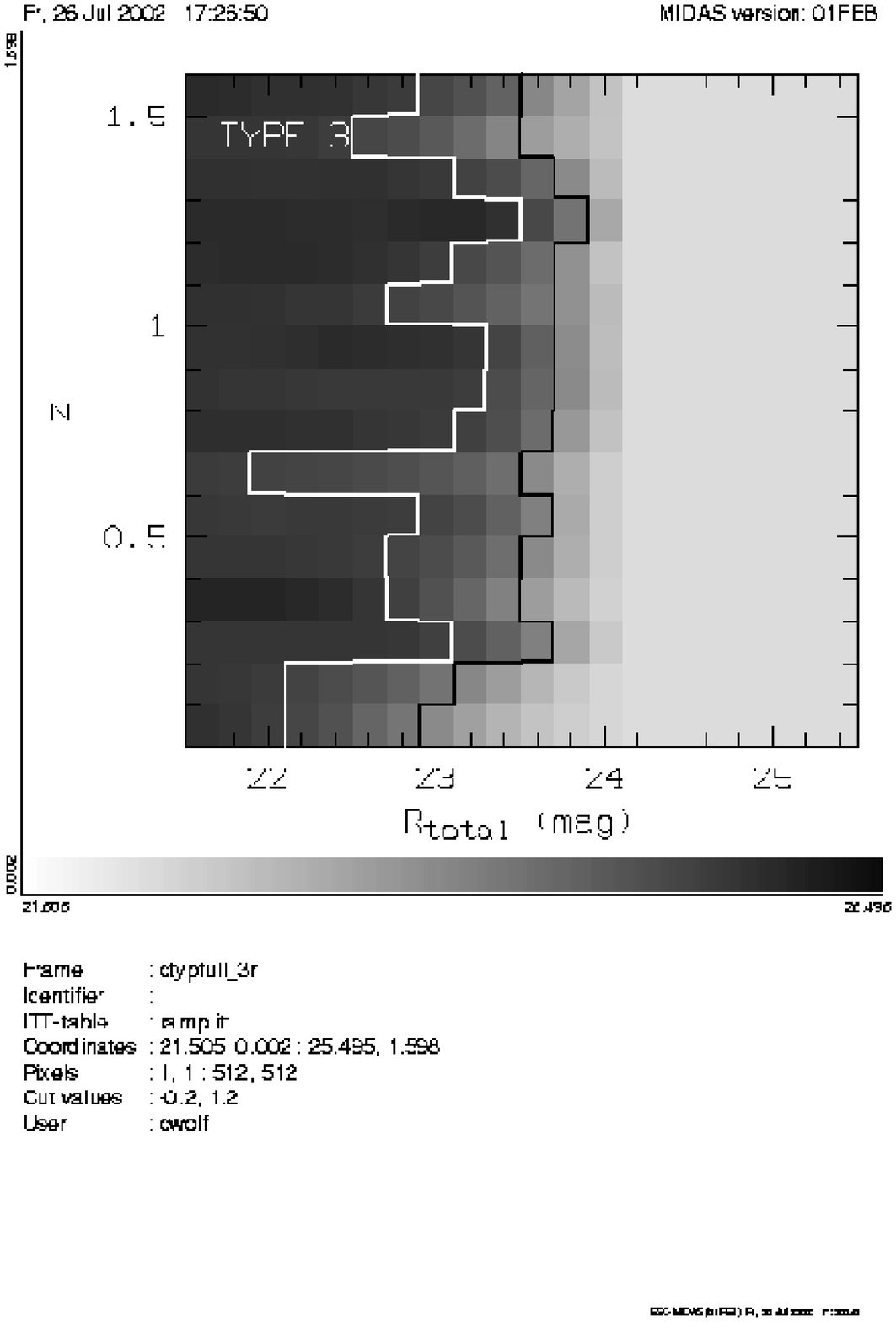,angle=0,clip=t,width=8cm}
\psfig{figure=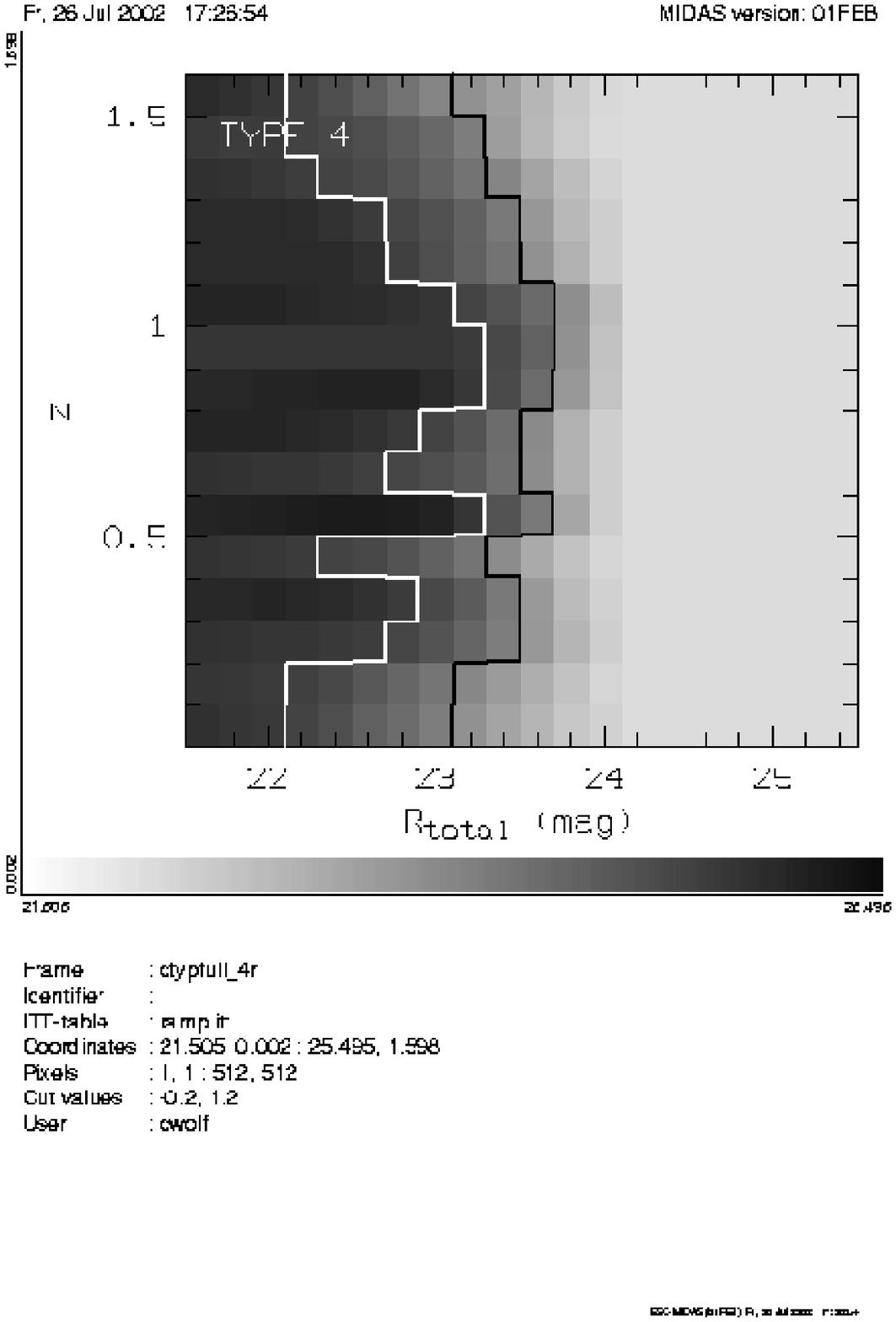,angle=0,clip=t,width=8cm}}}
}
\caption[ ]{Completeness maps for resolved galaxies: These full completeness 
maps for total magnitudes are generated from $C(\mu_\mathrm{central},z,SED)$ by 
selecting only objects of $R<24$ (survey criterion) and convolving the map 
with the distribution of surface brightnesses as found in the survey,
resulting in $C(R_\mathrm{tot},z,SED)$. Contour lines are drawn for 90\% (white) 
and 50\% completeness (black). Obviously, the selection function is fairly
uniform within the redshift interval from 0.2 to 1.2 and remains on average 
above 90\% for $R<23$. The limiting magnitude for 50\% completeness is roughly
at $R\approx 23.8$ and 23.5 for \tp 1 and \tp 4 galaxies, respectively. 
\label{ctypfull}}
\end{figure*}

The subsequent analysis will draw on galaxy catalogues, containing 
only objects with successful $z$ estimates
and SED classifications. It is therefore critical to understand for which
galaxies the data permit such a classification in which fraction of cases. 

The subject of this so-called completeness correction of a catalogue, or its 
resulting luminosity function, is indeed a fairly complex one: In order to 
correct a measurement of $\phi(M_\mathrm{tot},z,SED)$, we would like to have ideally
a completeness function $C(M_\mathrm{tot},z,SED)$ that obviously depends on the total 
absolute magnitude $M_\mathrm{tot}$. 

On the contrary, the observed survey is characterized by the signal-to-noise 
ratio of the photometry, which determines the classification performance and 
the completeness of the redshift catalogue. These all depend on the aperture 
photometry we use to establish the spectral shape, which essentially measures 
the central surface brightness of objects after convolving their appearance 
outside the atmosphere with an effective PSF of $1\farcs5$ diameter. 

\begin{figure}
\centerline{\hbox{
\psfig{figure=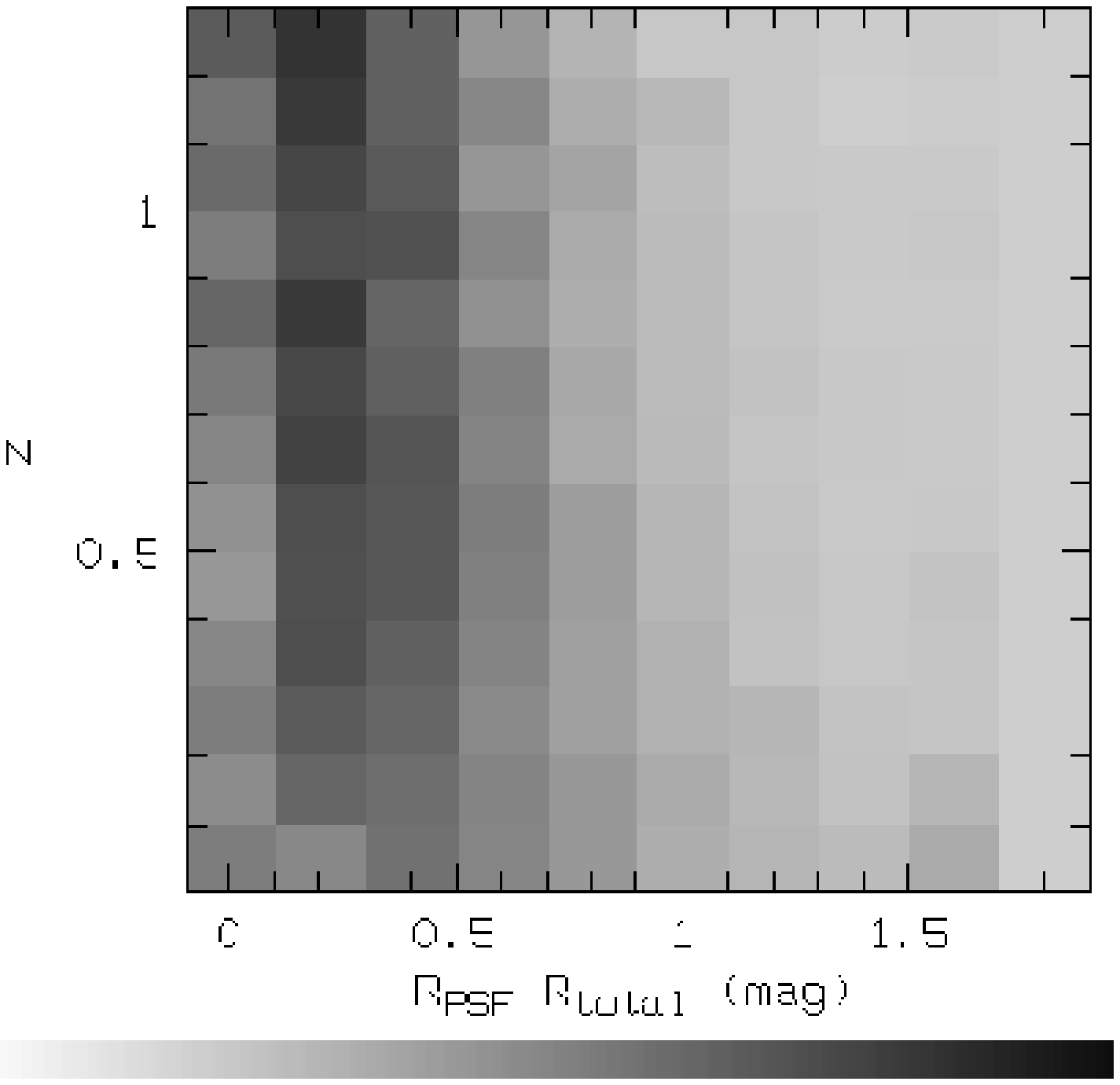,clip=t,width=8.7cm}}}
\caption[ ]{Distribution of aperture correction: The distribution of aperture 
correction values as a function of redshift. No correction applies to unresolved
galaxies. A more detailed investigation addressing issues of central surface 
brightness, concentration and size parameters of galaxies will be based on deep
HST/ACS images covering the entire CDFS field (cycle 12, PI Rix) and published
in forthcoming papers devoted to the subject.
\label{apcmap}}
\end{figure}

Given the survey parameters, the completeness of the classification and redshift
estimation can be derived from Monte-Carlo simulations as applied extensively 
and explained in Wolf, Meisenheimer \& R\"oser (2001) and as already used for 
the derivation of galaxy luminosity functions in CADIS \cite{Fri01}. The product 
of these simulations is a completeness map in fine bins of observed, convolved
central surface brightness, redshift and SED type $C(\mu_\mathrm{central},z,SED)$.
If we had proper knowledge of the distribution function $p(M_\mathrm{tot}|
\mu_\mathrm{central},z,SED)$, we could derive the function $C(M_\mathrm{tot},z,
SED)$ needed by a convolution:
\begin{eqnarray}
  C(M_\mathrm{tot},z,SED) & = & C(\mu_\mathrm{central},z,SED) * \\
                   &   & p(M_\mathrm{tot}|\mu_\mathrm{central},z,SED) \nonumber ~.
\end{eqnarray}
This approach is necessary, because galaxies of lower surface brightness which are 
near but below the total apparent aperture magnitude limit of the sample will drop 
out of the sample, although formally their overall total magnitude might suggest 
inclusion into the sample. Therefore, any completeness correction for the luminosity 
functions will ultimately be constrained by our limited knowledge on the distribution 
function $p(M_\mathrm{tot} | \mu_\mathrm{central},z,SED)$, which we try to estimate 
from the sample itself, while being affected by its very incompleteness (see 
Fig.~\ref{apcmap}).

For unresolved galaxies, our classification algorithm is more than 90\% complete 
across the redshift range considered here for most galaxies with $R\la24$. After
convolving the completeness map for point sources with the redshift-dependent 
distribution of typical corrections to total galaxy magnitude, we find that the
sample should still be 90\% complete at roughly $R\la 23$, although this number
depends in detail on redshift and galaxy type (see Fig.~\ref{ctypfull}). The 50\% 
completeness line ranges from $R<23.8$ to $R<23.5$ for \tp 1 to \tp 4, 
respectively.

\subsection{Influence of limited redshift accuracy}

\begin{figure*}
\centerline{\hbox{
\psfig{figure=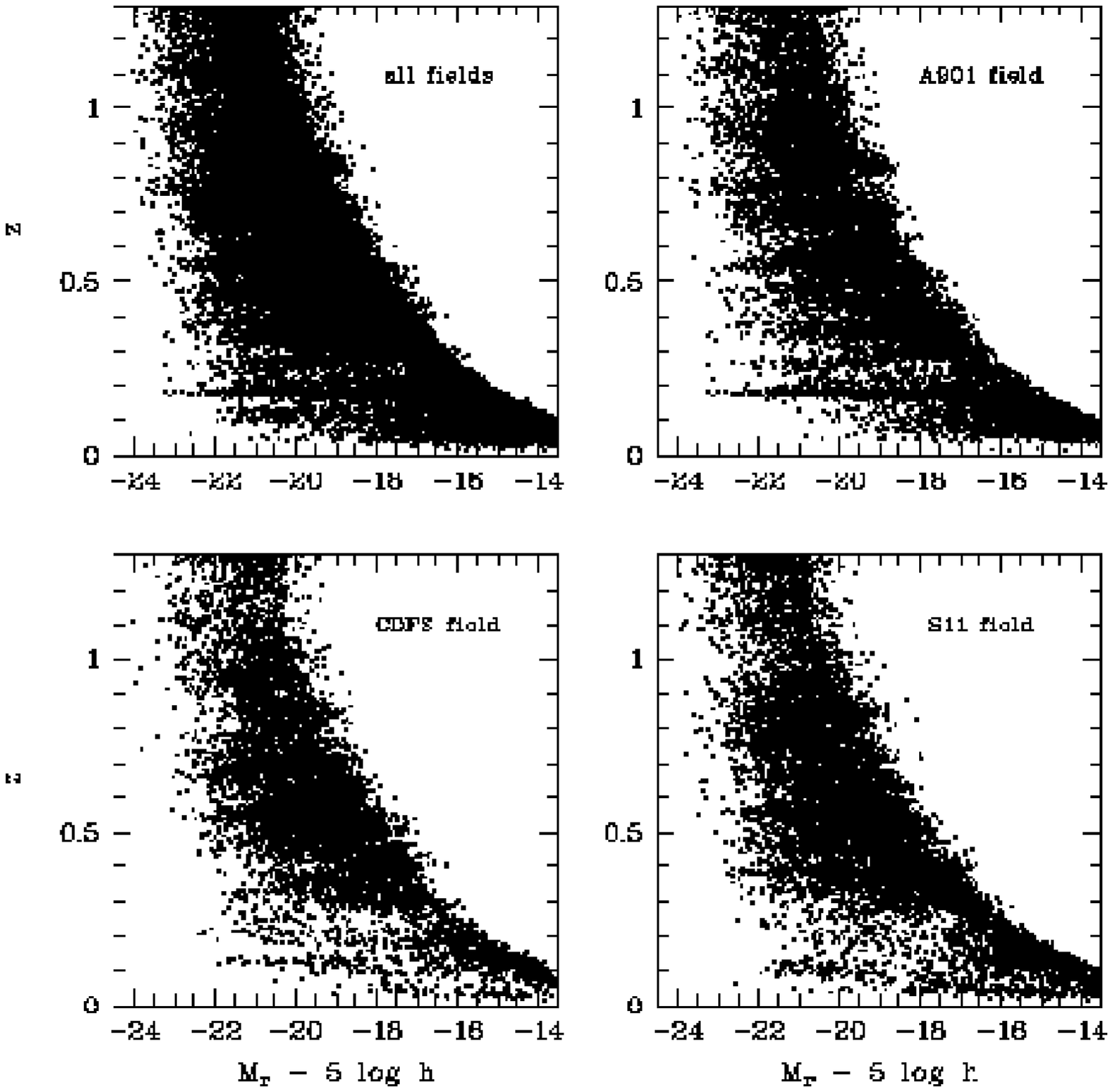,angle=270,clip=t,width=14cm}}}
\caption[ ]{The full galaxy sample split by field: Shown is 
redshift over total luminosity in the restframe SDSS r-band for $(\Omega_m, 
\Omega_\Lambda) = (0.3,0.7)$. All three fields combined contain 25,431 
galaxies in the range $0.2<z<1.2$ (upper left panel). The magnitude limit of 
the sample ($R<24$) appears as a fuzzy faint limit on the right side 
of the distribution. The fuzziness originates from applying the magnitude limit
not to the total magnitude but to the aperture magnitude of unresolved sources
(see also text). Several horizontal features can be identified which represent 
local overdensities. The most conspicuous feature is the concentration just 
below redshift 0.2 in the A901 field, which represents the Abell clusters 
901 and 902. Roughly 1000 cluster members are identified in the A901 field. 
The S11 field contains another rich cluster at $z=0.11$.
\label{samplefields}}
\end{figure*}

\begin{figure*}
\centerline{\hbox{
\psfig{figure=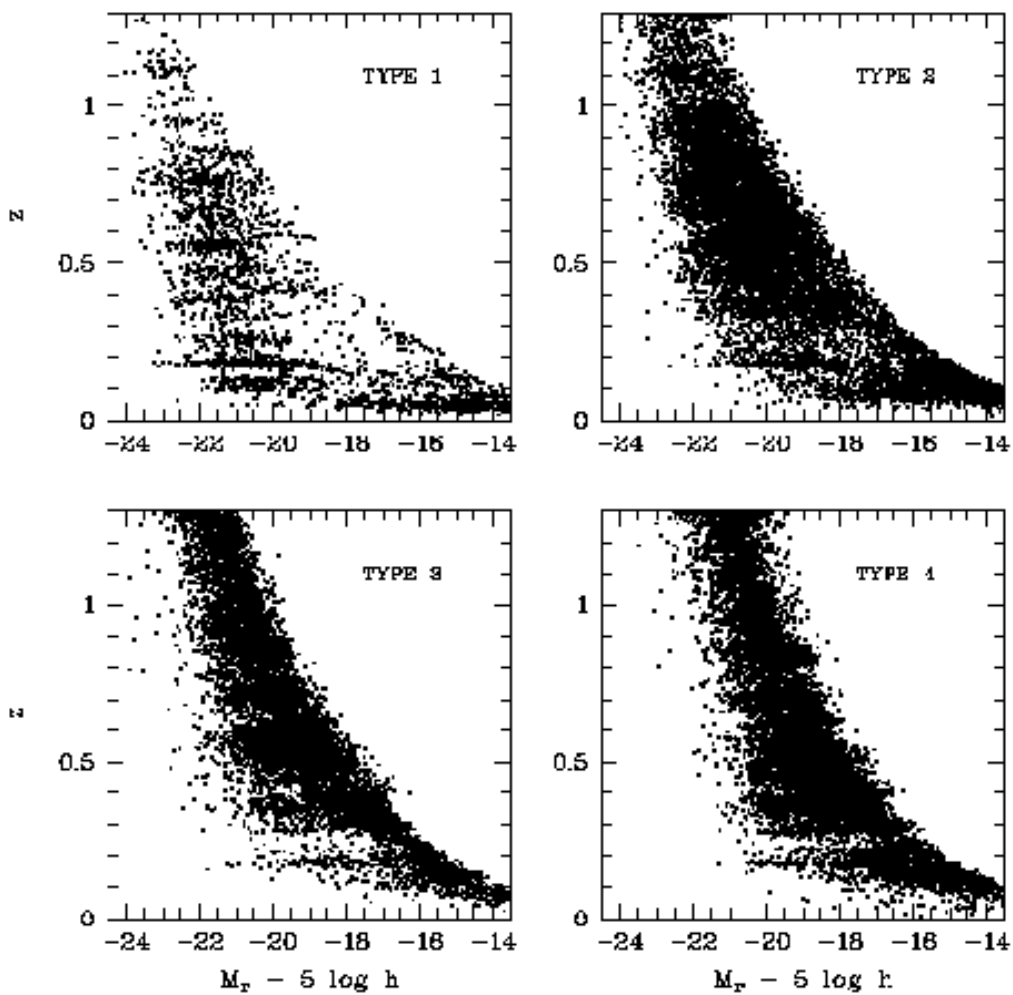,angle=270,clip=t,width=14cm}}}
\caption[ ]{The full galaxy sample split by spectral type: 
Shown is the redshift over the luminosity $M_\mathrm{r}$ in the restframe SDSS r-band. 
Several horizontal features represent local overdensities. The most 
conspicouos feature is the concentration just below redshift 0.2, which 
represents the Abell clusters 901 and 902. The $\sim$1000 cluster members
follow the known relationship of lower luminosity for bluer galaxy types. 
\label{sampletypes}}
\end{figure*}

For the subsequent analysis it is crucial to check to which extent the limited 
redshift accuracy ($\sigma_z \approx 0.03$) affects inferences about luminosity 
functions, compared to samples of high-resolution redshifts obtained from slit 
spectroscopy. 
Two major aspects need to be explored, {\it redshift aliasing}, and {\it 
catastrophic mistakes}, which are both irrelevant for interpreting 
well-exposed data from a spectrograph, but could play a role in our case: 
\begin{enumerate}
\item 
Aliasing results from the presence of structures finer than the resolution limit 
in violation of the sampling theorem, and appears as fake structure on a scale
that is typically slightly larger than the resolution. In this paper, we avoid 
dealing with the problem altogether by not trying to interpret structure close 
to the resolution limit ($\sigma_z \sim 0.01$ for bright \tp 1 galaxies and 
$\sigma_z \sim 0.1$ for faint \tp 4 galaxies), and instead choose our redshift 
bins wide enough. 

\item
Catastrophic mistakes occur in certain regions of colour space where different 
interpretations can be assigned to the same colour vector and probabilistic
assumptions are used to make a final redshift assignment. Obviously, a number
of cases will involve an assignment of the wrong redshift, but their effect on 
luminosity functions should be relatively small. They would only make a real
difference at the most luminous end, if many low-redshift objects of medium 
luminosity were wrongly assumed to reside at high redshift suggesting a high 
luminosity and consequently boosting the abundance of rare giant galaxies. 
From a simulation we inferred that the effects will be negligible even if
random redshifts were assigned to 10\% of all galaxies in the sample.
\end{enumerate}

At the moment a systematic cross check with significant samples of spectroscopic
redshifts is still pending for COMBO-17. Therefore, we estimate the redshift
accuracy drawing on the spectroscopic cross-check for CADIS using $\sim$100 
galaxies at $z \la 1.2$ \cite{Wolf01a}, which imply an average redshift error 
of $\sigma_z \approx 0.03$. Given a certain similarity of the CADIS filter set
and similar observing and reduction procedures, it is natural to assume that
the COMBO-17 redshifts would be at least as reliable and accurate as the CADIS
redshifts. Monte-Carlo simulations of the two filter sets argue in favour of a 
redshift accuracy that is slightly better in COMBO-17 \cite{WMR01,Wolf01a}. 
In CADIS, we have seen a fraction of catastrophic mistakes at $R<24$ of $\sim
10$\%. However, the classification and redshifts in the COMBO-17 fields are 
currently checked against spectra from 2dFGRS and 2QZ at the bright end, as well 
as against spectra of X-ray sources obtained in the CDFS at the faint end. Early 
indications are, that among bright galaxies at $z<0.3$ the accuracy is better than 
$\sigma_z<0.02$ and the rate of catastrophic mistakes well below 10\%. The Seyfert 
galaxies seen at fainter magnitudes in the CDFS are the most challenging objects
for the redshift estimation, but their rate of redshift outliers is still below
20\% and their accuracy is still on the order of $\sigma_z\sim0.05$.

Once, we have obtained better knowledge on the redshift error distribution,
we should also assess their impact on luminosities more carefully. A mean error
of 0.03 will lead to less than $0\fm1$ scatter in the luminosities at $z\sim$,
but up to $0\fm25$ scatter at $z\sim0.3$. The situation is more dramatic for faint
starburst galaxies, where we estimate redshift errors around 0.1, implying errors
in luminosity of $\pm0\fm25$ at $z\sim1$, increasing to $\pm0\fm75$ at $z\sim0.3$.
The latter errors potentially bias the steep luminosity function of starburst
galaxies to brighter $L^*$ values. However, we reserve a more rigorous treatment
of this issue to a later publication, following spectroscopic work for establishing
the error distribution.

\section{Results: redshifts and SEDs of $\sim$25,000 galaxies at $0.2<z<1.2$}

The immediate result of the above classification and redshift determination is shown
in Fig.~\ref{sampleredshift} and \ref{sampletypes}, where the results from the three 
fields (see Table~\ref{fields}) are combined, and in Fig.~\ref{samplefields}, where
the sample is split by field. We restrict our detailed analysis of the luminosity 
function as presented in Sect.~5 to the redshift interval z=[0.2,1.2], which 
contains 25,431 measured redshifts for galaxies at $R<24$.

\subsection{SED distribution}

Fig.~\ref{sampleredshift} shows the sample in terms of SED parameter over restframe 
SDSS r-band luminosity, split in four different redshift intervals of width $\Delta 
z=0.2$. In this diagram, the magnitude limit of the sample appears as a faint limit 
on the right side of the distribution. If a sample was selected on the basis of a
total SDSS r-band magnitude and observed at $z=0$, the selection limit would show 
up as a perfectly vertical border line. 

However, by selecting in the WFI R-band and -- more importantly -- by observing 
redshifted galaxies, the SDSS r-band luminosity corresponding to the selection limit 
depends on the restframe colours of the galaxy. Lower values of the SED parameter 
resemble redder galaxies, which are selected in their fainter restframe UV at higher 
redshift and therefore need to be more luminous to enter the sample. This explains 
the angle of the faint selection limit to the right side of the galaxy distribution. 

Another factor softening the border line is the fact that
the selection limit applies not to the total magnitude but to the aperture flux and
ultimately depends on the central surface brightness of the galaxy. The selection by
aperture flux corresponds only for unresolved objects to a fixed magnitude value, 
while extended galaxies will have brighter total magnitudes to a varying degree that 
depends on their morphology. The median correction for this effect is just below 
half a magnitude and gives the selection limit a fuzzy appearance. Quoting a limit 
for 99\% completeness strictly speaking requires 
knowledge about the true abundance of low-surface-brightness galaxies, which could 
always have escaped the object detection to some degree while featuring high total 
luminosities distributed over a large area.

The limit on the left side is produced by the steepness
of the bright end in the luminosity function. Here the figure clearly demonstrates
the well-known fact, that the highest luminosities are found among the reddest 
galaxies, and that there is a clear monotonic relationship of $L^*$ with SED type.
This apparently smooth relationship suggests, that parametric fits to the luminosity 
functions should be calculated in a bivariate fashion, depending both on luminosity 
and SED type or a suitable restframe colour.

\subsection{Luminosities and redshifts}

Fig.~\ref{samplefields} shows the sample in terms of redshift over restframe SDSS 
r-band luminosity, split by observed target field. In this diagram, the magnitude 
limit of the sample ($R<24$) appears again as a faint limit on the right side of 
the galaxy distribution. However, since the total luminosity corresponding to the 
selection limit depends on redshift, restframe colour and morphology, this border
line is again not defined as sharply as an aperture flux limit in the observed frame. 
The different panels show the imprint of large-scale structure in the individual 
fields. Narrow horizontal stripes point to clusters and sheets. While the 3-D 
positions $(x,y,z)$ of the galaxies can, of course, be used to find new clusters
out to redshifts of $z\la 1.2$, some clusters in the fields have already been
known. In fact, the A901 field has been selected for the very reason of studying
the clusters A901/902 in detail, which show up with their $\sim$1000 identified
cluster members just below $z=0.2$. The pencil beams of this survey contain a
natural mix of environments at all redshifts. In this paper, we do not attempt any
differentiation between field and cluster galaxies. Clusters do show up at small
numbers that change with redshift and fluctuate significantly. Therefore, our 
results for \tp 1 galaxies could be affected by peculiarities introduced by
clusters in certain redshift bins.

\begin{figure*}
\centerline{\hbox{
\psfig{figure=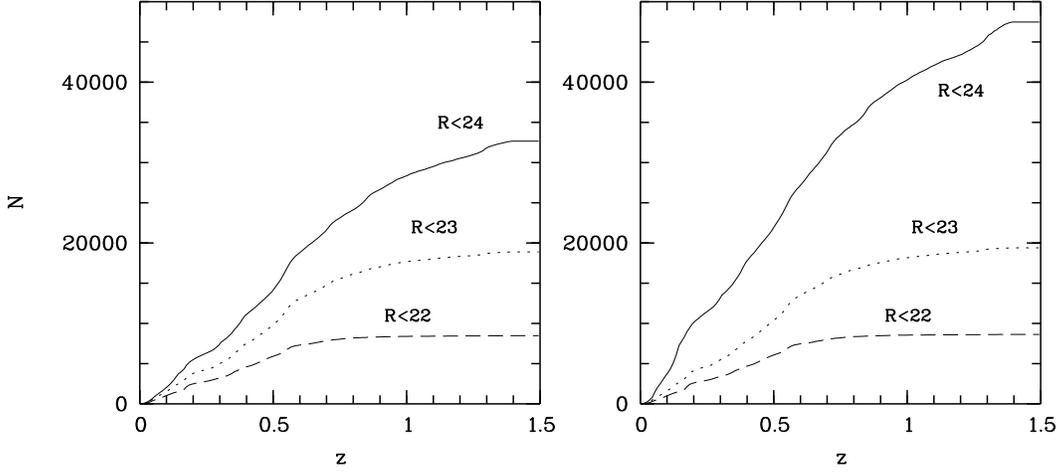,angle=270,clip=t,width=14cm}}}
\caption[ ]{Cumulative redshift histograms of our galaxy sample: {\it Left panel:} 
Uncorrected cumulative histograms of galaxies found in the catalogue at 
$R<[22,23,24]$. {\it Right panel:} Redshift histograms corrected for completeness. 
The median redshift for $R<24$ is $\sim0.55$ in either sample. In all cases, total 
magnitudes are used. The correction is based on the completeness maps in 
Fig.~\ref{ctypfull}. Our redshift accuracy at $z>1.2$ is currently unknown.
\label{zhist}}
\end{figure*}

Fig.~\ref{sampletypes} finally shows the sample in terms of redshift over restframe 
SDSS r-band luminosity, split by spectral type. Here, the dependence of restframe 
colour on the SED type affects the faint selection limit. At higher redshifts the
redder galaxies are observed in their fainter restframe UV region and therefore 
need to be intrinsically more luminous to be included in the sample. This explains
the flatter angle of the magnitude selection for \tp 1 as compared to the
steeper border line for the blue starburst galaxies in \tp 4. We are able to trace
starburst galaxies down to $M_\mathrm{r}\approx -20$ all the way out to $z\approx 1.2$,
but in contrast galaxies with restframe colours of present-day spheroids around $z
\sim 1$ are only identified at $M_\mathrm{r}\approx -21.5$. The bright cutoff on the left 
side of the galaxy distribution already demonstrates how the evolution with redshift 
depends on the spectral type. Luminous \tp 1 galaxies show basically no trend with
redshift while \tp 3 and 4 galaxies show a strong depletion of luminous objects when 
going from $z=1$ to $z=0$, either due to dimming or dropping density.

\subsection{The redshift distribution of $R<24$ galaxies}

Our subsequent discussion of the galaxies is phrased in terms of absolute 
magnitudes, i.e. luminosities, and redshifts. However, for many applications the 
redshift probability distribution of galaxies brighter than some {\it apparent} 
limit is of considerable interest. Foremost among such applications is perhaps 
gravitational lensing, in particular weak lensing, where the small image 
distortions of thousands of galaxies are combined to extract information about 
the intervening mass distribution (e.g. Mellier, 1999). In almost all practical 
cases there is no direct redshift information available about these numerous 
faint source galaxies.

In Fig.~\ref{zhist} (left panel) we show the redshift histogram of all galaxies 
from the catalogue discussed above for magnitude limits of $R<[22,23,24]$. These 
histograms can be corrected for incompleteness, using the estimates from Section 
3.4 (see right panel). The median redshift is $z\sim0.55$, with the 90\% lower and 
upper percentiles being $z=0.11$ and $z=1.20$, repectively. 

Redshift distributions to such faint apparent magnitudes alread exist over very
small fields (e.g. the HDF North and South; Cohen et al., 2000). However, there the
uncertainty in the resulting histogram is completely dominated by the field-to-field
variations, whose fractional contrast is near unity in $\Delta z\sim 0.2$ bins over 
such small fields. Here we can present for the first time the redshift distribution 
of galaxies to $R\la 24$ over fields large enough that the variance is low.

\subsection{Calculation of the luminosity function}

We follow the usual definition of the luminosity function as the number density 
of galaxies in an interval of luminosity or absolute magnitude, expressed in 
units of $h^3 {\rm Mpc}^{-3}$. Here, we use two estimators for its calculation
(see Willmer 1997 for a comprehensive overview):

\subsubsection{Non-parametric estimates of $\phi(M)$} 

We use the non-parametric $1/V_\mathrm{max}$ estimator \cite{Sch68} in the form proposed 
by Davis \& Huchra (1982) and modified by a completeness correction as outlined 
in Fried et al. (2001) to calculate the differential luminosity function, which
is just given by the sum of the density contributions of each individual galaxy
in the considered luminosity/redshift/SED-type bin: 
\begin{equation}
 \phi(M)dM = \sum_i \frac{1}{V_i (M,z,SED)}   ~.
\end{equation}

$V_i (M,z,SED)$ is the total comoving volume in which the galaxy $i$ could be 
located to be included in the sample. The boundaries of this volume are given by 
the redshift interval in question and by the selection function $C (M,z,SED)$ of 
the survey as discussed in Sect.~\ref{completeness}. Altogether, the volume is 
\begin{equation}
 V_i (M,z,SED) = \Delta\Omega \int_\mathrm{z_\mathrm{min}}^{z_\mathrm{max}}{C (M,z,SED) 
                 \frac{dV}{d\Omega dz}dz}
\end{equation}

The errors of $\phi(M)dM$ are just given by the square roots of the variances
$\sigma_\phi = \sqrt{\sum_i 1/V_i^2 (M,z,SED)}$, which means that we take only
statistical Poisson noise from the galaxy counts into account, but ignore any
errors in magnitude or redshift here. Furthermore, we ignore the effects
of clustering here, since the sample is fairly large and should not be strongly
affected. As we will see later, the Poissonian error bars appear to be a fair 
representation of the scatter in the data points around parametric fits of the
luminosity function, so we believe that we do not underestimate our errors.

However, due to biases and selection limits the binning process can lead to a 
suboptimal representation of the underlying measured luminosity function. As
Page \& Carrera (2000) have pointed out, the faintest bin containing the 
selection cutoff within its bin limits can be significantly harmed. Therefore,
we decided to ignore any $V_\mathrm{max}$ data points, where the magnitude cutoff of 
the survey shrank the accessible volume of the bin by more than 30\% compared 
to an infinitely deep survey.

\subsubsection{Parametric estimates of $M^*$ and $\alpha$} 

For a parametric maximum-likelihood fit to a Schechter function, we use the 
STY estimator \cite{STY79}. The Schechter function (1976) is
\begin{equation}
 \phi(L)dL = \phi^* (L/L^*)^\alpha e^{-L/L^*} ~dL
\end{equation}

or in terms of magnitudes using $x = 10^{-0.4(M-M^*)}$  
\begin{equation}
 \phi(M)dM = 0.4~ \ln{10} ~\phi^* x^{\alpha+1} e^{-x} ~dM       ~.
\end{equation}

In the STY formalism, the free parameters of an underlying parent distribution 
(here $M^*$ and $\alpha$) are tested by calculating the probability that the 
observed magnitude distribution is consistent with the parent distribution 
$\phi(M)$, given the selection function $C (M,z,SED)$ of the survey. The 
probability that a single galaxy of luminosity $M$ will be detected is given by
\begin{equation}
 p(M,z,SED) = \frac{\phi(M) C (M,z,SED)}
        {\int{\phi(M) C (M,z,SED) ~dM}} ~,
\end{equation}

where the integral is effectively limited by the selection function containing
also upper and lower magnitude cutoffs. The joint probability that all galaxies
in the sample belong to the same parent distribution is then
\begin{equation}
 {\cal L} = \prod_\mathrm{i} p(M,z,SED)
\end{equation}

and this likelihood needs to be maximized by varying $M^*$ and $\alpha$. As shown
later, we observe an upturn of the luminosity function for \tp 1 galaxies, just 
as the 2dFGRS has seen it before. We treat this upturn as an extra component and
restrict the luminosity range for the STY fit (only for \tp 1 galaxies) to obtain 
the parameters of the dominant, luminous component, using magnitude limits of 
$(M_\mathrm{280},M_\mathrm{B},M_\mathrm{r}) < (-15,-17,-18)$. At low redshift, the parameters $M^*$ and 
$\alpha$ are well constrained and their errors are estimated from the error 
ellipsoid defined as
\begin{equation}
 \ln {\cal L} = \ln {\cal L}_\mathrm{max} - \frac{1}{2} \chi_\beta^2(N)
\end{equation}

where $\chi_\beta^2(N)$ is the $\beta$ point of the $\chi^2$ distribution with
$N$ degrees of freedom ($\Delta \chi^2=2.30$ and $6.17$ for 1$\sigma$-and 
2$\sigma$-limits corresponding to confidence intervals of 68.3\% and 95\%,
repectively). Individual errors on $M^*$ and $\alpha$ are quoted by ignoring
the covariance but measuring their confidence interval within projected contours 
of $\Delta \chi^2=1.0$. 

For most restframe passbands and galaxy types we do not constrain the knee of the
luminosity function out to the highest redshifts. The covariance between $\alpha$
and $M^*$ makes it virtually impossible to obtain a well constrained measurement.
We therefore decided to measure $\alpha_\mathrm{local}$ at low redshift and assume it 
does not vary with redshift. We then look only at sections of the likelihood map 
in the interval $\alpha_\mathrm{local}$(type)$ \pm \delta\alpha$(type),
providing us now with a more constrained estimate for $M^*$ which is valid only 
under the assumption made, {\it that $\alpha$ does not vary with redshift}. The
error on this $M^*$ estimate is obtained after rescaling the $\chi^2$-map such that 
the best fit $M^*$ value found at $\alpha_\mathrm{local}$(type) has $\Delta \chi^2=1.0$ 
with respect to the discarded global minimum of the full map over the unlimited 
range in $\alpha$. The error we quote is given by the confidence interval within
the projected contours of $\Delta \chi^2=1.0$ above $\chi^2(M^*,\alpha_\mathrm{local})$
found in the map constrained in $\alpha$.

\subsubsection{Determination of $\phi^*$} 

Since the normalisation $\phi^*$ of the luminosity function cancels in 
our implementation of the STY fit
procedure and only the shape of the distribution is subject to the test,
we calculate  $\phi^*$ afterwards by
\begin{equation}
 \phi^* = \frac{\bar{n}}{\int_\mathrm{M_1}^{M_2}{\phi(M)/\phi^* ~dM}}
\end{equation}

where the average galaxy density $\bar{n}$ within a certain volume is given by
\begin{equation}
 \bar{n} = \frac{n_\mathrm{gal}}{\int dV}      ~.
\end{equation}

There are two different error sources for $\phi^*$, one arising from cosmic
variance, i.e. large-scale structure, and the other from the covariance with $M^*$.
All errors quoted for $\phi^*$ in this paper are determined from field-to-field 
variation among our three fields, whatever redshift bin or SED type we consider;  
this turns out to be the dominant source of uncertainty.
However, we do list the covariance with $M^*$ as well, but note, that the latter
does not dominate over the error from structure under the assumption made that 
$\alpha$ does not vary with redshift.

\subsubsection{Determination of j} 

We furthermore calculate the luminosity density of galaxies, $j$, after 
integration over the luminosity axis. The full integrals can easily be obtained 
from the STY fits as, e.g., 
\begin{equation}
 j = \int dL \phi(L) L = \phi^* L^* \Gamma(\alpha+2)	~,  
\end{equation}

where $\Gamma$ is the Gamma-function. Later we show explicit results 
for a range of finite lower luminosity limits to illustrate the
effect of extrapolating the luminosity function with constant $\alpha$ for the
different types. The very assumption of $\alpha$ remaining constant to infinitely 
low lumniosities is the largest source of error for $j$-estimates at high redshift.
It is not reflected in any error estimates we quote, but can be assessed from the
convergence of the integral with decreasing luminosity limits as shown in the
following section. If we believe the STY fits and use their formal errors to 
derive an error estimate on $j$, we find the error in $\phi^*$ originating from 
field-to-field variation to dominate and derive $j$ errors directly from these.
As solar luminosities we use (Vega-normalised) values of $M_\mathrm{280,\sun}=6.66$, 
$M_\mathrm{B,\sun}=5.30$ and $M_\mathrm{r,\sun}=4.47$.

\section{Discussion}

\subsection{The ``quasi-local'' luminosity function at $z\sim 0.3$}

With a total field size of 0.78~$\sq\degr$ our survey volume at $z<0.2$ is too 
small for a sensible derivation of a luminosity function. Also, the present 
redshift errors can have a larger impact onto the luminosity measurement in this
very local regime. Our lowest redshift bin therefore covers the range $z=[0.2,0.4]$ 
and contains 5,674 galaxies from three target fields. Already the median redshift 
in this bin is $\sim0.34$ in the luminosity range of $M_\mathrm{r}=[-23,-17]$. This 
value is higher than for recent, large surveys of the ``local", or present-day, 
galaxy population, such as SDSS or 2dFGRS. We therefore refer to our lowest $z$, 
reference bin as the quasi-local sample.

\subsubsection{The field-to-field variation of the luminosity function}

\begin{figure*}
\centerline{\hbox{
\psfig{figure=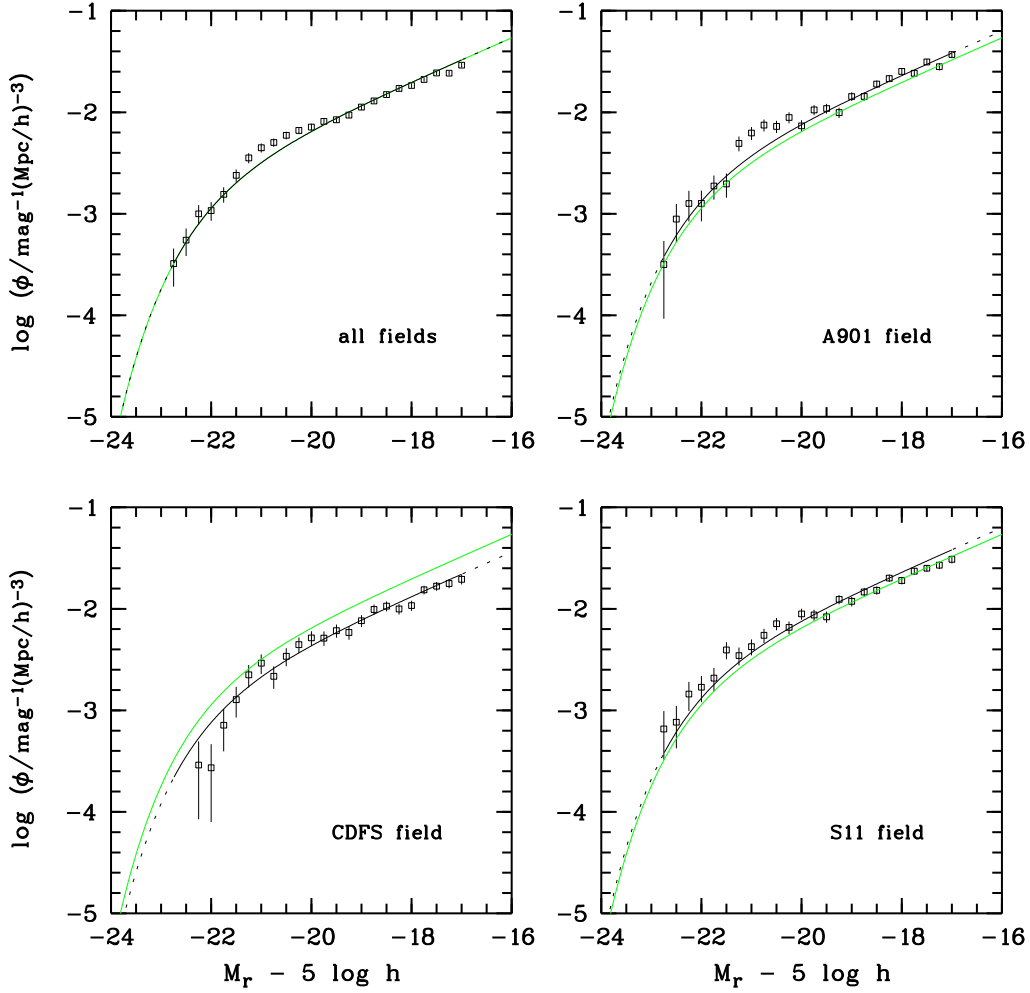,angle=270,clip=t,width=14cm}}}
\caption[ ]{Field-to-field variation:
Comparison of the luminosity function $\phi(M_\mathrm{r})$ of galaxies from the lowest 
COMBO-17 redshift bin ($z=[0.2,0.4]$) among the three disjoint survey fields. 
Error bars are 1-$\sigma$-Poissonian. Grey line: STY fit of Schechter function
to the whole sample. Black line: fit to the panel-specific field sample. 
By choice, some galaxies in the A901 field should be subject to magnification by 
gravitational lensing. We estimate, that this brightening corresponds to $\Delta m 
\sim -0\fm25$ in the central density peak of the cluster, but less in its outer 
regions. At this point, we neglect the effect for the field as a whole and assume
that is does not alter the conclusions of this paper. The hump around $M\sim
-21$ in the LF of the A901 field appears to originate from large-scale structure,
as it is related to a group of bright galaxies at similar redshift near the Eastern 
edge of the field.  
\label{c17_3f_comp}}
\end{figure*}

In our lowest redshift bin, $z=[0.2,0.4]$, the co-moving volumes are smallest 
and the structure in the galaxy population has developed furthest. Therefore, 
this redshift bin is the most critical to explore to which extent our inferences 
about the galaxy distribution are affected by field-to-field variations. 
Fig.~\ref{c17_3f_comp} shows the luminosity functions in the restframe SDSS 
r-passband derived separately from the three individual fields and for the whole
sample combined. All error bars shown reflect only the 1$\sigma$ Poisson 
variance from the finite number of galaxies in the bin. It is apparent that due 
to large scale structure the results differ somewhat more from field to field 
than implied by Poisson errors. However, for the most part, these differences are
merely fluctuations in $\phi^*$, which we use to estimate errors on $\phi^*$.

The CDFS field appears to have a lower galaxy density than the other two fields. 
We note, that observations of the Chandra observatory have revealed that the 
CDFS also shows number counts of X-ray sources lower by a factor of two compared 
to the CDF-North \cite{Norman02}. The implication of this field-to-field variance 
is complicated as the fields were not chosen at random: while the CDFS is very 
``empty", the A901 field was chosen to contain a cluster (though not in the 
redshift bin shown). This test shows, that even with a $0.5\degr\times0.5\degr$ 
field size the field-to-field variations are still noticeable and must be accounted 
for, but they are much less pronounced than in tiny fields, such as the Hubble Deep 
Fields.

\subsubsection{The SED type and passband dependence of the luminosity function}

\begin{figure*}
\centerline{\hbox{
\psfig{figure=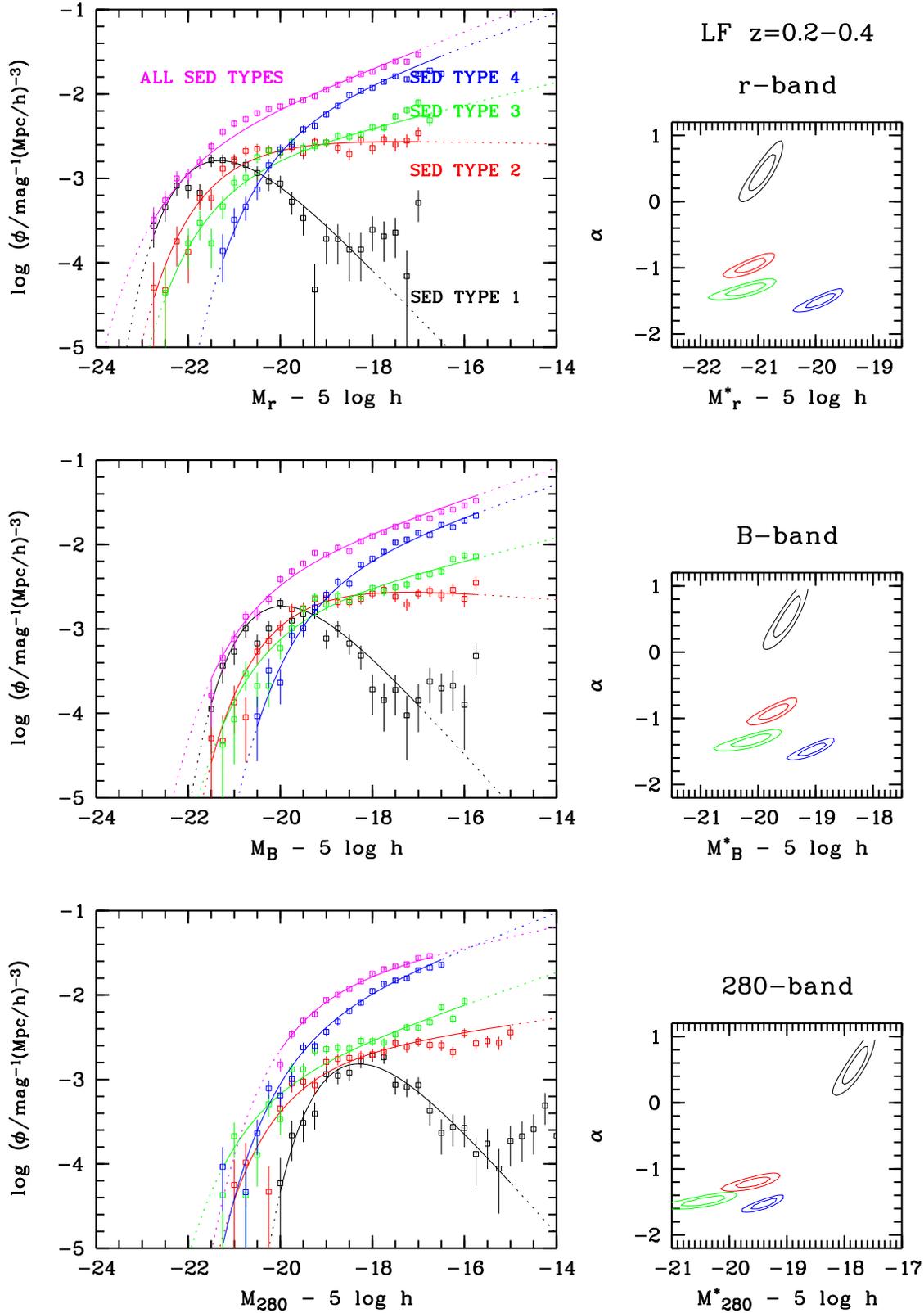,angle=0,clip=t,width=16cm}}}
\caption[ ]{SED dependence of the luminosity function: Results are shown for 
quasi-local sample at $z=[0.2,0.4]$ in three different restframe passbands, 
SDSS-r, $B$, and at 280~nm, from top to bottom respectively, and split 
by SED types.
{\it Left panels:} $V_\mathrm{max}$ datapoints with Poissonian error bars and STY 
fits of Schechter functions.
{\it Right panels:} Likelihood contours for parameters $M^*$ and $\alpha$ with 
1$\sigma$ and 2$\sigma$ error contours. Differences in $M^*$ between the bands 
reflect mean restframe colours of the type samples. The upturning faint end of
\tp 1 has been excluded from the STY fits. Note the 
relatively fainter $M^*$ of \tp 4 (starburst) galaxies compared to \tp 3.
\label{vmaxlocal}}
\end{figure*}

\begin{figure*}
\centerline{\hbox{
\psfig{figure=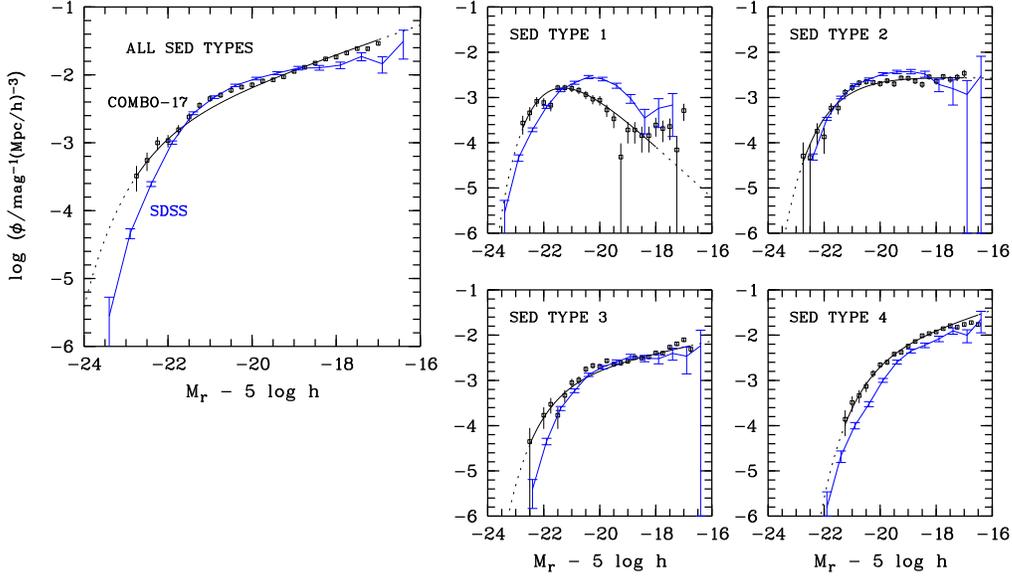,angle=270,clip=t,width=14cm}}}
\caption[ ]{COMBO-17 vs. SDSS: the luminosity function $\phi(M_\mathrm{r})$ of galaxies 
from the quasi-local COMBO-17 sample at $z\sim 0.3$ compared with the SDSS 
results by Blanton et al. (2001), which are represented as error bars connected 
by a dashed line. The original AB magnitudes are converted to Vega magnitudes, 
and the types have been matched by adjusting the SDSS $(g-r)$-colour cuts,
although this will not lead to an emulation of our type definition (see text).
For the combined sample of all SED types, the basic match around the knee of 
the function is quite good, but differences appear at the bright and faint end.
While faint-end differences relate to evolution of faint blue galaxies between
the different median redshifts of the two faint end samples, $\sim 0.34$ and 
$\sim 0.05$, bright-end differences might result from different sensitivity 
to faint surface brightness envelopes in giant spheroids, as they are expected 
between 5-hour and 1-minute exposures. 
\label{c17sdss_comp}}
\end{figure*}

\begin{figure*}
\centerline{\hbox{
\psfig{figure=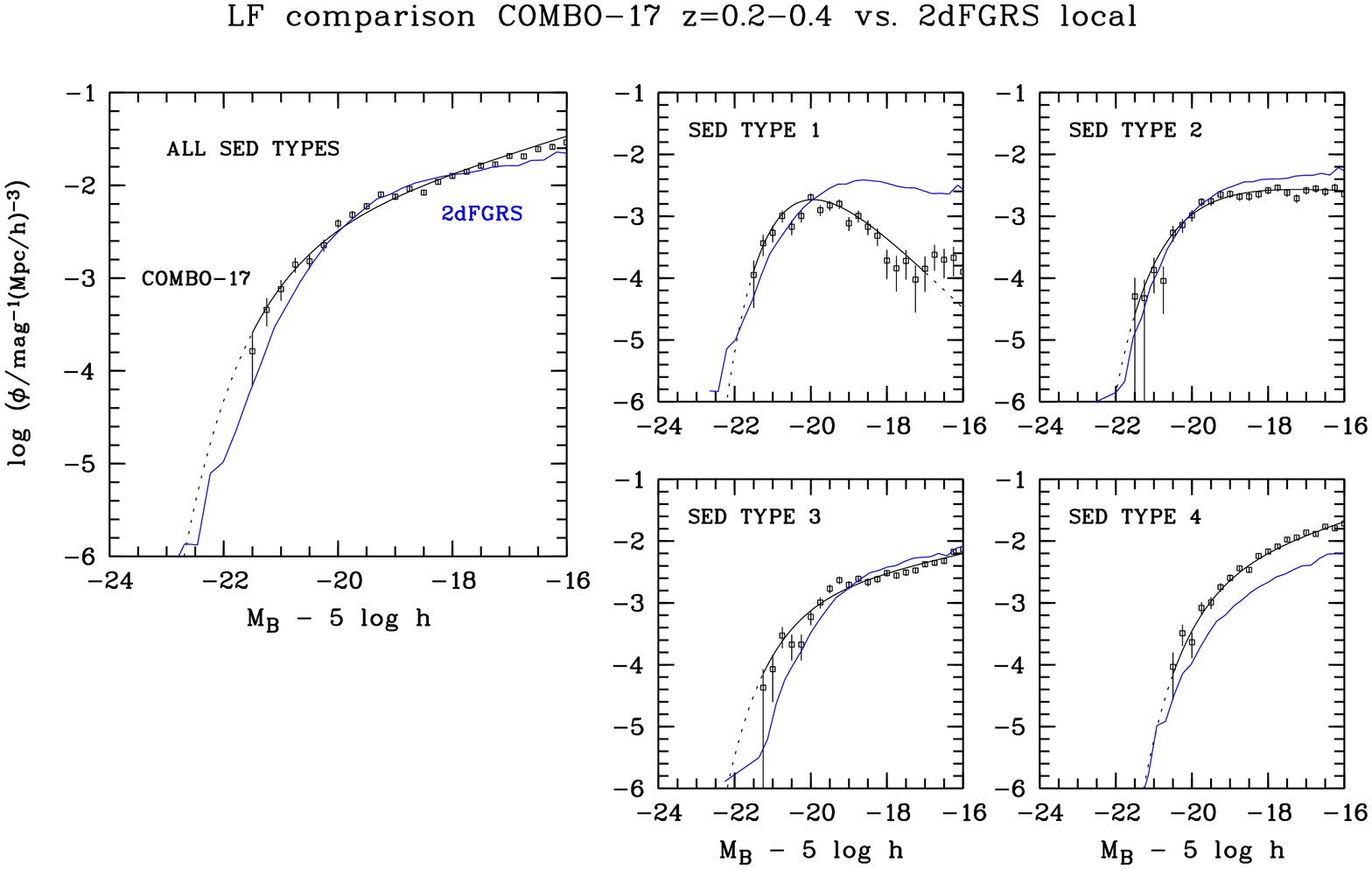,angle=270,clip=t,width=14cm}}}
\caption[ ]{COMBO-17 vs. 2dFGRS: the luminosity function $\phi(M_\mathrm{B})$ of 
galaxies from the quasi-local COMBO-17 sample compared with the 2dFGRS results by 
Madgwick et al. (2002), which are transformed into the Johnson-B system and
represented by a dashed line (error bars are vanishing in size). The original 2dF
galaxy types have been adopted without change and therefore correspond only very
roughly to our types. The agreement of the combined sample is rather good with 
slight differences at the faint end arising from evolution of faint-blue galaxies.
\label{c17_2df_comp}}
\end{figure*}

The $\sim$5,700 galaxies in the quasi-local bin are enough to draw up a 
quite comprehensive statistical picture of the 
$\langle z \rangle \approx 0.3$ galaxy population, which is summarized
in Fig.~\ref{vmaxlocal}. We have enough galaxies that we can study 
the LF in the four broad SED classes defined in Section 3.2. We have enough
wavelength coverage that we can construct the LF in three wavebands, 280~nm,
b$_j$, and $r$. Further, with $R\la 24$ our data reach well below the knee 
in the LF, $M^*$, so that the faint end slope $\alpha$ is well constrained. 

Is is obvious from Fig.~\ref{vmaxlocal} how much the faint end of the luminosity 
function depends on SED type: the later or bluer the SED type, the steeper the 
faint end of the luminosity function. This effect, quantified in the right hand 
panels of Fig.~\ref{vmaxlocal}, is present in all three wavebands. We remind the 
reader that the SED types are defined by redshift- and luminosity-independent 
restframe colour, not by morphological or evolutionary type.

For the most part, these SED types reflect a mean stellar age sequence, although 
metallicity effects complicate the relationship between age and colour: in 
particular for early types the colour-luminosity relation will place fainter 
(presumably more metal-poor and hence bluer) galaxies of a given stellar age 
preferentially into a later SED bin. This effect can contribute to a downturn at
the faint end of \tp 1, but it will by far not be the dominant effect leading to 
the positive $\alpha$.  

Not surprisingly, the characteristic absolute magnitude, $M^*$, of the LF depends 
both on SED type and on the observed waveband. In the r-band
$M^*$ is nearly independent of the SED for \tp 1 to 3, while the starburst 
galaxies of \tp 4 are significantly fainter. When looking at the other bands,
the mean restframe colours of the galaxies belonging to the different types 
(as shown in Fig.~\ref{restcol_UBR}) naturally shift the $M^*$ values such 
that bluer galaxies appear relatively more luminous in the bluer bands. In 
the b$_j$ band and at 280~nm, the brightest $M^*$ is found for \tp 3 objects
and probably originating form large, strongly star-forming spiral galaxies. 

Obviously, we do not observe the entire galaxy population, but only galaxies 
with $M\la-16$, i.e. galaxies to SMC luminosity, in the quasi-local sample.
While our parametric fits and forthcoming density calculations assume the LF to 
continue to infinitely faint levels at constant $\alpha$, the LF might turn over 
at some lower luminosity or simply have a more complicated shape as given by the 
Schechter function, especially when we look at the combined sample of all SED 
types. The sum of Schechter functions for individual types with different
$(M^*,\phi^*,\alpha)$-parameters will not in general give a Schechter function 
again, unless the steepest function dominates the sum everywhere as it is almost
the case in the 280-band. In Fig.~\ref{vmaxlocal} we can see the bright end of 
the B-band and r-band LFs being dominated by \tp 1 and the faint 
end by \tp 4. At intermediate luminosities we get almost equal contributions 
from all types and the result does not resemble a Schechter function so clearly
anymore. Indeed we can see a variation of $\alpha$ with the luminosity domain 
since the \tp 4 LF contributes significantly only in the faintest domain. 
If \tp 4 galaxies continue to rise at their rate fainter
than $M_\mathrm{280}\ga-16$, they will drive up the $\alpha$-value for the combined
sample as well. The very reason for our observed $\alpha$-values, which are 
steeper than those for the 2dFGRS- and SDSS-samples (see following section), is 
the stronger prominence of starburst galaxies at the higher redshifts we look at.

\subsubsection{Comparison of COMBO-17 with 2dFGRS and SDSS}

Figures \ref{c17sdss_comp} and \ref{c17_2df_comp} present a comparison of the 
COMBO-17 quasi-local sample to the yet more nearby samples from SDSS \cite{Bla01}
and 2dFGRS \cite{Mad02}, which both have $\langle z \rangle \approx 0.1$. For the
2dFGRS we plot the luminosity function just for their published types
without any attempt to adjust the respective SED type definitions. Therefore,
differences are expected to some degree. For the SDSS, we have obtained the
values of the luminosity function in fine bins over the $(g-r)$-colour axis,
and have chosen type limits with the aim of matching the shapes of their LF
to ours. However, a brief look at Fig.~\ref{restcol_UBR} should remind the reader
that cutting the sample by intervals on the $(B-r)$-colour axis does not allow
to tune the limits such that truely similar type samples are obtained as with our
own definition. 

When taking all SED types (leftmost panel), the agreement with SDSS and 2dFGRS 
is quite good around the knee of the Schechter function. The $V_\mathrm{max}$ results from 
COMBO-17 are shown as squares with errors bars having no ticks, and the STY fits 
as solid lines following a Schechter function. The comparison surveys are shown
as another solid line connecting their $V_\mathrm{max}$ data points. For the SDSS, the 
individual $V_\mathrm{max}$ data points are further shown as error bars with ticks, but
for the 2dFGRS the error bars are omitted because their sample of $\sim$100,000 
galaxies renders them virtually invisible. 

The luminosity axis has been adjusted for both surveys, by converting from 
ABmag to Vega-mag in case of the SDSS, and by transforming from the photographic 
$b_\mathrm{J}$ system to Johnson-B in case of the 2dFGRS. Using synthetic photometry
on the galaxy templates, we derived a mean magnitude offset $B-b_\mathrm{J}$ of
0.12, 0.11, 0.07, 0.06 and 0.09 for \tp1, \tp2, \tp3, \tp4 and the combined sample,
respectively.

There are significant differences to SDSS and 2dFGRS, for very faint and for
very bright galaxies. These differences could be physical or results of the data
treatment, and we cannot give conclusive explanations. However, two plausible 
reasons come to mind: the mean redshift of dwarf galaxies in the SDSS and 2dFGRS 
sample is $z\approx 0.05$, while it is $z\approx 0.34$ in the quasi-local 
COMBO-17 sample. I.e. there is a difference of over two Gigayears in the mean 
look-back time. Given the strong redshift evolution of dwarf galaxies (see also 
Sect. 5.2), this difference reflects most likely the physical evolution of 
faint star-forming galaxies found by Ellis et al. (1996) and Heyl et al. (1997). 

For the brightest galaxies, in large part early type galaxies, there may be a 
completely different explanation for the difference: all magnitudes involved 
are ultimately some version of isophotal magnitudes. The luminosities in
COMBO-17 are based on the multi-hour R-band exposure, which reaches much fainter 
surface brightness limits (the SDSS exposure is shorter than 1 min), and may 
therefore measure a larger fraction of the low-surface brightness, outer parts 
of early type galaxies.

While these effects might have caused differences between COMBO-17 on one side
and 2dFGRS as well as SDSS on the other side, we would not conclude to see major
differences between SDSS and 2dFGRS as their deviation from the COMBO-17 LF at 
$z\sim0.3$ is quite similar. Early claims of significant differences between SDSS
and 2dFGRS \cite{Bla01} have since been reconciled by Norberg at al. (2002), who 
demonstrated that applying necessary colour transformations and further data 
adjustments with due diligence produces compatible results.

\subsection{The redshift evolution of the luminosity function since $z\sim 1.2$}

\begin{figure*}
\centerline{\hbox{\vbox{
\psfig{figure=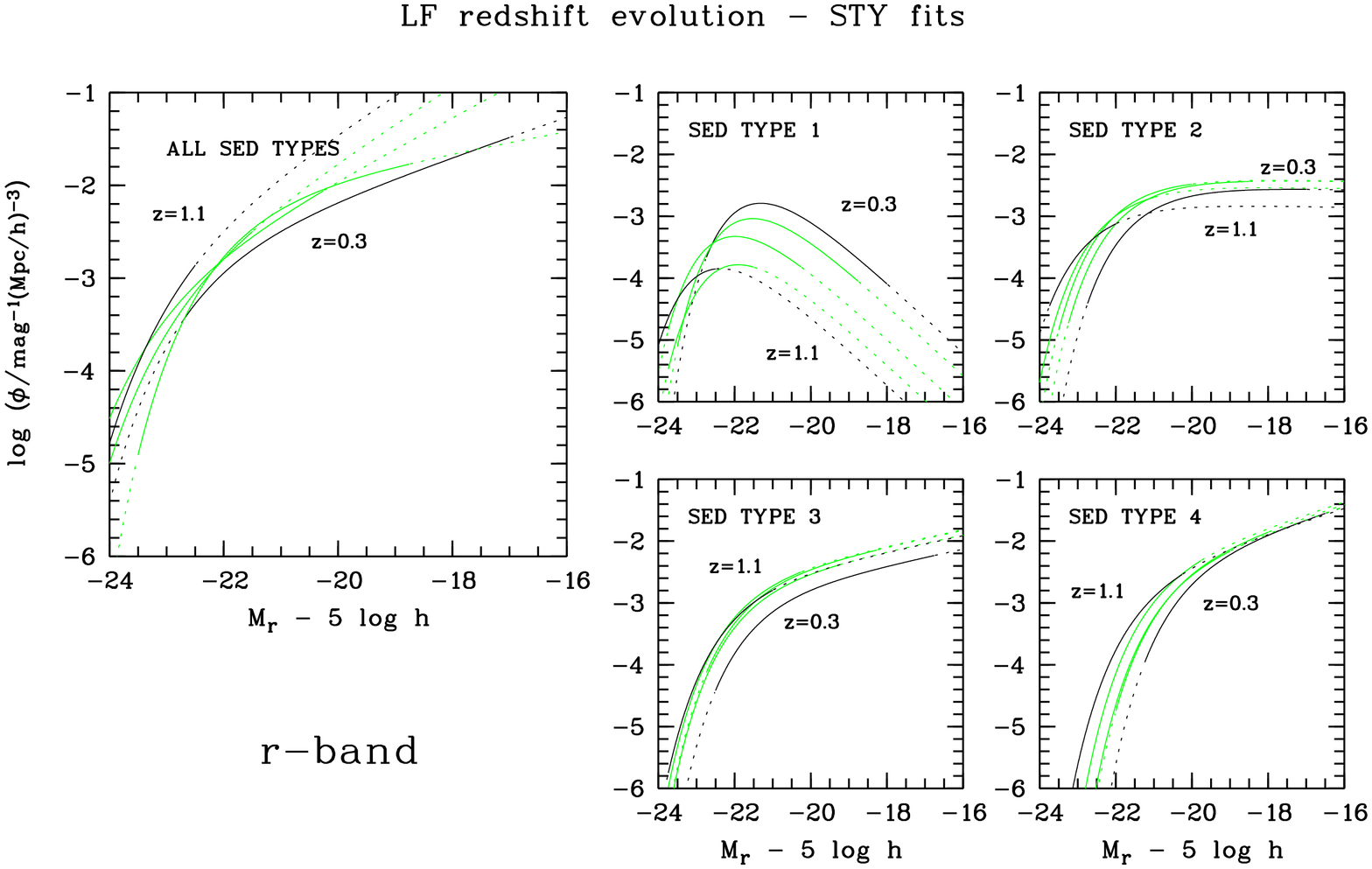,angle=270,clip=t,width=13cm}
\psfig{figure=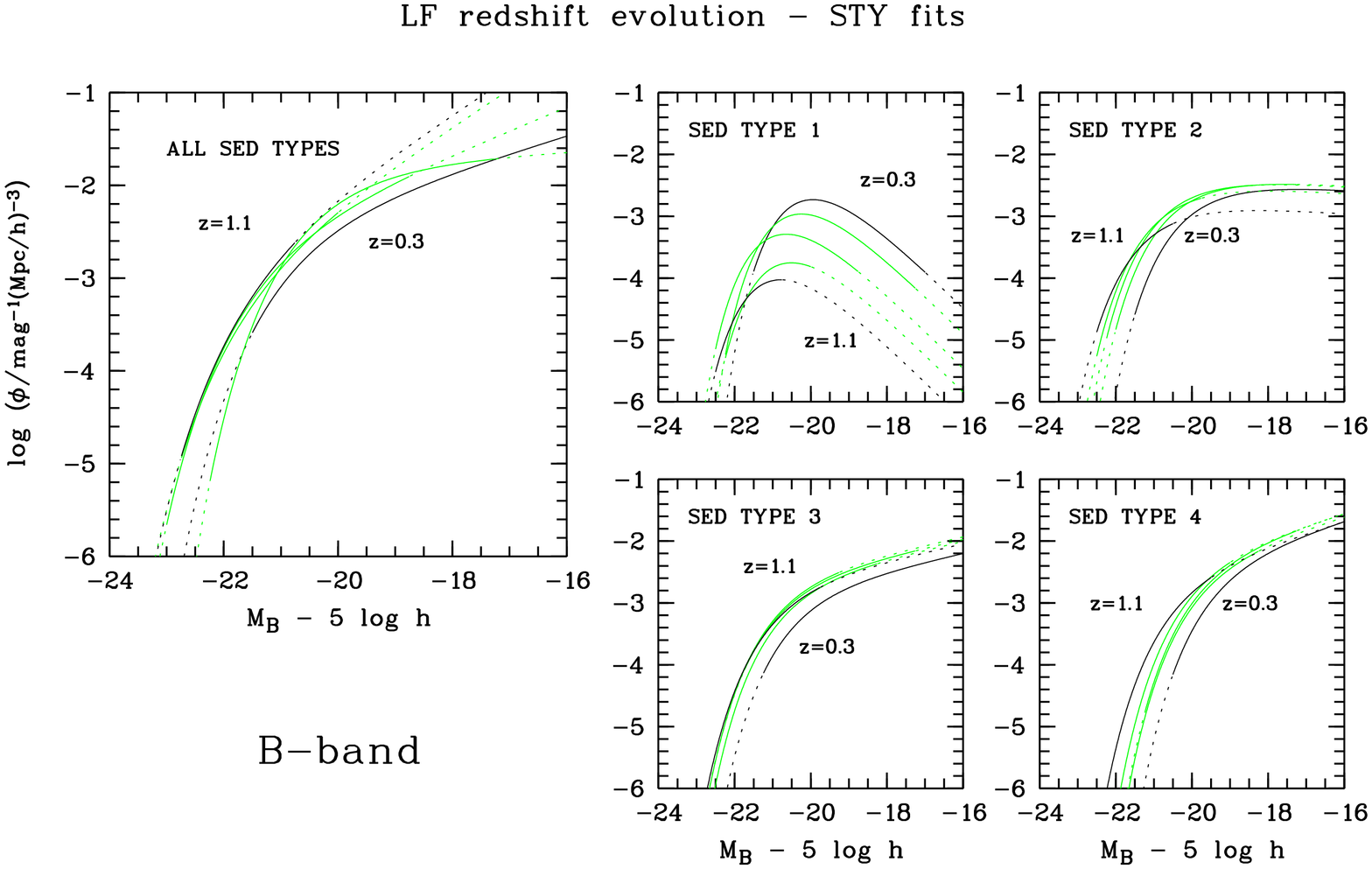,angle=270,clip=t,width=13cm}
\psfig{figure=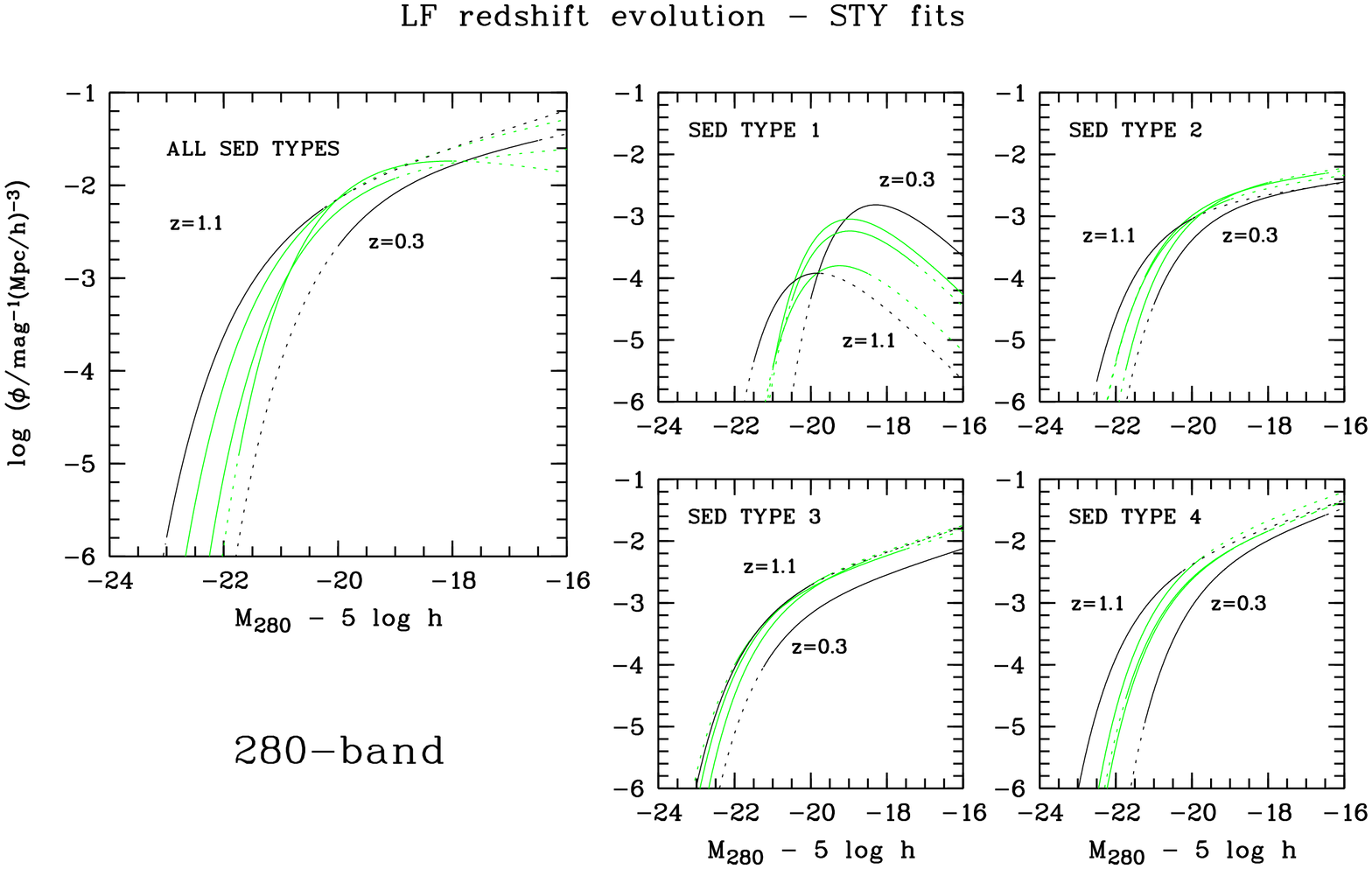,angle=270,clip=t,width=13cm}
}}}
\caption[ ]{Redshift evolution of the luminosity functions $\phi(M_\mathrm{r})$, 
$\phi(B)$ and $\phi(M_\mathrm{280})$: STY fits of Schechter functions 
in five redshift intervals of width $\Delta z=0.2$, centered at 
$z=[1.1,0.9,0.7,0.5,0.3]$. The first and last of these fits is plotted in black, 
and the three enclosed intervals are shown in grey. For the individual SED types 
the faint-end slope $\alpha$ was determined in the ``quasi-local'' sample and 
kept constant for all redshifts (for $\alpha$ values see Appendix, 
Tab.~\ref{lfpars_rs}, \ref{lfpars_bj} and \ref{lfpars_uc}).
\label{lf_evo1}}
\end{figure*}

\begin{figure*}
\centerline{\hbox{
\psfig{figure=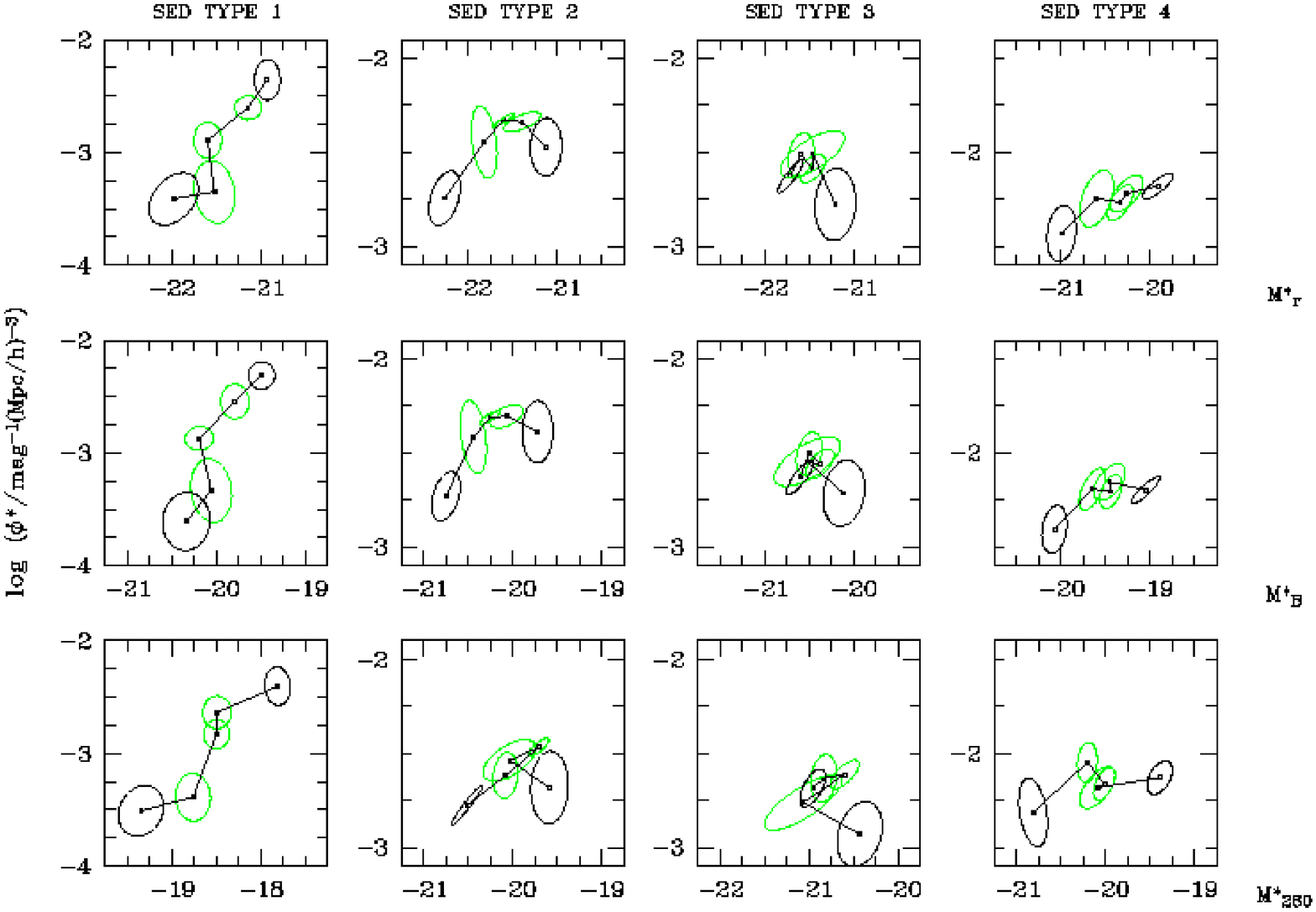,angle=270,clip=t,width=17cm}}}
\caption[ ]{Redshift evolution of the parameters $M^*$ and $\phi^*$:
Results of STY fits in five redshift intervals centered at $z=[1.1,0.9,0.7,
0.5,0.3]$. For individual SED types the faint-end slope $\alpha$ was determined 
in the quasi-local sample and fixed for all redshifts. Plotted are $M^*_\mathrm{band}-
5$~log~$h$ values and 1-$\sigma$ error contours derived from the error in $M^*$, 
the field-to-field variation in $\phi^*$ and the covariance between $M^*$ and 
$\phi^*$. The most distant sample ($z=1.1$) and the quasi-local sample ($z=0.3$)
are plotted in black, and the redshifts in between in grey. For orientation we
note, that in every panel the quasi-local sample is found at the faintest 
(right-most) $M^*$ position.
\label{lf_evo1_all}}
\end{figure*}

In this section we extend the discussion to the redshift interval z=[0.2,1.2],
which contains 25,431 measured redshifts for galaxies at $R<24$. In this section, 
we show only STY fits (Fig.~\ref{lf_evo1}) for the luminosity functions, and use  
them as a basis for our discussion since they form a good representation of the 
$V_\mathrm{max}$ data points. For reference, we show the full set of plots depicting 
$V_\mathrm{max}$ data points together with STY fits for all redshifts, all SED types and 
all restframe bands into an Appendix, where the interested reader can look at 
finer details and find tables with all parameters.

Luminosity function fits to different redshift bins of the same SED type, where
we allowed $M^*$ and $\alpha$ to vary freely, showed that the faint end slope
$\alpha$ does not significantly evolve with cosmic epoch. We therefore
adopted the hypothesis that the faint end slope only depends on SED type, but not 
on cosmic epoch. This approach avoids the ill-defined covariances with the $M^*$ fit, 
as less and less of the sub-$L^*$ regime is fit towards higher redshifts. In practice, 
the STY fits were obtained by measuring $\alpha$ in the quasi-local sample and
keeping it fixed for all other redshifts. While there seems to be an indication 
of a flatter $\alpha$ at $z=[0.4,0.6]$, the quasi-local $\alpha$ again fits
better at $z>0.6$ (see Appendix). At this point we suspect that the apparent, slight
flattening of $\alpha$ in the epoch around to $z\approx0.5$ originates from an 
unknown and still uncorrected feature of our data. 

Using a fixed $\alpha$ for all redshifts works quite well for the luminosity
functions of individual SED types. But due to the change in the type mix with
redshift, this approach is useless for the combined sample of all types. There,
$\alpha$ has been determined in each redshift interval independently. But at 
higher redshift the flatter low-luminosity regime in the LF is less constrained 
by the observations and a correct determination of $\alpha$ is difficult.

The evolution of the galaxy luminosity function is best observed by type. The four 
different SED types show different evolutionary patterns although there are trends 
linking them. Given that the Schechter function 
contains just three free parameters, $(M^*,\phi^*,\alpha)$, keeping $\alpha$ 
constant with redshift means, that we can observe the evolution of the LF as an 
$(M^*,\phi^*)$-vector evolving with redshift $z$ (see Fig.~\ref{lf_evo1_all}). 

If any colour evolution lead only to migration of objects between types, but left
the distribution of restframe colours within an SED type sample unchanged with 
redshift, it would suffice to look at only one restframe passband. The other two 
would then contain purely redundant information, because our SED types are 
defined by restframe colour. However, due to finite width of the colour intervals 
associated with the types, slow migration can shift the mean colour within types, 
and still cause slightly different results among the three passbands. In
principle, such a migration could also lead to a shift in $\alpha$, which again
motivates a bivariate luminosity function depending on restframe colour directly.

As shown in Fig.~\ref{lf_evo1_all}, all types have some luminosity fading with 
cosmic time in common but show quite different trends in density evolution.
For \tp 1, we see a consistent $\ga$10-fold increase in density for all three 
bands when going from $z=1.1$ to $z=0.3$. The steep increase suggests that we
have not yet seen the epoch of maximum density for \tp 1 galaxies, and will
see a continued increase in the future of the universe. Furthermore, all three 
bands show evidence of strong fading --- from $\Delta m\approx 1.0$~mag in the 
r-band to $\Delta m\approx 
1.5$~mag at 280~nm across the redshift range from $z\sim1.1$ to $z\sim0.3$.
The stronger fading in the UV is basically a consequence of the mean colour in
the \tp 1 sample getting redder with time as the member galaxies age further,
a trend also reflected in the drift of the mean template for \tp 1 objects
with redshift (see Fig.~\ref{SEDtemps}).

For \tp 2, we again find uniform evidence from all three bands for fading at the
$\sim1$~mag level with no indication of a change in mean restframe colour. The 
density trend of \tp 2 galaxies is positive at higher redshift and turns slightly
negative towards more recent epochs, suggesting an enclosed epoch of maximum 
density for \tp 2 galaxies around $z\approx 0.5 \ldots 0.7$.

The luminosity function of \tp 3 galaxies shows basically no evolution at $z>0.5$, 
but a later reduction in density of $\sim0.25$~dex, meaning that almost half of 
the galaxies have disappeared from the \tp 3 bin by $z\sim0.3$. 

The starburst galaxies of \tp 4 show a stronger decrease in luminosity, ranging 
from $\Delta m\approx 1.0$~mag in the r-band to $\Delta m\approx 1.5$~mag at 
280~nm across the redshift range from $z\sim1.1$ to $z\sim0.3$. The stronger 
fading in the UV is again a consequence of the mean colour in \tp 4 getting 
redder with time. This is not reflected in any change of the mean template 
with redshift, probably due to the non-monotonic behaviour of the starburst 
templates. The trend in colour is consistent with the starbursts getting 
relatively less prominent within the given underlying galaxy. Alternatively, one 
could speculate about increased dust reddening. The \tp 4 galaxies show no strong
trend in density in any band. The steep faint end of the \tp 4 luminosity 
function leads to a strong covariance between $M^*$ and $\phi^*$ and an increased 
difficulty for perfectly disentangling density evolution from fading. However,
in terms of total luminosity, the epoch of maximum starburst activity certainly 
occured somewhere beyond our redshift limit of $z=1.2$. 

On the whole, it appears that these patterns can be understood as a decrease in 
starburst activity with time following an activity maximum at $z\ga1$ beyond the
limits of this work, in combination with a propagation of the galaxies through
the types as the mean age of the stellar population increases. As the starburst
activity continues to drop, the fraction of galaxies with only old populations
continuously increases.    

The comparison with local surveys from the previous subsection suggests a simple
continuation of the trends found between $\langle z\rangle =1.1$ and $\langle z
\rangle =0.3$, such that the density of \tp 1 galaxies does continue to rise and 
\tp 4 galaxies continue to vanish or fade. \Tp 2 and 3 galaxies remain almost 
unchanged, but we remind the reader that our comparison with local surveys is 
only of limited value due to significant differences in the type definition. 

If the differentiation between fading and density trends is too unreliable in
places, given only limited constraints on the knee of the Schechter function at
higher redshift, the way out is to look at figures of integrated luminosity 
density, which we do in the following section, where we continue to include the 
local samples into the discussion.

\subsection{Evolution of the co-moving luminosity density at different 
 rest-wavelengths since $z\sim 1.2$}

As an alternative to discussing luminosity functions, we can describe the
redshift evolution of the co-moving luminosity density stemming from galaxies 
of different SEDs. I.e. we can explore the
evolution of the integral over the luminosity function, which avoids
the problems arising from the $\phi^*$, $M^*$ and $\alpha$ co-variances.

Conceptually, this approach suffers from the fact that the faint end of the 
luminosity function eventually becomes unobservable for a given redshift, and that 
an extrapolation for this unobserved galaxy contribution becomes necessary. However, 
in the present context this is not a serious limitation: For one, our data show 
directly that the faint end slope covers $-1.5\le \alpha \le 0.5$, depending on 
SED type, so the LF integral will converge quite rapidly for luminosities below 
$M^*$. Further, our data reach considerably fainter than previous deep surveys, 
so for any given $\alpha$ much less extrapolation is necessary.

\begin{figure*}
\centerline{\hbox{\vbox{
\psfig{figure=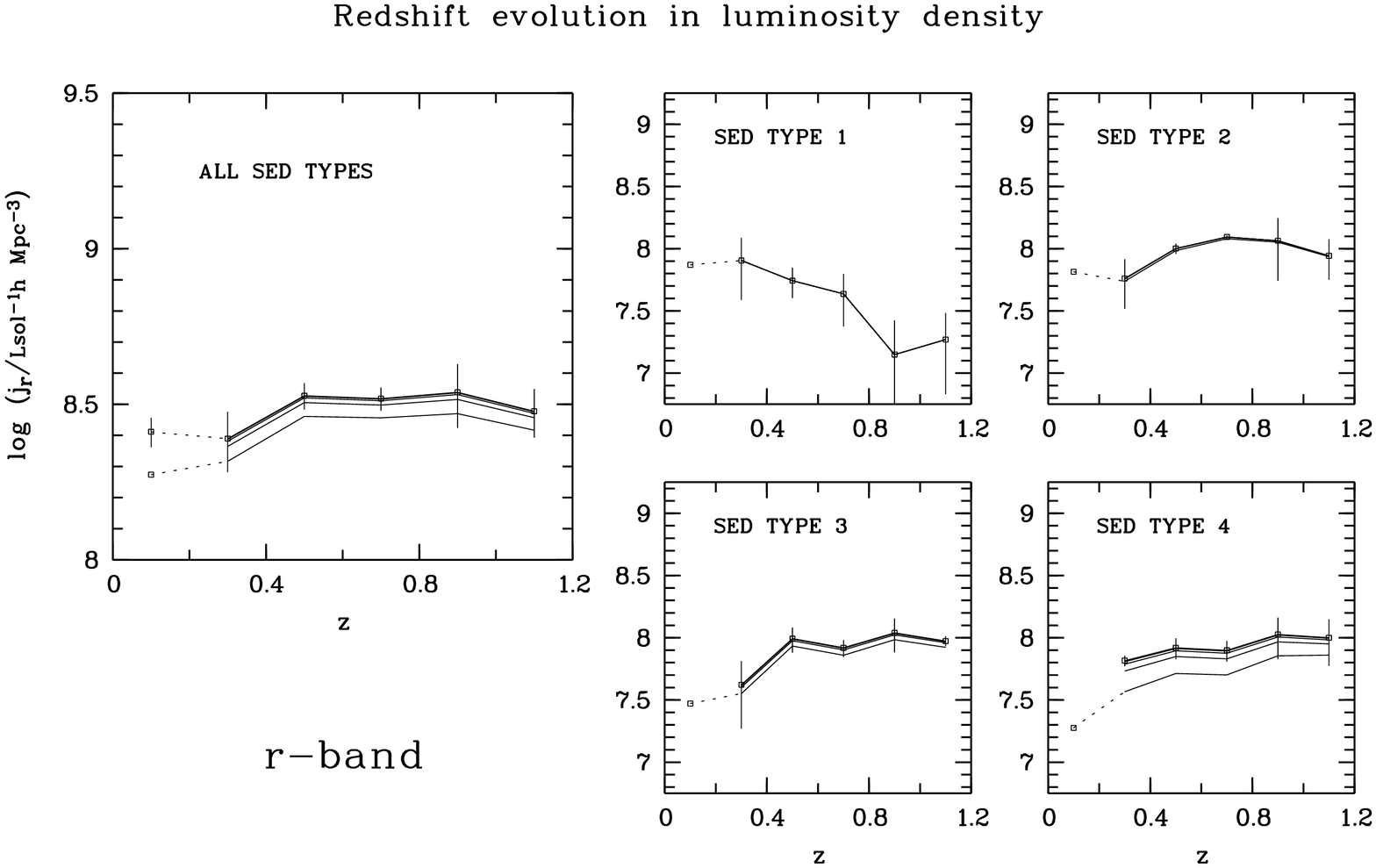,angle=270,clip=t,width=13cm}
\psfig{figure=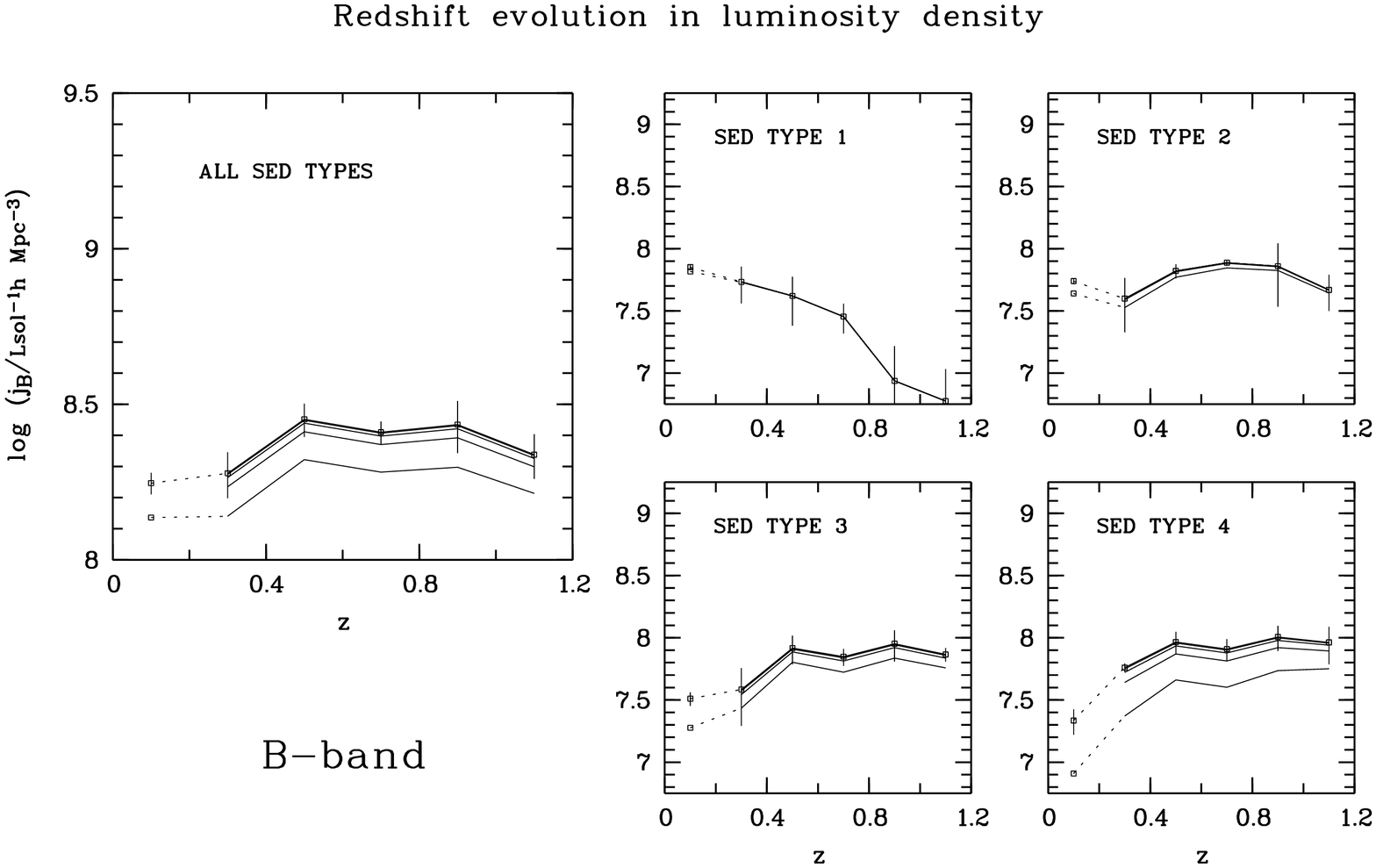,angle=270,clip=t,width=13cm}
\psfig{figure=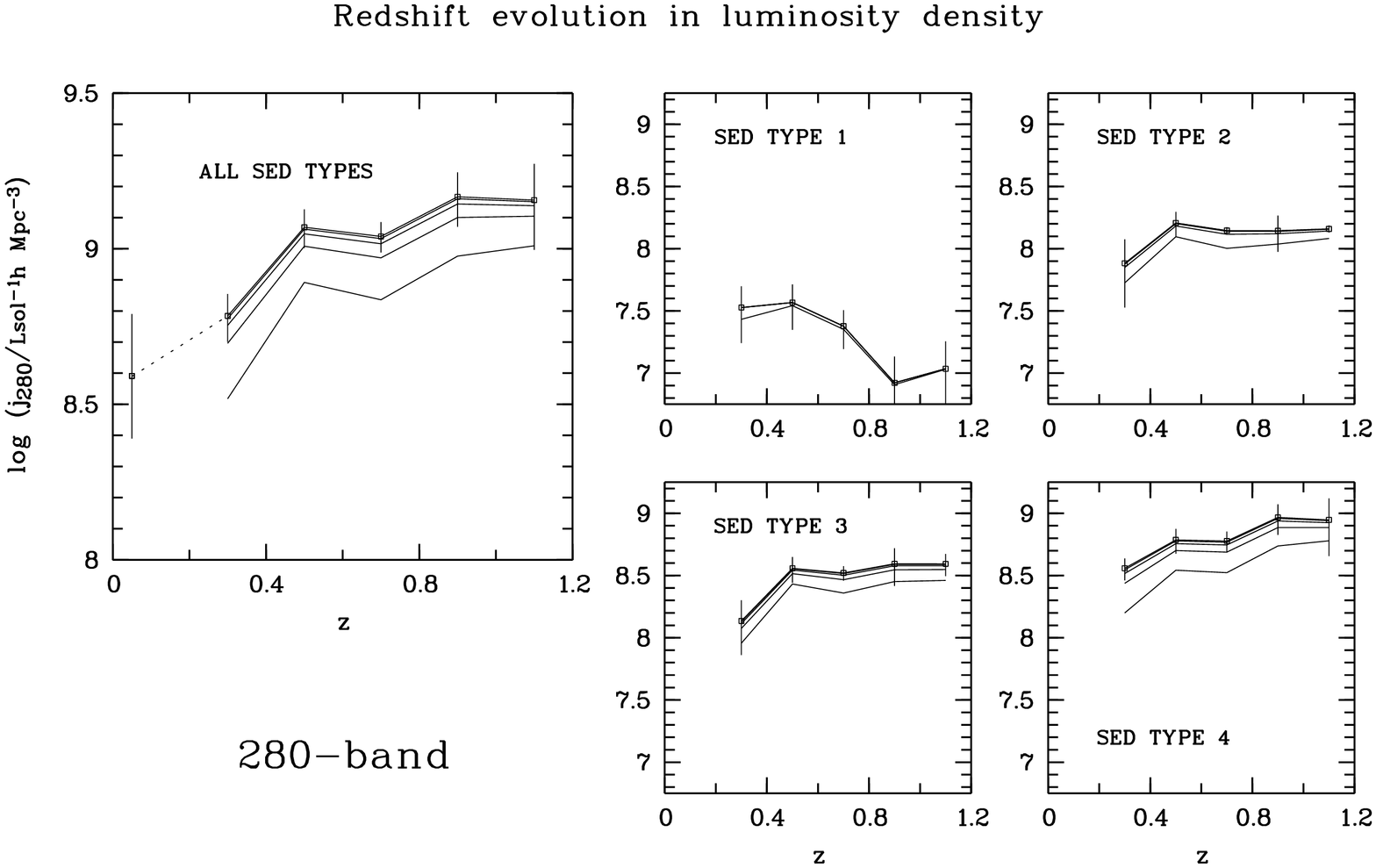,angle=270,clip=t,width=13cm}
}}}
\caption[ ]{Redshift evolution in luminosity density $j_\mathrm{r}$, 
$j_\mathrm{B}$ and $j_\mathrm{280}$:
$j$ integrals calculated from STY fits. The five lines belong to different values 
of the integration limit of $M<[-18,-16,-14,-12,-10]$ and converge to a total 
luminosity for every value of $\alpha>-2$ (see Eq.~11). For \tp 1 $j_\mathrm{r}$ 
and $j_\mathrm{B}$ have already converged at $M<-18$ rendering the five
lines indistinguishable. With $\alpha\approx-1.5$ for \tp 4 galaxies, the integral 
to the faintest levels contains up to 2.5$\times$ the luminosity measured to $M<-18$.
Local data from SDSS (r-band), 2dFGRS (transformed into B-band) and from Lilly et al.
(1996) as based on Loveday et al. (1992) in the 280-band. 
\label{ld_evo}}
\end{figure*}

Fig.~\ref{ld_evo} shows that the integrated luminosity density in the different 
wavebands is quite well constrained by direct observations. Five different lower
luminosity limits have been applied for the integration, ranging from $M<-18$ to
$M<-10$ to show the effect of the extrapolation for different types. In any 
bandpass the luminosity density integral has largely converged if the integration 
interval extends to $M\la-16$.

In all three bands we see roughly the same picture for the individual types but 
significant differences for the combined luminosity of all galaxies, as the mix
of types changes with redshift. As $j \propto L^*\phi^*$, the former statement is 
just a consequence of the observation, that already the evolution of $M^*$ and 
$\phi^*$ was consistently seen among the different bands. The B-band 
displays most clearly the overall trend in luminosity density from $z\sim1$ to now: 
a decrease among strongly star-forming galaxies is accompanied by perhaps a maximum 
of \tp 2 galaxies at intermediate redshifts and a ten-fold increase in \tp 1 objects.

\begin{figure*}
\centerline{\hbox{
\psfig{figure=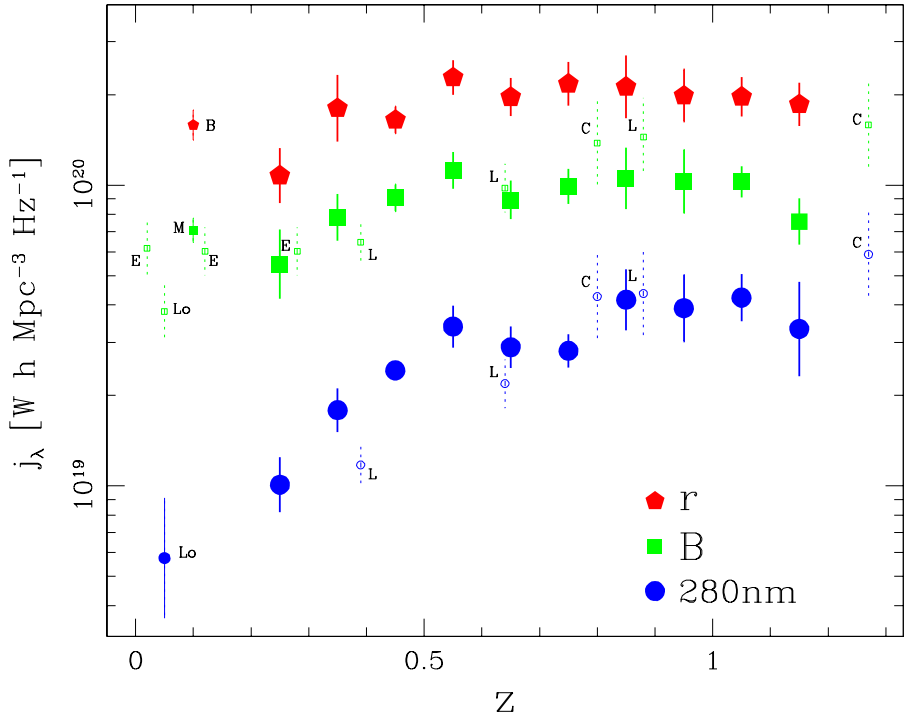,clip=t,width=12cm}}}
\caption[ ]{Overall redshift evolution of the luminosity density in all bands:
The data points represent the total luminosity density (integrated to faintest 
luminosities) for 280~nm (circles), B (squares) and r (pentagons). The large 
solid symbols are the values from this paper, but split into finer redshift bins
of $\Delta z=0.1$ for clarity, the small solid symbols at the low 
redshifts represent the local comparison surveys \cite{Bla01,Mad02}, and small 
open symbols represent the results from previous redshift surveys (Madau, 
Pozzetti and Dickinson 1998 listed values for Ellis et al. 1996, Connolly et al. 
1997, and Lilly et al. 1996 also listed values for Loveday et al. 1992).
Luminosities of $M_\lambda=0$ objects for conversion into physical units: 
$L_\mathrm{r}  =3.84\times10^{13}$~W/Hz, $L_\mathrm{B}  =4.93\times10^{13}$~W/Hz
and$L_\mathrm{280}=1.18\times10^{13}$~W/Hz.
\label{j_of_z}}
\end{figure*}

The overall evolution of the luminosity density is summarized in Fig.~\ref{j_of_z},
showing our own results (large filled symbols) in comparison to earlier work 
(open symbols) and large, contemporary surveys of the local universe (small filled 
symbols). The values of older work have, of course, been adjusted to the
cosmological parameters used here. 

In the B- and r-band the luminosity density remains virtually unchanged 
from $z=1.1$ to $z=0.5$, and even j$_\mathrm{280}$ drops only little down to $z=0.5$. 
All bands show a subsequent decrease in luminosity density to the present epoch, 
most strongly at 280~nm. Between $z\sim0$ and $z\sim1$ we observe $j_\mathrm{280}$ to 
drop by a factor of $\sim6$, while $j_\mathrm{B}$ and $j_\mathrm{r}$ drop only little.
If we considered our data at $z=0.5$ to be outliers towards the top, the flat 
domain would reduce to $z=0.7\ldots1.1$ in the B- and r-band. The 280-band
would then show a more gradual change in the slope from the flatter domain 
around $z\sim1$ to the steeper gradient in the local universe.

We can not quite confirm the steep increase oberved by Lilly et al. (1996), that
leads to apparently inconsistent data at high redshift. By using the Loveday (1992)
as a local reference, their data suggest an increase by a factor of 4.6 from $z=0$ 
to $z\sim1$ in $j_\mathrm{B}$, which is much stronger than what we see. However, 
we believe the results are consistent within their error bars.

\begin{figure*}
\centerline{\hbox{
\psfig{figure=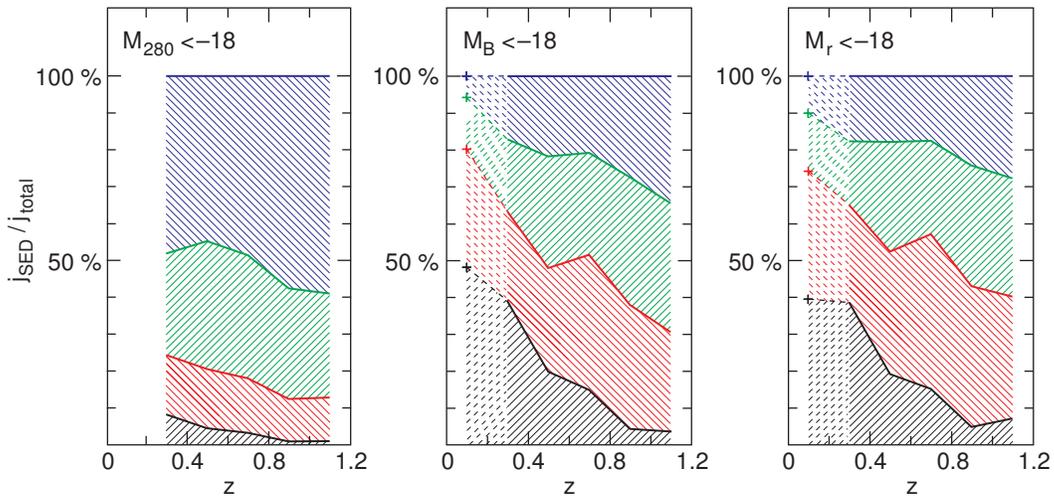,angle=0,clip=t,width=14cm}}}
\caption[ ]{Redshift evolution in luminosity density fraction by type:
Fraction of the total luminosity density contributed by a single SED type
(\tp 1 at the bottom, \tp 4 at the top). The total luminosity integrals are 
calculated from the STY fits. Local data: 2dFGRS and SDSS.}
\label{SED-fractions}
\end{figure*}

Finally, we show what fraction of the luminosity density at a given wavelength 
arises from which SED type. This is illustrated in Fig.~\ref{SED-fractions} for 
all three wavelengths. While the strongly star-forming galaxies of \tp 3+4
produced the bulk of the radiation in all three bands at $z\sim 1$, they nowadays 
contribute only an important fraction of the luminosity when considering the 
near-UV (280~nm). On the contrary, the fraction of light arising from early-type 
SEDs (\tp 1) has increased by large factors in each of the bands. 

\subsection{Comparison with other higher-redshift surveys}

\begin{table}
\caption{Characteristics of galaxy redshift surveys compared here. {\it Note:} 
We ignore the local CFRS sample at $z<0.2$ due to its small size, although 
results have been published. The most distant bin of its red sample is 
actually at median redshift $\sim 0.87$.
\label{surveycomp} }
\begin{tabular}{lcrrc}
Survey   & Median of  bins		& objects	& Types	& $\alpha(z)$ \\
	 & min$\ldots$max	\\
\noalign{\smallskip} \hline \noalign{\smallskip} 
CFRS     & $\sim0.35\ldots1.1$	& $\sim   600$ 	& 2	&  \\ 
CNOC2    & $\sim0.2 \ldots0.45$	& $\sim 2,200$ 	& 3	& const \\ 
CADIS    & $\sim0.4 \ldots0.9$	& $\sim 2,800$ 	& 3	&  \\
COMBO-17 & $\sim0.33\ldots1.1$	& $\sim25,000$ 	& 4	& const \\
\noalign{\smallskip} \hline
\end{tabular}
\end{table}

In this section, we would like to compare the findings on the type-dependent 
evolution of galaxies at $z<1.2$ with other recent surveys, i.e. the Canadian
French Redshift Survey, CFRS \cite{Lilly95}, the CNOC2 survey of the Canadian
Network for Observational Cosmology \cite{Lin99}, and the Calar Alto Deep
Imaging Survey, CADIS \cite{Fri01}. Our comparison will be done qualitatively,
taking into account the differences arising from incompatible type definitions.

We summarize beforehand, that altogether the observations of all four surveys 
are consistent with each other and no significant contradictions have been seen 
exceeding the size of the error bars. However, COMBO-17 constrains the evolution
of galaxies much more clearly than any of the comparison surveys, due to larger 
sample size and/or wider redshift coverage. Especially, the smaller samples of
previous surveys did not allow to see any second-order effects as demonstrated 
by the peak in \tp 2 density. Instead they had to rely on fitting first-order 
models with redshift to the data.  

Whenever seemingly contrary results were derived, their origin is mostly to be 
found in different type definitions or assumptions on parameter evolution. We 
therefore believe, that it is not worth the effort to reproduce all the 
different type definitions between the surveys in order to do a precise 
quantitative comparison, and omit this exercise.

\subsubsection{Characteristics of comparison surveys}

All three comparison surveys derived the luminosity function in the restframe 
B-band. Most relevant for the following discussion are these characteristics 
(see Tab.~\ref{surveycomp}):
\begin{itemize}
\item The CFRS covers the same wide redshift range as COMBO-17, $z=[0.2,1.2]$,
 and features the same median redshift of its sample ($\langle z\rangle=0.56$),
 but provides the smallest of the samples considered here with 591 objects. The
 results are derived for two galaxy types with $\alpha = \alpha(z)$, a red sample
 corresponding roughly to our \tp 1+2, and a blue sample resembling our \tp 3+4.

\item The CNOC2 reaches only out to $z=0.55$, the median reshift of COMBO-17 and
 the CFRS, and also less deep, being limited to $R_\mathrm{C}<21.5$. However, it 
 contains over 2000 galaxies and has derived results with fairly small error bars
 and $\alpha = const$ for three types of galaxies. Assuming that $\alpha$ depends
 monotonically on restframe colour, and comparing the values found for the LF by
 type, we could very crudely identify their three types with our \tp ``1-1.5'',
 \tp ``1.5-2'', and \tp 3+4.

\item Finally, CADIS covers almost the full redshift range of COMBO-17 and CFRS,
 although only three redshift bins were used with median redshifts ranging from
 0.4 to 0.9 as opposed to 0.3 to 1.1 for COMBO-17. Like CNOC2 it contains more
 than 2000 galaxies, and results were obtained with $\alpha = \alpha(z)$ for three
 types, corresponding roughly to our \tp 1, \tp 2 and \tp 3+4. 
\end{itemize}

\subsubsection{Redshift evolution of $\alpha$?}

The CNOC2 survey considers $\alpha$ to be not changing with redshift, but the
CFRS and CADIS allow $\alpha$ to change while investigating a much wider redshift
range than CNOC2. Although COMBO-17 studies a wide redshift range as well, we 
found that the assumption of $\alpha$ being constant in redshift is reasonable,
provided that the types are not too broadly defined. 

Indeed, we are compelled to
believe, that $\alpha$ is a monotonic function of restframe colour, but does not
change with redshift at least at $z\la 1$. Any galaxy type encompassing a wide 
range of colours can obviously experience a strong evolution in its sub-type mix. 
In that case, the $\alpha$ of the whole type sample is dominated by different 
sub-types at different epochs and therefore reflects their respective, different 
$\alpha$ values. 

We like to stress again, that if the steep LF of starburst 
galaxies does not turn over at some low luminosity, it will always drive the 
$\alpha$ of any whole sample to a steep value at the faintest end, which should 
even be observable in local samples if only galaxies of sufficiently low 
luminosity were observed.

It is this assumption of $\alpha$ being constant, combined with a fairly dense
sample, that allows the CNOC2 and COMBO-17 to derive the clearest evolutionary
patterns. All three previous surveys and possibly even COMBO-17 might have too
small and shallow samples to measure positively any evolution in $\alpha$. 
Allowing it to vary freely in CFRS and CADIS has then greatly reduced the 
constraints on the evolutionary patterns they derived.

\subsubsection{CFRS vs. COMBO-17}

When looking towards high redshift, the CFRS reports to see: \begin{itemize}
\item  no evolution of the luminosity function (neither 
in $\phi^*$ nor in M$^*$) in its red sample
\item  a significant amount of brightening and steepening in the
luminosity function of the blue sub-sample.
\end{itemize}

Their red smaple is at all redshifts dominated by our \tp 2 galaxies which show
only little evolution that could easily be invisible in a small sample. The fading 
with time observed by us might have been compensated by an increasing density of
luminous \tp 1 objects, rendering any changes on the luminous side and around the 
knee of the combined \tp 1+2 luminosity function hard to detect.

Their blue sample is dominated by our \tp 3 galaxies at $M_B<-20$ and by our
\tp 4 galaxies at $M_B>-20$. 
When looking at the steep, luminous end of the LF the
density increase of \tp 3 objects observed by us appears like a brightening as
well. At the faint end, the brightening of \tp 4 objects seen by us causes an
increasing fraction (with redshift) of \tp 4 objects being seen in the faint 
domain of the combined \tp 3+4 luminosity function. This is an illustrative case 
of a change in sub-type mix mimicking a steepening.

\subsubsection{CNOC2 vs. CFRS and COMBO-17}

The CNOC2 concluded to see (when looking back in redshift): \begin{itemize}
\item  strong rise in luminosity but a reduction in density for its early-type sample
\item  gentle rise in luminosity but no trend in density for its intermediate sample
\item  no trend in luminosity but a strong rise in density for its late-type sample
\end{itemize}

They fit an evolutionary model to their samples whereby $M^*(z)-M^*(0) = -Qz$ and 
$\log{\rho(z)/\rho(0)} = 0.4Pz$ with type-dependent evolutionary parameters $P$ and
$Q$. They derive values supporting the clear evolutionary pattern from above, albeit
with large error bars due to the short redshift baseline. In principle, we can 
predict changes $M^*(1.1)-M^*(0.3)$ and $\log{\rho(1.1)/\rho(0.3)}$ with respective 
errors, as we should see them in COMBO-17. After trying to adjust the mismatch in 
type definitions, the results become rather consistent within large errors.

For its blue sample CNOC2 found a clear $\phi^*$ drop with little $M^*$ change, 
while the starburst galaxies in COMBO-17 appear to evolve only in $M^*$, but there 
is indeed no contradiction, as the CNOC2 blue sample is dominated by our \tp 3 
objects, which are more prominent at low redshift and $R<21.5$.

The apparent differences between the CNOC2 and CFRS results are best explained
by the difference in type definition. For the red CFRS sample, also a partial 
cancellation of opposing trends in density and luminosity plays a role. When
combining part of the intermediate CNOC2 sample with its late-type sample, it
is easy to imagine that on the whole brightening and steepening is seen, again.

\subsubsection{CADIS vs. CFRS, CNOC2 and COMBO-17}

CADIS concluded to see (when looking back in redshift): \begin{itemize}
\item  no trend in luminosity but a reduction in density for its red sample
\item  no trend in luminosity and only a slight reduction in density for its
 intermediate sample
\item  a significant amount of brightening and steepening in its blue sample.
\end{itemize}

The CADIS results have been claimed to be mostly consistent with the CFRS results
while improving on accuracy due to an increased sample size. They appear to be in
disagreement with CNOC2, involving a dispute on the justification of keeping 
$\alpha$ fixed with redshift. We believe, however, that the observations are 
consistent, and the conclusions could become comparable after matching up the type 
definitions and agreeing on the behaviour of $\alpha$. Keeping $\alpha$ free also
makes it more difficult to separate trends in $M^*$ and $\phi^*$.

Again, the brightening and steepening observed for the blue sample resembles the
combined evolution of \tp 3+4 with the $\alpha$ value of \tp 4 being more 
important at higher redshift and that of \tp 3 being dominant at low redshift 
--- at least within the luminosity interval observed. We repeat here our
claim that the COMBO-17 observations do not support any change of $\alpha$ with
redshift for a sample with any given narrow interval in restframe colour. It is
simply changes in type mix occuring in broadly defined types that lead to the 
observation of $\alpha$ changing in the typical luminous domains as they are 
currently observed in high-redshift surveys.

\section{Summary and Conclusions}

Based on photometry in 17 (mostly medium-band) filters obtained by the COMBO-17 
survey on three independent fields of 0.26~$\sq\degr$ each, we have derived 
redshifts and SED classifications for $\sim 25,000$ galaxies to $R\la 24$. 
The redshifts may be viewed either as high-precision photometric redshift 
estimates or very low resolution (R$\sim12$) spectral redshifts. They have a 
precision of $\sigma_z\approx 0.03$ and lie mostly in the range $0.2<z<1.2$ 
with a median redshift of 0.55. The upper limit of galaxy redshifts considered 
here is set by the availability of spectroscopic cross-checks confirming their
precision and accuracy. Our results are unaffected by random redshift errors of 
$\sim0.03$, and are robust even for somewhat larger, systematic errors.

Compared to previously published large-sample surveys with better redshift 
accuracy, COMBO-17's flux limit is nearly two magnitudes fainter and the number 
of galaxies is more than an order of magnitude larger. Taken together, the 
properties of the survey allow us to draw up a comprehensive picture of how the 
luminosity function and the SEDs of galaxies have evolved over the last half of 
cosmic evolution. The survey is deep enough that the bulk of the stellar light 
over this redshift range is directly observed. As far as possible, we complement 
our data by ``local'' information from low-redshift surveys (e.g. 2dFGRS and SDSS), 
as our survey volume is too small for $z<0.2$.

The observed galaxy SEDs cover a wavelength range of $\lambda_\mathrm{obs} = 350 \ldots
930$~nm and allows us to determine restframe properties of the population at 
280~nm and at $B$ with hardly any extrapolation in the range $0.2<z<1.2$. We 
also derived restframe properties in the SDSS r-band, by applying a type-dependent 
extrapolation of the observed SED at $z\ga0.5$.

The goal of this present paper is to present an empirical picture of the evolution 
of the galaxy population as a function of luminosity, SED and redshift, as a solid 
constraint for any models of statistical galaxy evolution. Direct comparison with 
cosmolgical models will be reserved for future papers. Our main findings are as 
follows:
\begin{enumerate}

\item Survey areas approaching an area of 1~$\sq\degr$ are necessary to reduce 
field-to-field variations at L$_*$ below a factor of two in redshift bins of 
$\Delta z \sim 0.2$.

\item The faint end slope of the luminosity function depends quite strongly on 
SED-type at all observed redshifts. However, within a given (non-evolving) SED 
type there is no evidence for evolution of the luminosity function shape with 
redshift (to $z\sim1$).

\item A comparison of our ``quasi-local'' sample at $z=[0.2,0.4]$ with 2dFGRS 
and SDSS shows consistent results around the knee of the luminosity function,
but a steeper faint end in COMBO-17, probably reflecting the evolution of
faint blue galaxies between the median redshifts $\sim0.05$ and $\sim0.34$.

\item If we define SED type via a fixed, present-day spectral sequence for 
galaxies, from ellipticals to  young starbursts, then the evolution of the 
luminosity function from $z\sim 1.2$ to now depends strongly on SED type, 
both in its density normalization, $\phi^*$, and in its characteristic 
luminosity, $M^*$:

(a) The faint end slope is $\alpha = 0.50\pm 0.20$ for early type galaxies 
(SED \tp 1), steepening to $\alpha = 1.50\pm 0.06$ for galaxies with the 
bluest starburst colours (SED \tp 4).

(b) For early type galaxies (defined by z=0 colours bluer than Sa galaxies) 
$\phi^*$ increases by more than an order of magnitude from $z=1.2$ to now, 
while M$_*$ gets fainter by more than a magnitude.

(c) For the latest type galaxies $\phi^*$ remains roughly constant over the 
redshift range probed, while M$_*$ gets fainter by almost two magnitudes.

\item When averaging over all SED types, the evolution of the luminosity 
density depends considerably on the rest-wavelength considered:

(a) In the restframe B- and r-bands, the integrated luminosity density 
remains constant from $z=1.1$ to $z=0.5$, dropping subsequently to the present 
epoch by perhaps 30\%.

(b) In the restframe near-UV (280~nm) the integrated luminosity density drops 
by a factor of six from $z\sim1$ to now, where it appears that much of that 
drop occurs at redshifts below $z\sim 0.6$. Our more accurate estimates of the 
near-UV luminosity density imply a considerably shallower evolution than 
indicated by the data of Lilly et al. (1995), but is consistent within their 
1.5$\sigma$ confidence limits.

\item The ratio of the stellar luminosity density contributed by early and 
late SED type galaxies at different wavelengths changes quite drastically with 
redshift: At redshift $z\sim 1$ strongly star-forming galaxies (\tp 3 and 4) 
make up the bulk of the radiation in all three restframe bands with a 
contribution near 90\% at 280~nm. At the present epoch, in particular the 
extreme starbursts (\tp 4) play only a significant role at 280~nm, providing 
about half of the luminosity there. At the same time, the fraction of light 
from galaxies with early-type spectra (\tp 1) have increased from below
10\% to 40\% in the B- and r-bands over the same period.

\end{enumerate}

We reserve any rigorous modeling of these results for a later paper, as it appears that
the data warrant a cosmologically motivated interpretation, such as a comparison 
with semi-analytic or N-Body/SPH models rather than any empirical speculation about
``luminosity evolution" or ``density evolution" of various sub-populations.
Nonetheless, it seems approriate to make a few qualitative remarks in closing:
our results confirm earlier findings that the epoch of galaxy formation, or star
formation in galaxies, has been waning over the second half of the present cosmic age,
leading to a strong decrease of star-burst galaxies (both luminous and faint) accompanied
by the strong emergence of ``old" galaxies, whose overall SED is redder than an Sa
template spectrum. 
The COMBO-17 data set permits to study this evolution differentially for different
SED types and still provides much smaller error bars, due to the enlarged survey
area and sample increase. Compared to earlier work, the decline of cosmic star formation
activity appears gentler, i.e. the near-UV luminosity density as a proxy for the
density of young stars, is found to drop only by a factor of five from $z=1.1$ to now.

Beyond questions of interpretation, there are a number of observational issues that 
still need to be addressed. Calibration of our 17-band redshifts through spectra for 
samples of many hundred objects is forthcoming in the course of the ESO-GOODS key 
project, and will increase the reliability of our estimates even further. The 
addition of several other fields will reduce the impact of the field-to-field
variation, now still a limiting factor in several aspects of the analysis.
Finally, the full analysis of the SEDs and the addition of near-IR data will
allow us to extend the analysis to include direct observational estimates
of the stellar mass.

\begin{acknowledgements}
We would like to thank our referee, Dr Matthew Colless, for a large number of 
important comments improving the manuscript, and for his fast response in the
process. We also thank Dr Lutz Wisotzki for observing the standard stars for
COMBO-17. Furthermore, we thank Dr Mark Dickinson for comments on the evolution
of the luminosity density. This work was supported by the DFG--SFB 439 and by 
the PPARC rolling grant in Observational Cosmology at University of Oxford. 

\end{acknowledgements}

\appendix
\section{Luminosity functions plots and data}

The appendix contains detailed tables and plots of luminosity functions
for all restframe passbands, all SED types and all redshift intervals. 
Whenever, the reader likes to refer to a parametric luminosity function 
for all galaxy types combined we strongly suggest to use a sum of the 
type-specific functions, rather than the Schechter fit to the combined
sample, which is not entirely appropriate, especially in the extrapolated
faint domain. For this reason, the luminosity density of all galaxy types
combined is given as the sum of the four types rather than calculated from
the inappropriate Schechter fit.

\begin{table}[h]
\caption{Field-to-field variation of $\phi^*$ in the ``quasi-local'' sample 
of 5,674 galaxies at $z=[0.2,0.4]$.
\label{phi_var}}
\begin{tabular}{lrrrr@{$\pm$}r}
band  	& $\phi^*_\mathrm{CDFS}$ & $\phi^*_\mathrm{A901}$ & $\phi^*_\mathrm{S11}$ &
	  \multicolumn{2}{c}{ $\phi^*_\mathrm{aver}$ } \\
	&  \multicolumn{5}{c}{ ($10^{-4}~h^3$~Mpc$^{-3}$) } \\
\noalign{\smallskip} \hline \noalign{\smallskip}
r	& 18.72	& 32.69	& 32.56 & 27.99&  8.03 \\
B	& 38.57	& 63.15	& 58.09 & 53.27& 12.98 \\
280	&100.02	&190.47	&184.75 &158.41& 50.65 \\
\noalign{\smallskip} \hline
\end{tabular}
\end{table}

\clearpage

\begin{figure*}
\centerline{\hbox{
\psfig{figure=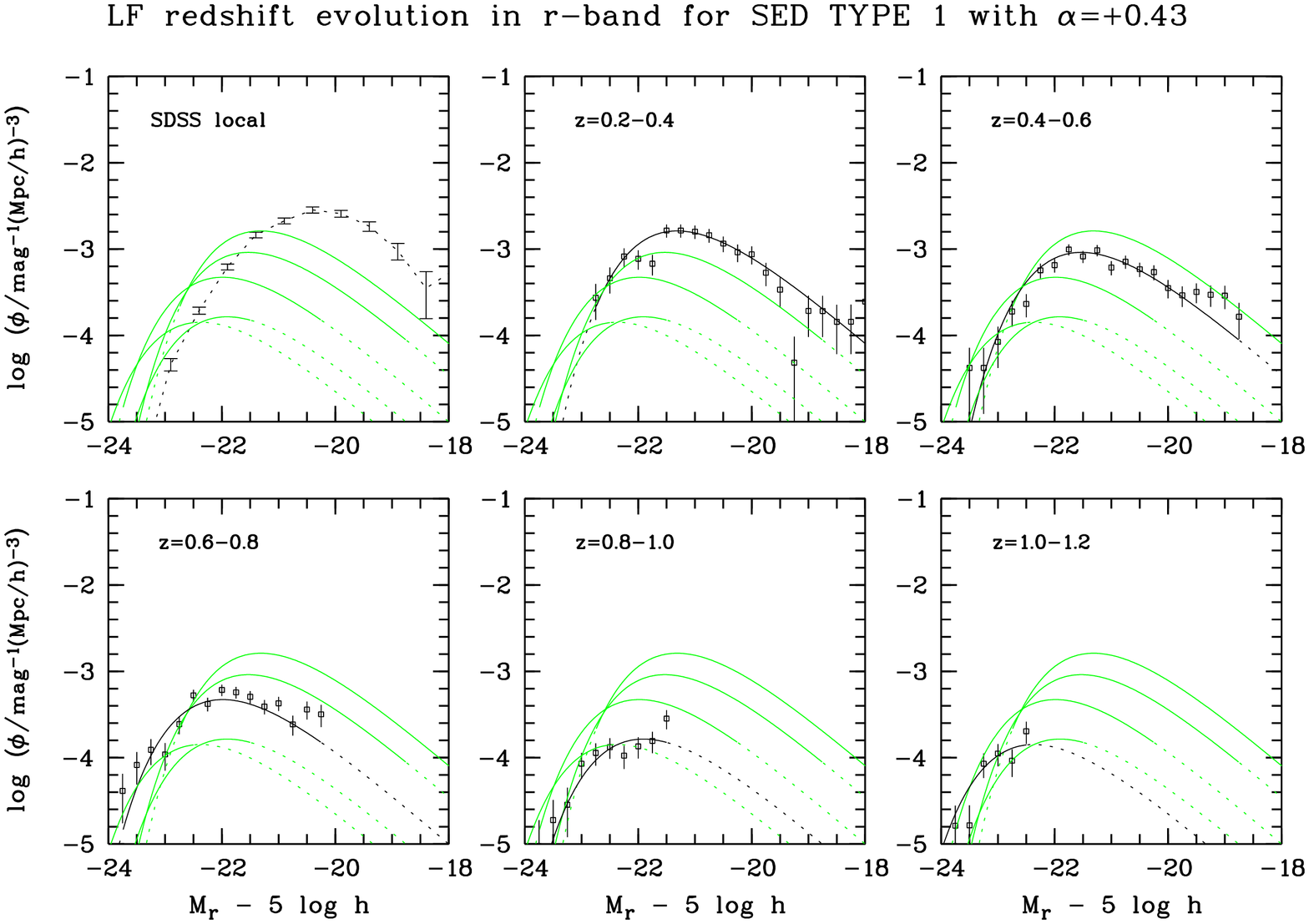,angle=270,clip=t,width=15.5cm}}}
\caption[ ]{Redshift evolution of $\phi(M_\mathrm{r})$ for type 1 galaxies:
$V_\mathrm{max}$ data points are shown with error bars for one redshift interval per panel.
The corresponding STY fit is plotted as a black line, while the fits for the other
redshifts are shown as grey lines. Local reference: SDSS data.
\label{lf_evo1r}}
\end{figure*}

\begin{figure*}
\centerline{\hbox{
\psfig{figure=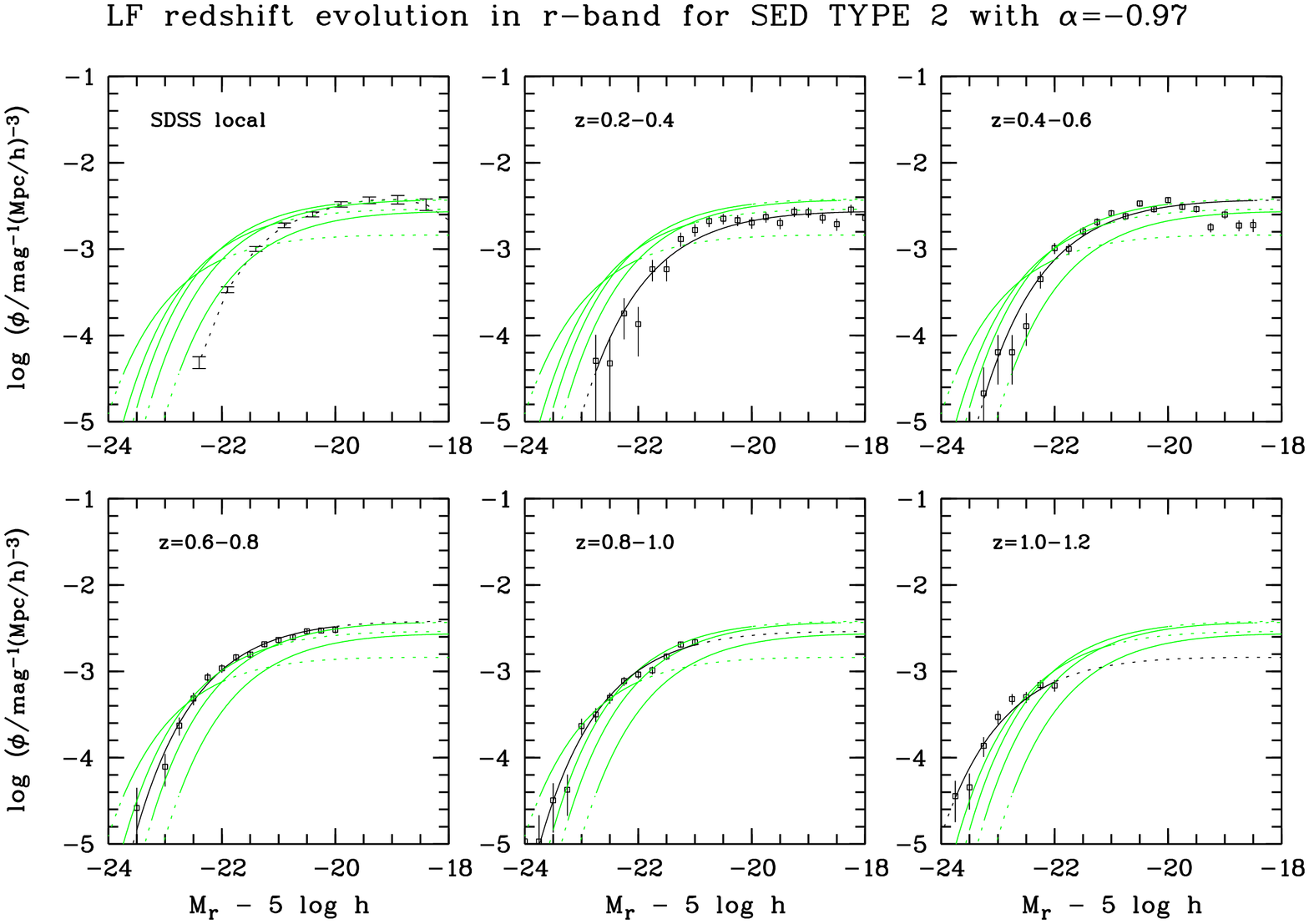,angle=270,clip=t,width=15.5cm}}}
\caption[ ]{Redshift evolution of $\phi(M_\mathrm{r})$ for type 2 galaxies:
$V_\mathrm{max}$ data points are shown with error bars for one redshift interval per panel.
The corresponding STY fit is plotted as a black line, while the fits for the other
redshifts are shown as grey lines. Local reference: SDSS.
\label{lf_evo2r}}
\end{figure*}

\begin{figure*}
\centerline{\hbox{
\psfig{figure=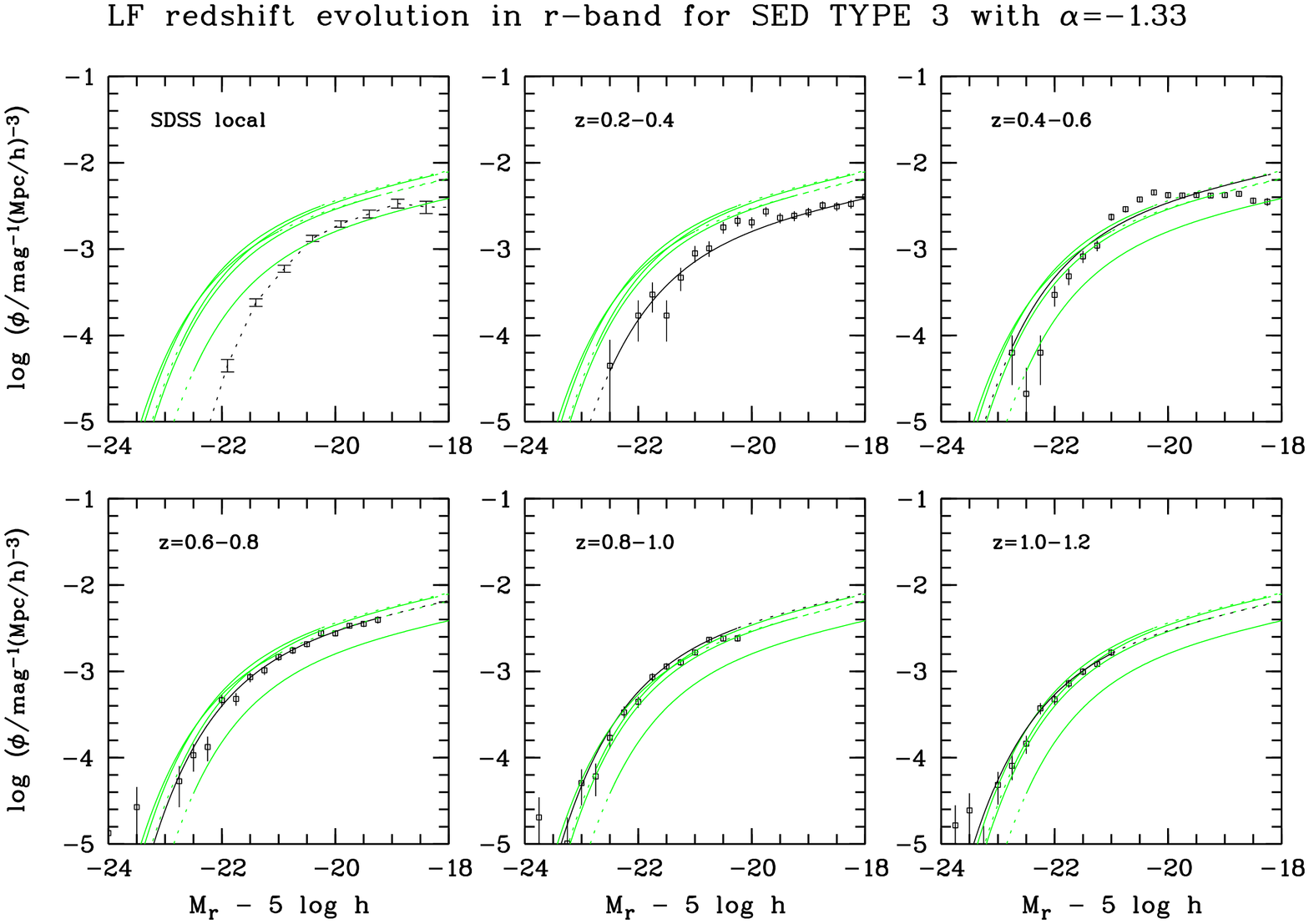,angle=270,clip=t,width=15.5cm}}}
\caption[ ]{Redshift evolution of $\phi(M_\mathrm{r})$ for type 3 galaxies:
$V_\mathrm{max}$ data points are shown with error bars for one redshift interval per panel.
The corresponding STY fit is plotted as a black line, while the fits for the other
redshifts are shown as grey lines. Local reference: SDSS.
\label{lf_evo3r}}
\end{figure*}

\begin{figure*}
\centerline{\hbox{
\psfig{figure=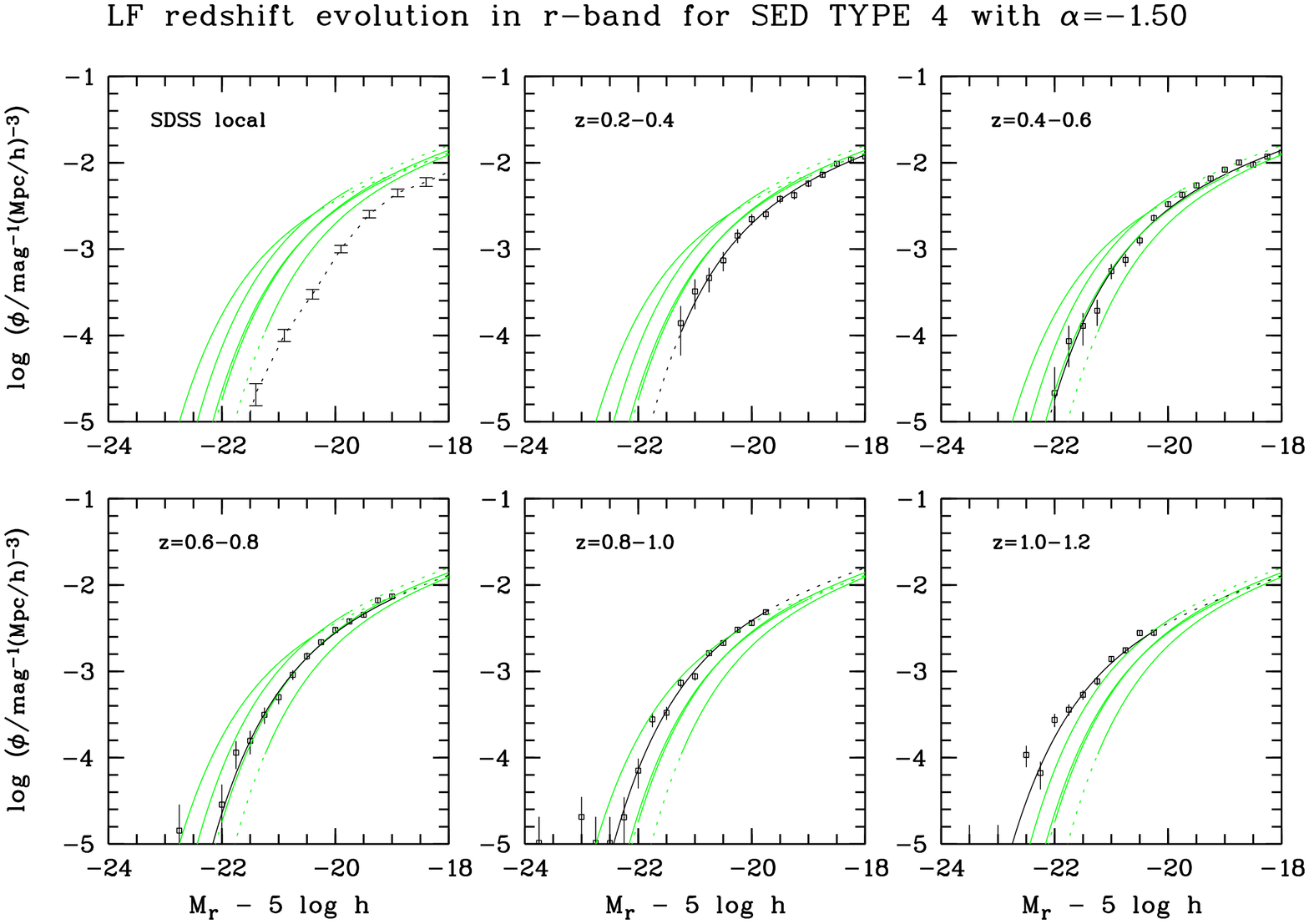,angle=270,clip=t,width=15.5cm}}}
\caption[ ]{Redshift evolution of $\phi(M_\mathrm{r})$ for type 4 galaxies:
$V_\mathrm{max}$ data points are shown with error bars for one redshift interval per panel.
The corresponding STY fit is plotted as a black line, while the fits for the other
redshifts are shown as grey lines. Local reference: SDSS.
\label{lf_evo4r}}
\end{figure*}

\begin{figure*}
\centerline{\hbox{
\psfig{figure=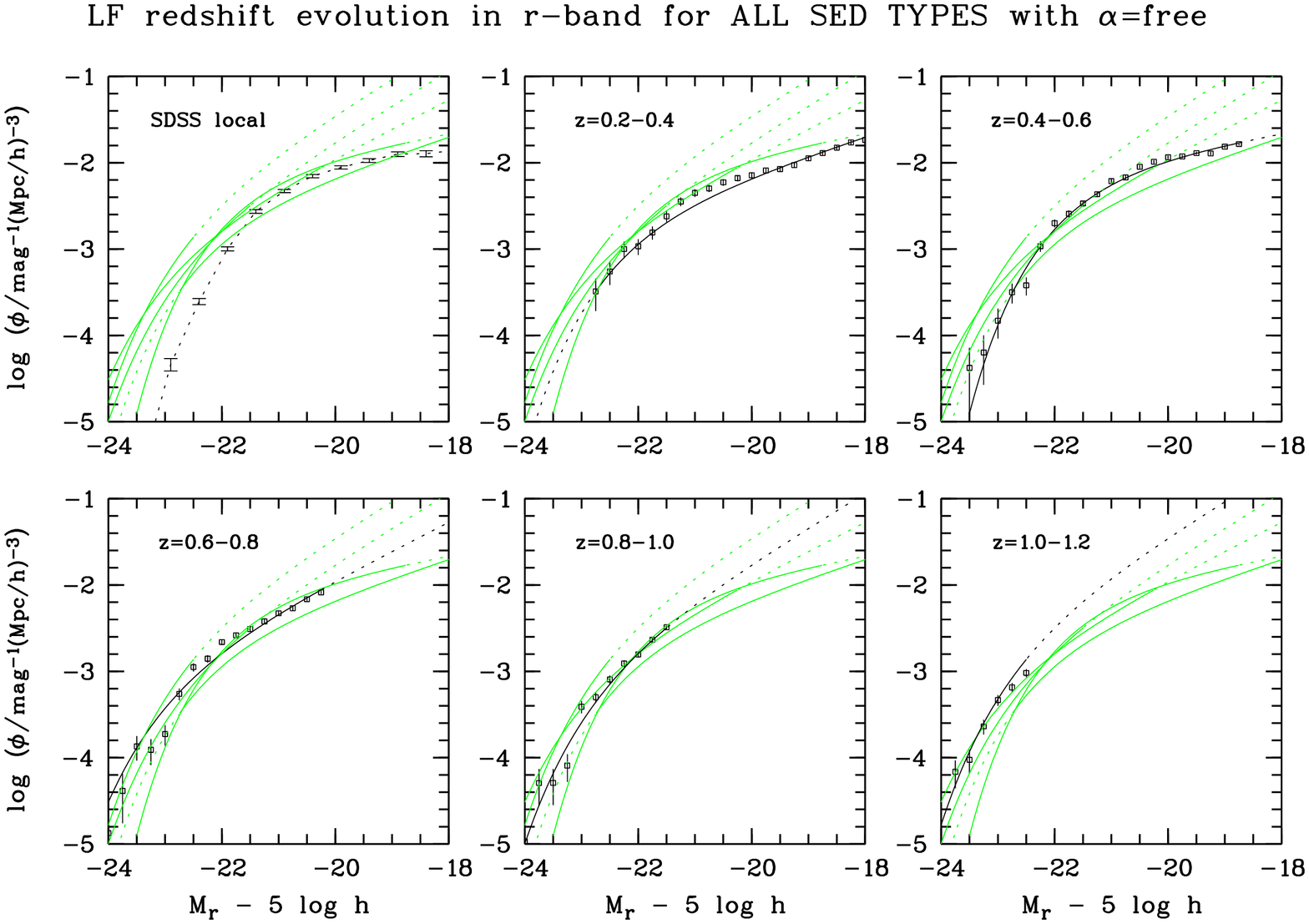,angle=270,clip=t,width=15.5cm}}}
\caption[ ]{Redshift evolution of $\phi(M_\mathrm{r})$ for all galaxies:
$V_\mathrm{max}$ data points are shown with error bars for one redshift interval per panel.
The corresponding STY fit is plotted as a black line, while the fits for the other
redshifts are shown as grey lines. Local reference: SDSS.
\label{lf_evo6r}}
\end{figure*}

\begin{table*}
\caption{Redshift evolution of parameters for $\phi(M_\mathrm{r})$:
Results of STY fits in five redshift intervals centered at $z=[1.1,0.9,0.7,
0.5,0.3]$. The faint-end slope $\alpha$ was determined in the quasi-local 
sample and fixed for all redshifts. Listed are $M^*$, $\phi^*$ with its 
field-to-field variation, $j$ and the covariance between $\phi^*$ and $L^*$.
The $j$ values for the combined sample are actually sums over the $j_{type}$.
\label{lfpars_rs}}
\begin{center}
\scriptsize
\begin{tabular}{lcr@{$\pm$}rr@{$\pm$}rr@{$\pm$}rr@{$\pm$}rc}
parameter &$<z>$ & \multicolumn{2}{c}{$M^*-5$~log~$h$} & 
  \multicolumn{2}{c}{$\phi^* \times 10^{-4}$} & 
  \multicolumn{2}{c}{$\alpha$} & 
  \multicolumn{2}{c}{$j \times 10^{7} L_{\odot}$} & $c_{\phi^*,L^*}$ \\
 & & \multicolumn{2}{c}{(Vega mag)} & \multicolumn{2}{c}{$(h/$Mpc)$^{-3}$} &
  \multicolumn{2}{c}{} & \multicolumn{2}{c}{$(h/$Mpc$^3$)}  \\
\noalign{\smallskip} \hline \noalign{\smallskip}
        & 0.3 & $-20.92$ & $ 0.14$ & $ 44.09$ & $22.81$ & $ 0.43$ & $ 0.18$ & $  8.04$ & $ 4.16$ & $-0.009$ \\
        & 0.5 & $-21.14$ & $ 0.15$ & $ 24.88$ & $ 6.77$ &         &         & $  5.53$ & $ 1.50$ & $-0.051$ \\
type 1  & 0.7 & $-21.59$ & $ 0.16$ & $ 12.81$ & $ 5.80$ &         &         & $  4.33$ & $ 1.96$ & $-0.185$ \\
        & 0.9 & $-21.52$ & $ 0.21$ & $  4.47$ & $ 3.95$ &         &         & $  1.41$ & $ 1.24$ & $-0.867$ \\
        & 1.1 & $-21.98$ & $ 0.26$ & $  3.86$ & $ 2.46$ &         &         & $  1.86$ & $ 1.19$ & $-1.577$ \\
\noalign{\smallskip} \hline \noalign{\smallskip}
        & 0.3 & $-21.12$ & $ 0.17$ & $ 33.83$ & $14.48$ & $-0.97$ & $ 0.07$ & $  5.76$ & $ 2.46$ & $-0.317$ \\
        & 0.5 & $-21.40$ & $ 0.20$ & $ 45.97$ & $ 4.38$ &         &         & $ 10.03$ & $ 0.96$ & $-0.435$ \\
type 2  & 0.7 & $-21.60$ & $ 0.12$ & $ 47.12$ & $ 1.34$ &         &         & $ 12.41$ & $ 0.35$ & $-0.742$ \\
        & 0.9 & $-21.82$ & $ 0.12$ & $ 35.88$ & $18.75$ &         &         & $ 11.58$ & $ 6.05$ & $-1.114$ \\
        & 1.1 & $-22.26$ & $ 0.16$ & $ 18.03$ & $ 6.47$ &         &         & $  8.76$ & $ 3.15$ & $-1.559$ \\
\noalign{\smallskip} \hline \noalign{\smallskip}
        & 0.3 & $-21.20$ & $ 0.22$ & $ 16.68$ & $ 9.25$ & $-1.33$ & $ 0.06$ & $  4.18$ & $ 2.32$ & $-0.541$ \\
        & 0.5 & $-21.46$ & $ 0.36$ & $ 31.02$ & $ 7.09$ &         &         & $  9.84$ & $ 2.25$ & $-0.625$ \\
type 3  & 0.7 & $-21.47$ & $ 0.14$ & $ 26.04$ & $ 4.09$ &         &         & $  8.30$ & $ 1.30$ & $-0.807$ \\
        & 0.9 & $-21.59$ & $ 0.14$ & $ 30.60$ & $ 9.25$ &         &         & $ 10.94$ & $ 3.31$ & $-1.100$ \\
        & 1.1 & $-21.71$ & $ 0.14$ & $ 23.55$ & $ 2.21$ &         &         & $  9.41$ & $ 0.88$ & $-1.458$ \\
\noalign{\smallskip} \hline \noalign{\smallskip}
        & 0.3 & $-19.91$ & $ 0.17$ & $ 66.14$ & $ 6.43$ & $-1.50$ & $ 0.07$ & $  6.54$ & $ 0.64$ & $-0.839$ \\
        & 0.5 & $-20.27$ & $ 0.19$ & $ 60.36$ & $11.56$ &         &         & $  8.30$ & $ 1.59$ & $-0.923$ \\
type 4  & 0.7 & $-20.33$ & $ 0.14$ & $ 54.03$ & $10.22$ &         &         & $  7.92$ & $ 1.50$ & $-1.232$ \\
        & 0.9 & $-20.60$ & $ 0.18$ & $ 56.73$ & $20.74$ &         &         & $ 10.63$ & $ 3.89$ & $-1.498$ \\
        & 1.1 & $-21.00$ & $ 0.14$ & $ 37.03$ & $15.01$ &         &         & $ 10.00$ & $ 4.05$ & $-1.568$ \\
\noalign{\smallskip} \hline \noalign{\smallskip}
        & 0.3 & $-22.13$ & $ 0.16$ & $ 27.99$ & $ 8.03$ & $-1.54$ & $ 0.02$ & $ 24.52$ & $ 5.40$ & $-0.639$ \\
        & 0.5 & $-21.55$ & $ 0.08$ & $ 93.99$ & $14.51$ & $-1.29$ & $ 0.03$ & $ 33.70$ & $ 3.28$ & $-0.658$ \\
all     & 0.7 & $-22.85$ & $ 0.15$ & $ 14.32$ & $ 2.16$ & $-1.83$ & $ 0.03$ & $ 32.96$ & $ 2.81$ & $-1.116$ \\
        & 0.9 & $-22.54$ & $ 0.09$ & $ 19.46$ & $10.21$ & $-2.00$ & $ 0.00$ & $ 34.56$ & $ 8.01$ & $-1.787$ \\
        & 1.1 & $-22.48$ & $ 0.09$ & $ 42.07$ & $16.33$ & $-2.00$ & $ 0.01$ & $ 30.03$ & $ 5.34$ & $-2.747$ \\
\noalign{\smallskip} \hline
\end{tabular}
\end{center}
\end{table*}

\clearpage

\begin{figure*}
\centerline{\hbox{
\psfig{figure=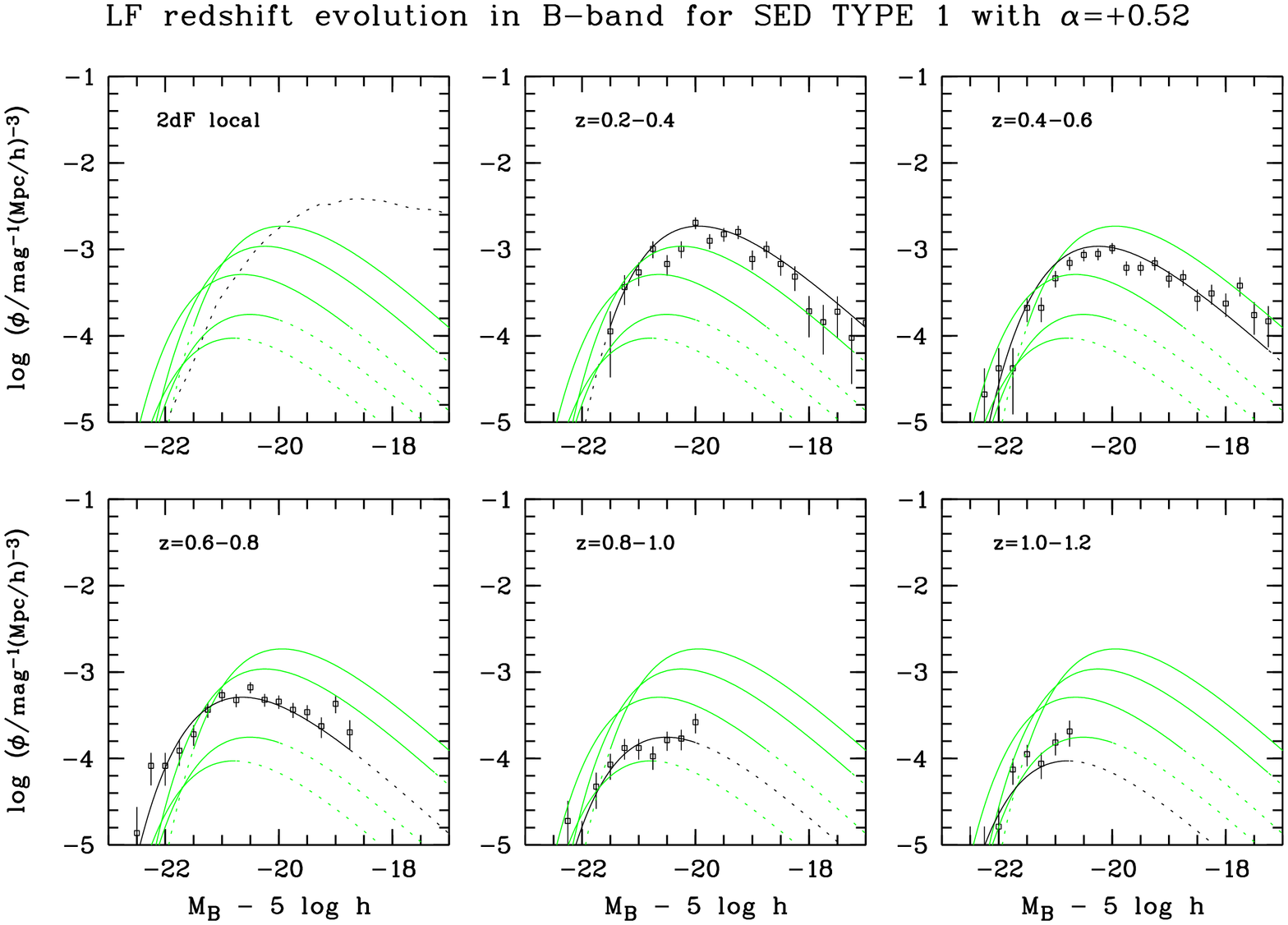,angle=270,clip=t,width=15.5cm}}}
\caption[ ]{Redshift evolution of $\phi(M_\mathrm{B})$ for type 1 galaxies:
$V_\mathrm{max}$ data points are shown with error bars for one redshift interval per panel.
The corresponding STY fit is plotted as a black line, while the fits for the other
redshifts are shown as grey lines. Local reference: 2dFGRS.
\label{lf_evo1b}}
\end{figure*}

\begin{figure*}
\centerline{\hbox{
\psfig{figure=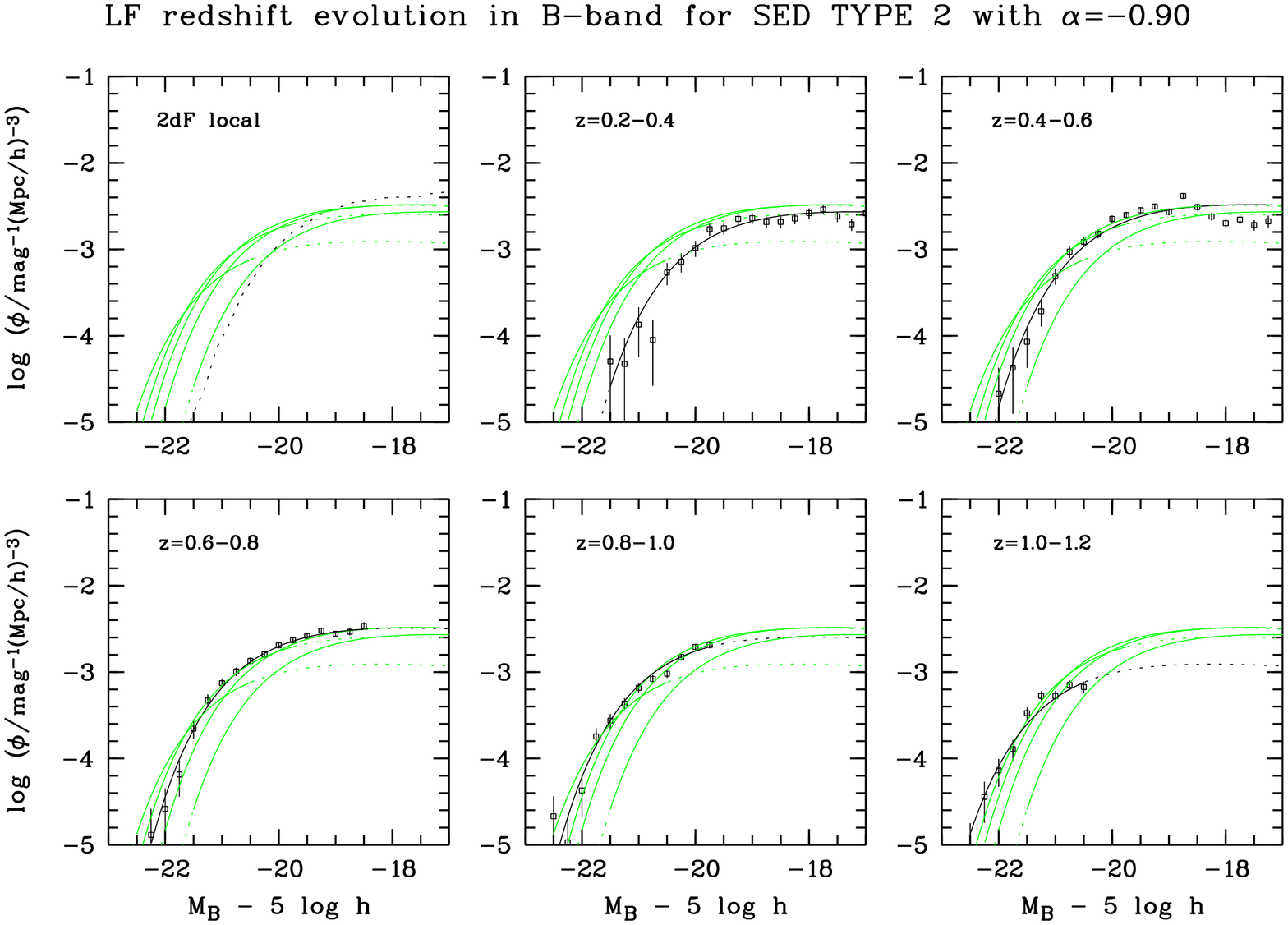,angle=270,clip=t,width=15.5cm}}}
\caption[ ]{Redshift evolution of $\phi(M_\mathrm{B})$ for type 2 galaxies:
$V_\mathrm{max}$ data points are shown with error bars for one redshift interval per panel.
The corresponding STY fit is plotted as a black line, while the fits for the other
redshifts are shown as grey lines. Local reference: 2dFGRS.
\label{lf_evo2b}}
\end{figure*}

\begin{figure*}
\centerline{\hbox{
\psfig{figure=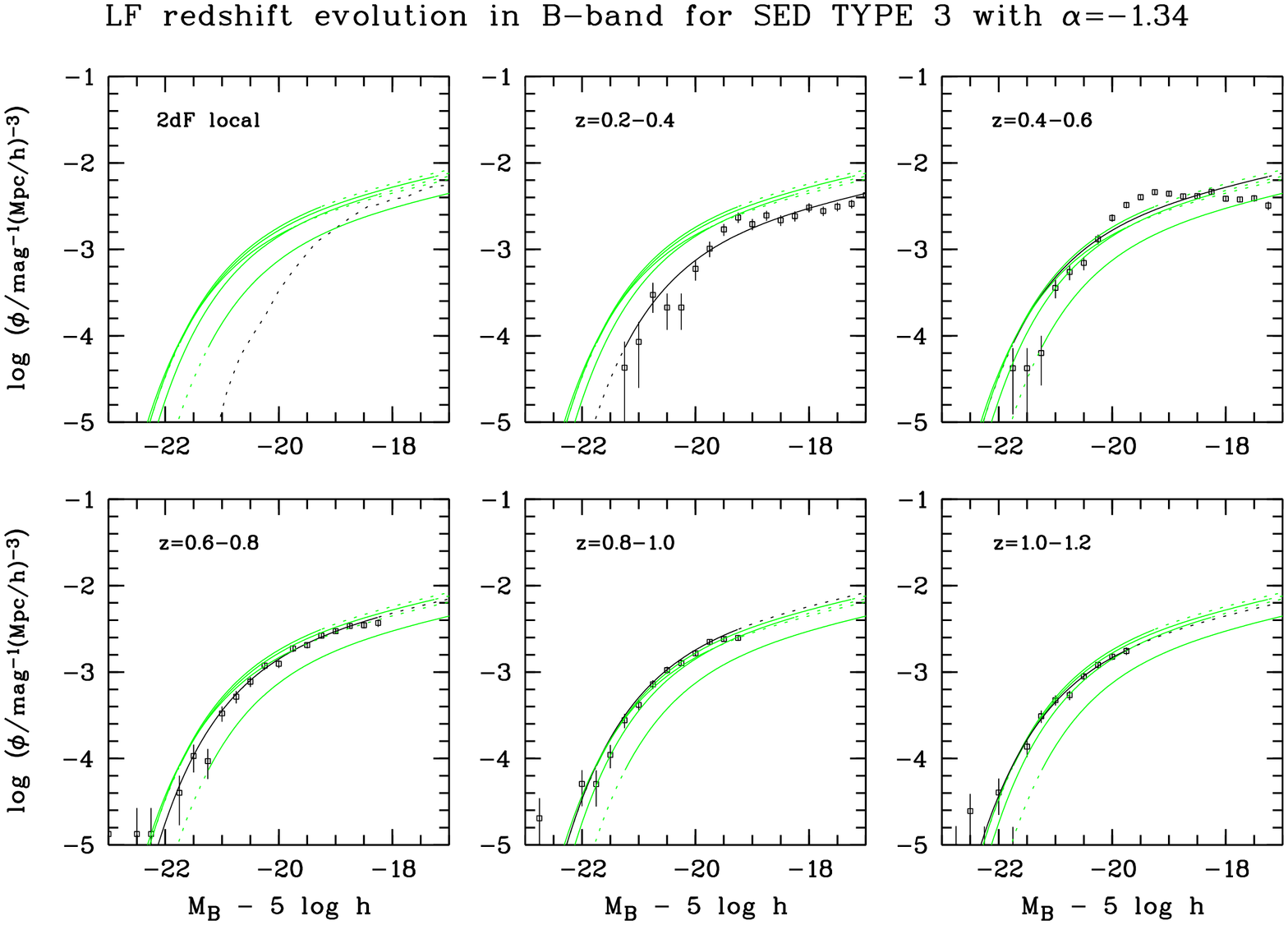,angle=270,clip=t,width=15.5cm}}}
\caption[ ]{Redshift evolution of $\phi(M_\mathrm{B})$ for type 3 galaxies:
$V_\mathrm{max}$ data points are shown with error bars for one redshift interval per panel.
The corresponding STY fit is plotted as a black line, while the fits for the other
redshifts are shown as grey lines. Local reference: 2dFGRS.
\label{lf_evo3b}}
\end{figure*}

\begin{figure*}
\centerline{\hbox{
\psfig{figure=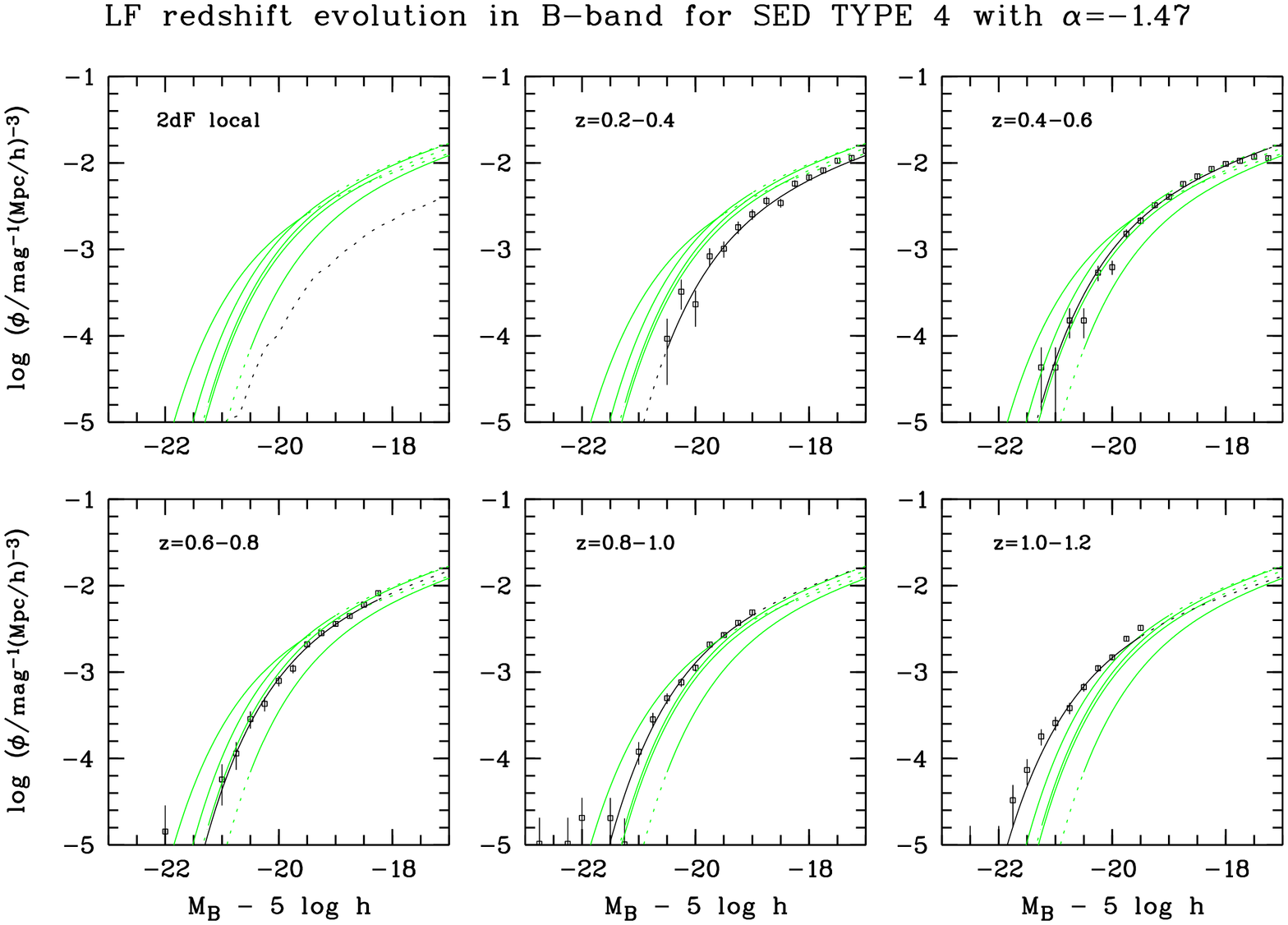,angle=270,clip=t,width=15.5cm}}}
\caption[ ]{Redshift evolution of $\phi(M_\mathrm{B})$ for type 4 galaxies:
$V_\mathrm{max}$ data points are shown with error bars for one redshift interval per panel.
The corresponding STY fit is plotted as a black line, while the fits for the other
redshifts are shown as grey lines. Local reference: 2dFGRS.
\label{lf_evo4b}}
\end{figure*}

\begin{figure*}
\centerline{\hbox{
\psfig{figure=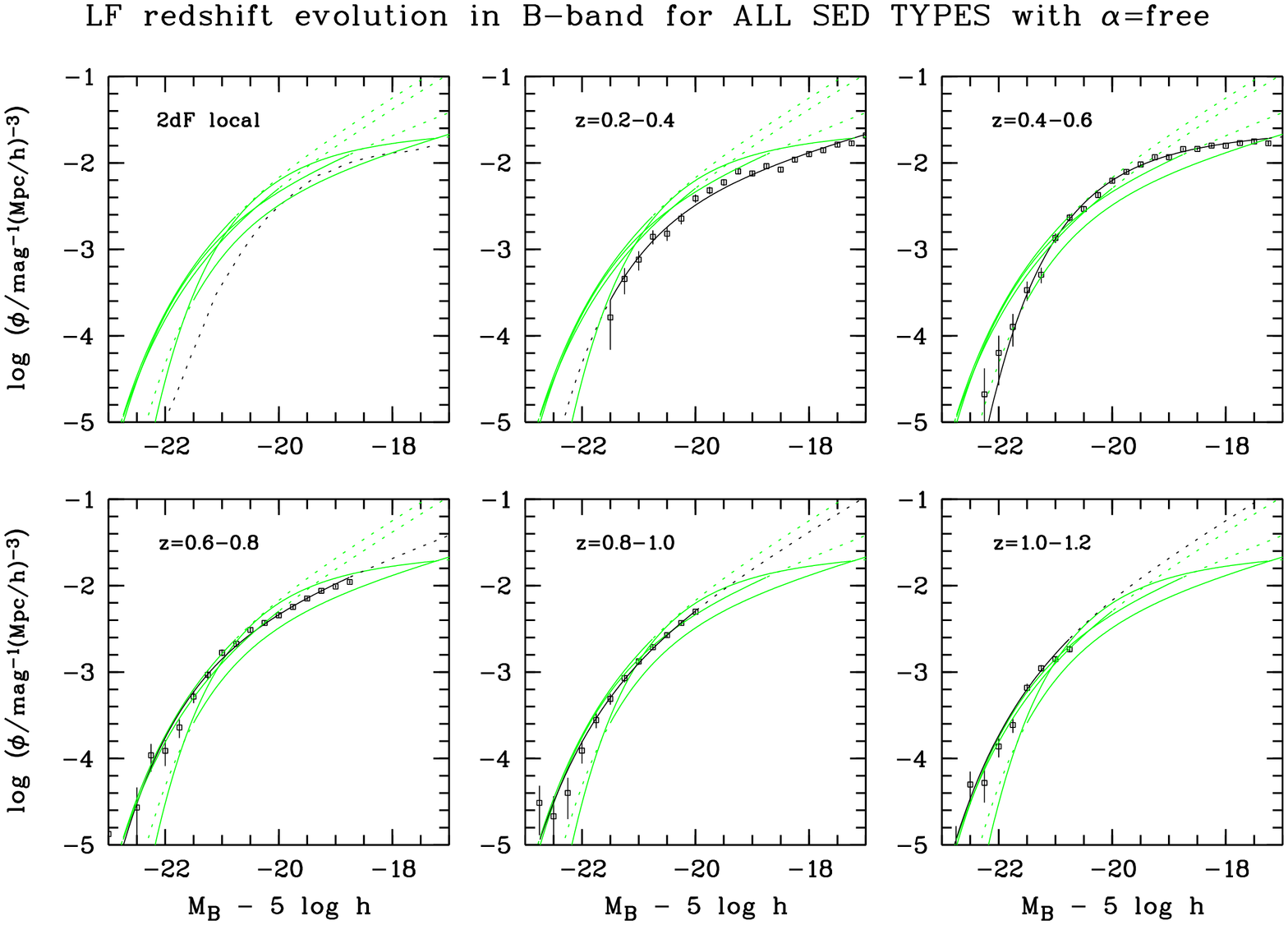,angle=270,clip=t,width=15.5cm}}}
\caption[ ]{Redshift evolution of $\phi(M_\mathrm{B})$ for all galaxies:
$V_\mathrm{max}$ data points are shown with error bars for one redshift interval per panel.
The corresponding STY fit is plotted as a black line, while the fits for the other
redshifts are shown as grey lines. Local reference: 2dFGRS.
\label{lf_evo6b}}
\end{figure*}

\begin{table*}
\caption{Redshift evolution of parameters for $\phi(M_\mathrm{B})$:
Results of STY fits in five redshift intervals centered at $z=[1.1,0.9,0.7,
0.5,0.3]$. The faint-end slope $\alpha$ was determined in the quasi-local 
sample and fixed for all redshifts. Listed are $M^*$, $\phi^*$ with its 
field-to-field variation, $j$ and the covariance between $\phi^*$ and $L^*$.
The $j$ values for the combined sample are actually sums over the $j_{type}$.
\label{lfpars_bj}}
\begin{center}
\scriptsize
\begin{tabular}{lcr@{$\pm$}rr@{$\pm$}rr@{$\pm$}rr@{$\pm$}rc}
parameter &$<z>$ & \multicolumn{2}{c}{$M^*-5$~log~$h$} & 
  \multicolumn{2}{c}{$\phi^* \times 10^{-4}$} & 
  \multicolumn{2}{c}{$\alpha$} & 
  \multicolumn{2}{c}{$j \times 10^{7} L_{\odot}$} & $c_{\phi^*,L^*}$ \\
 & & \multicolumn{2}{c}{(Vega mag)} & \multicolumn{2}{c}{$(h/$Mpc)$^{-3}$} &
  \multicolumn{2}{c}{} & \multicolumn{2}{c}{$(h/$Mpc$^3$)}  \\
\noalign{\smallskip} \hline \noalign{\smallskip}
        & 0.3 & $-19.49$ & $ 0.15$ & $ 48.68$ & $15.96$ & $ 0.52$ & $ 0.20$ & $  5.41$ & $ 1.77$ & $-0.049$ \\
        & 0.5 & $-19.79$ & $ 0.16$ & $ 28.49$ & $12.04$ &         &         & $  4.17$ & $ 1.76$ & $-0.035$ \\
type 1  & 0.7 & $-20.19$ & $ 0.16$ & $ 13.44$ & $ 3.62$ &         &         & $  2.85$ & $ 0.77$ & $-0.144$ \\
        & 0.9 & $-20.06$ & $ 0.21$ & $  4.63$ & $ 4.16$ &         &         & $  0.87$ & $ 0.78$ & $-0.788$ \\
        & 1.1 & $-20.34$ & $ 0.23$ & $  2.46$ & $ 1.99$ &         &         & $  0.60$ & $ 0.48$ & $-1.363$ \\
\noalign{\smallskip} \hline \noalign{\smallskip}
        & 0.3 & $-19.72$ & $ 0.17$ & $ 40.97$ & $19.05$ & $-0.90$ & $ 0.08$ & $  3.97$ & $ 1.85$ & $-0.468$ \\
        & 0.5 & $-20.07$ & $ 0.19$ & $ 49.56$ & $ 6.48$ &         &         & $  6.63$ & $ 0.87$ & $-0.407$ \\
type 2  & 0.7 & $-20.25$ & $ 0.11$ & $ 48.67$ & $ 2.43$ &         &         & $  7.72$ & $ 0.39$ & $-0.646$ \\
        & 0.9 & $-20.44$ & $ 0.12$ & $ 38.48$ & $20.29$ &         &         & $  7.24$ & $ 3.82$ & $-1.151$ \\
        & 1.1 & $-20.75$ & $ 0.15$ & $ 18.63$ & $ 6.00$ &         &         & $  4.66$ & $ 1.50$ & $-1.522$ \\
\noalign{\smallskip} \hline \noalign{\smallskip}
        & 0.3 & $-20.11$ & $ 0.23$ & $ 19.30$ & $ 9.45$ & $-1.34$ & $ 0.07$ & $  3.84$ & $ 1.88$ & $-0.684$ \\
        & 0.5 & $-20.52$ & $ 0.37$ & $ 28.44$ & $ 7.33$ &         &         & $  8.25$ & $ 2.13$ & $-0.625$ \\
type 3  & 0.7 & $-20.38$ & $ 0.16$ & $ 27.61$ & $ 4.31$ &         &         & $  7.04$ & $ 1.10$ & $-0.836$ \\
        & 0.9 & $-20.50$ & $ 0.14$ & $ 31.43$ & $ 8.82$ &         &         & $  8.95$ & $ 2.51$ & $-1.149$ \\
        & 1.1 & $-20.60$ & $ 0.15$ & $ 23.51$ & $ 2.94$ &         &         & $  7.34$ & $ 0.92$ & $-1.371$ \\
\noalign{\smallskip} \hline \noalign{\smallskip}
        & 0.3 & $-19.05$ & $ 0.16$ & $ 63.44$ & $ 4.93$ & $-1.47$ & $ 0.06$ & $  5.76$ & $ 0.45$ & $-1.030$ \\
        & 0.5 & $-19.45$ & $ 0.17$ & $ 69.93$ & $14.50$ &         &         & $  9.24$ & $ 1.92$ & $-0.914$ \\
type 4  & 0.7 & $-19.44$ & $ 0.12$ & $ 62.25$ & $12.51$ &         &         & $  8.11$ & $ 1.63$ & $-1.278$ \\
        & 0.9 & $-19.65$ & $ 0.14$ & $ 64.13$ & $14.68$ &         &         & $ 10.15$ & $ 2.32$ & $-1.614$ \\
        & 1.1 & $-20.08$ & $ 0.12$ & $ 39.11$ & $13.04$ &         &         & $  9.17$ & $ 3.06$ & $-1.672$ \\
\noalign{\smallskip} \hline \noalign{\smallskip}
        & 0.3 & $-20.50$ & $ 0.13$ & $ 53.27$ & $12.98$ & $-1.47$ & $ 0.03$ & $ 18.97$ & $ 3.21$ & $-0.721$ \\
        & 0.5 & $-20.03$ & $ 0.06$ & $179.34$ & $31.08$ & $-1.09$ & $ 0.03$ & $ 28.30$ & $ 3.47$ & $-0.526$ \\
all     & 0.7 & $-21.00$ & $ 0.11$ & $ 41.74$ & $ 3.88$ & $-1.63$ & $ 0.04$ & $ 25.72$ & $ 2.15$ & $-1.028$ \\
        & 0.9 & $-21.27$ & $ 0.09$ & $ 23.34$ & $10.91$ & $-2.00$ & $ 0.02$ & $ 27.21$ & $ 5.18$ & $-1.672$ \\
        & 1.1 & $-21.20$ & $ 0.09$ & $ 33.81$ & $10.84$ & $-2.00$ & $ 0.03$ & $ 21.78$ & $ 3.56$ & $-2.225$ \\
\noalign{\smallskip} \hline
\end{tabular}
\end{center}
\end{table*}

\clearpage

\begin{figure*}
\centerline{\hbox{
\psfig{figure=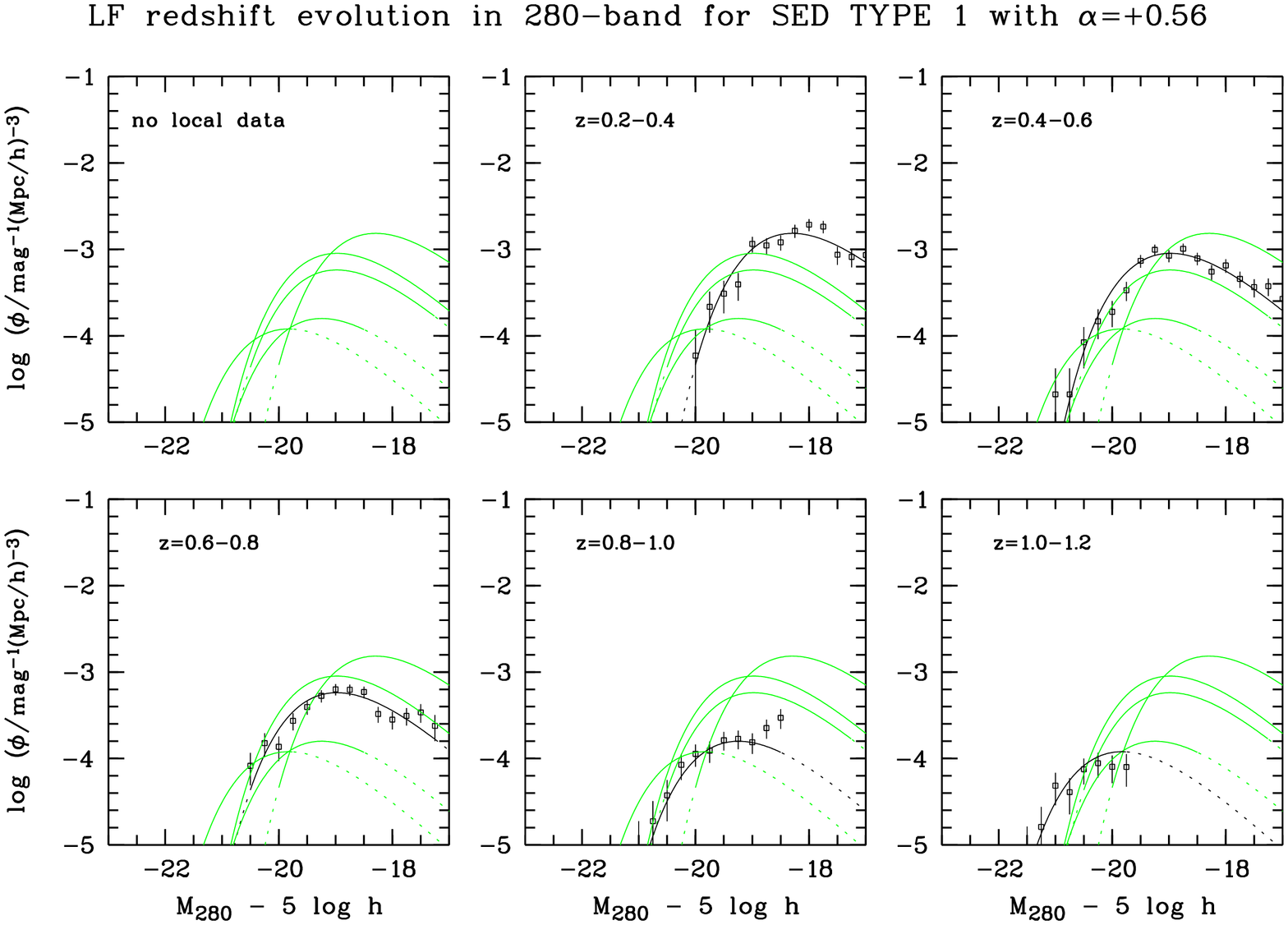,angle=270,clip=t,width=15.5cm}}}
\caption[ ]{Redshift evolution of $\phi(M_\mathrm{280})$ for type 1 galaxies:
$V_\mathrm{max}$ data points are shown with error bars for one redshift interval per panel.
The corresponding STY fit is plotted as a black line, while the fits for the other
redshifts are shown as grey lines. No local reference available.
\label{lf_evo1u}}
\end{figure*}

\begin{figure*}
\centerline{\hbox{
\psfig{figure=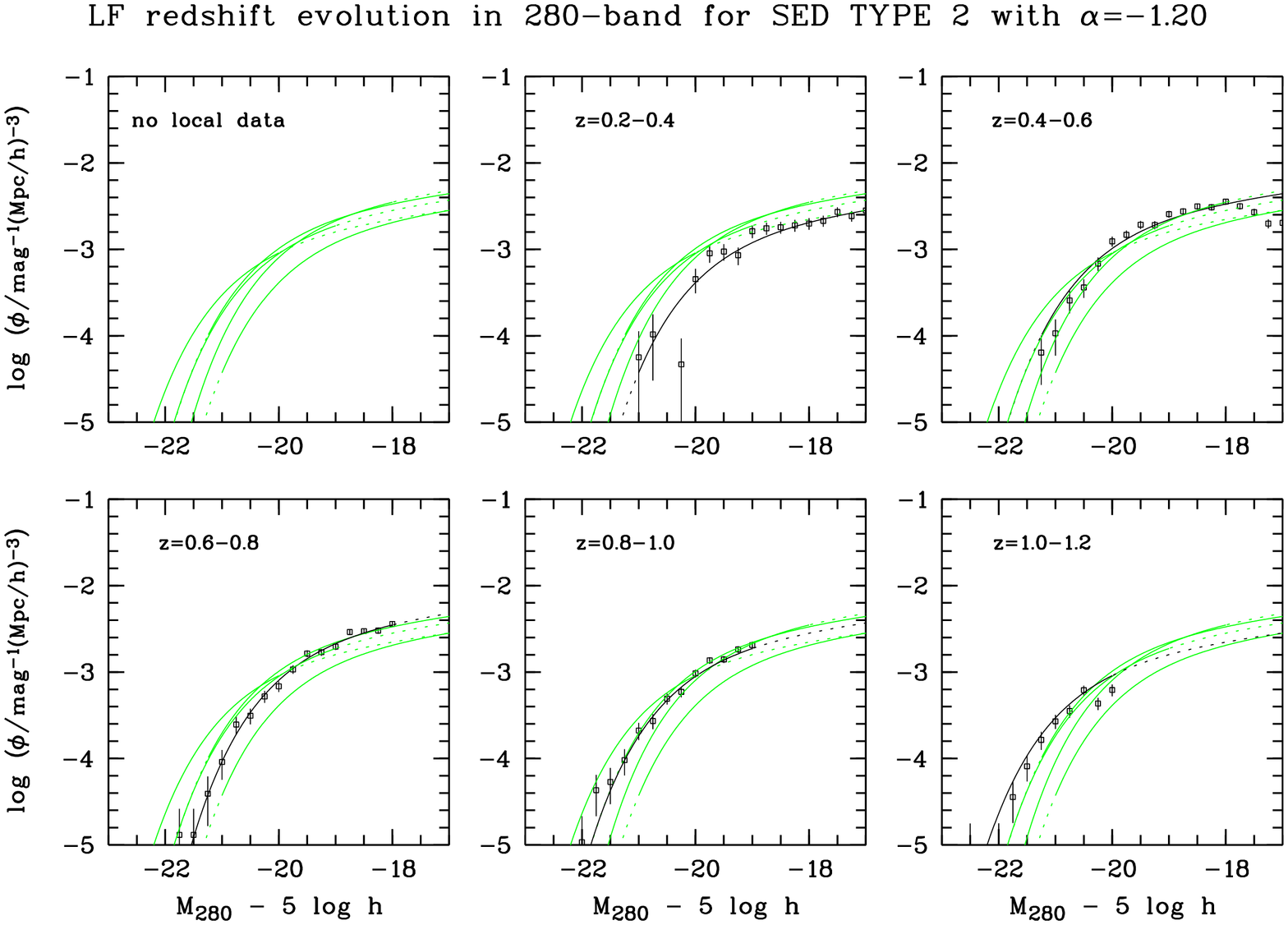,angle=270,clip=t,width=15.5cm}}}
\caption[ ]{Redshift evolution of $\phi(M_\mathrm{280})$ for type 2 galaxies:
$V_\mathrm{max}$ data points are shown with error bars for one redshift interval per panel.
The corresponding STY fit is plotted as a black line, while the fits for the other
redshifts are shown as grey lines. No local reference available.
\label{lf_evo2u}}
\end{figure*}

\begin{figure*}
\centerline{\hbox{
\psfig{figure=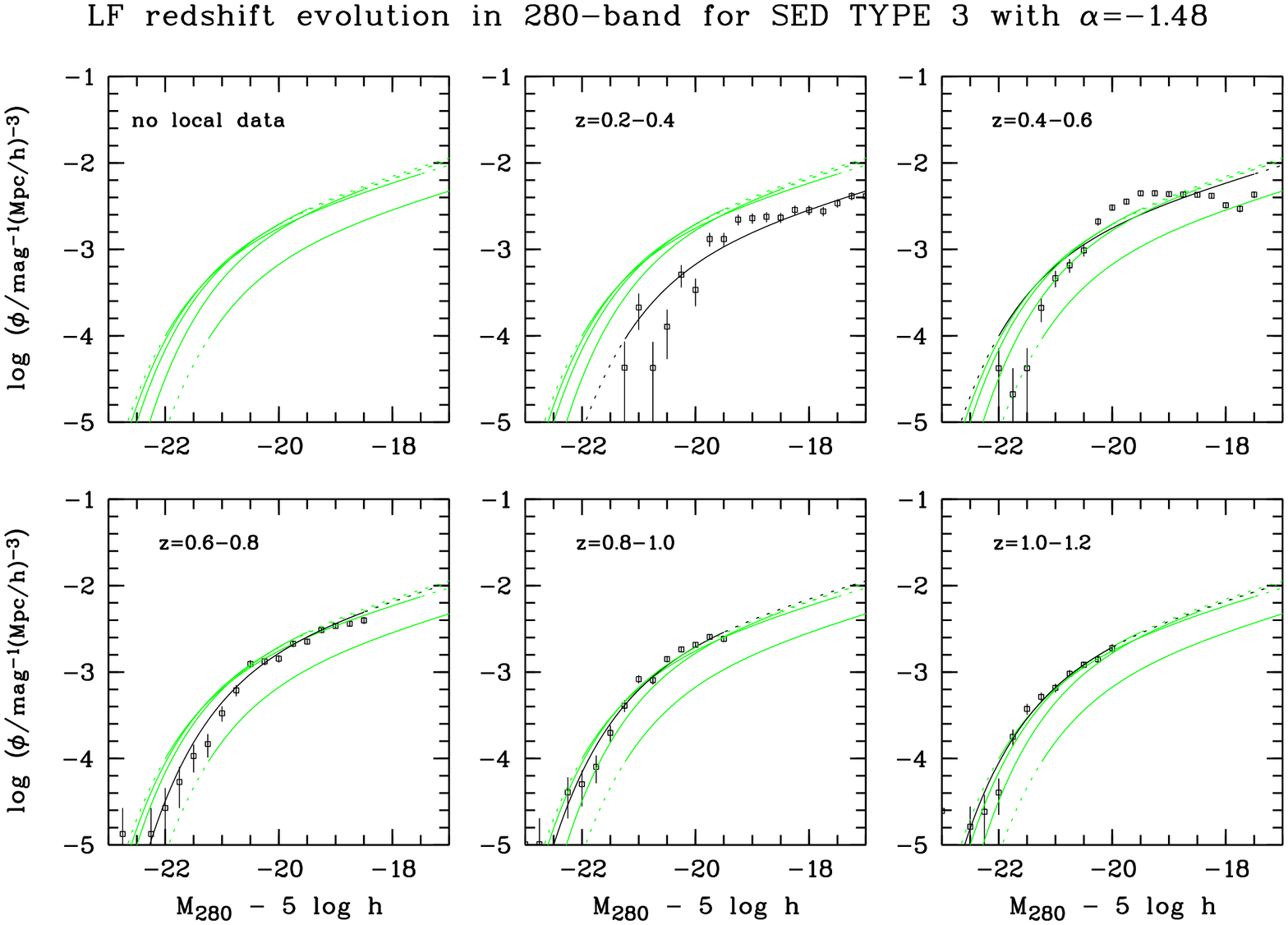,angle=270,clip=t,width=15.5cm}}}
\caption[ ]{Redshift evolution of $\phi(M_\mathrm{280})$ for type 3 galaxies:
$V_\mathrm{max}$ data points are shown with error bars for one redshift interval per panel.
The corresponding STY fit is plotted as a black line, while the fits for the other
redshifts are shown as grey lines. No local reference available. 
\label{lf_evo3u}}
\end{figure*}

\begin{figure*}
\centerline{\hbox{
\psfig{figure=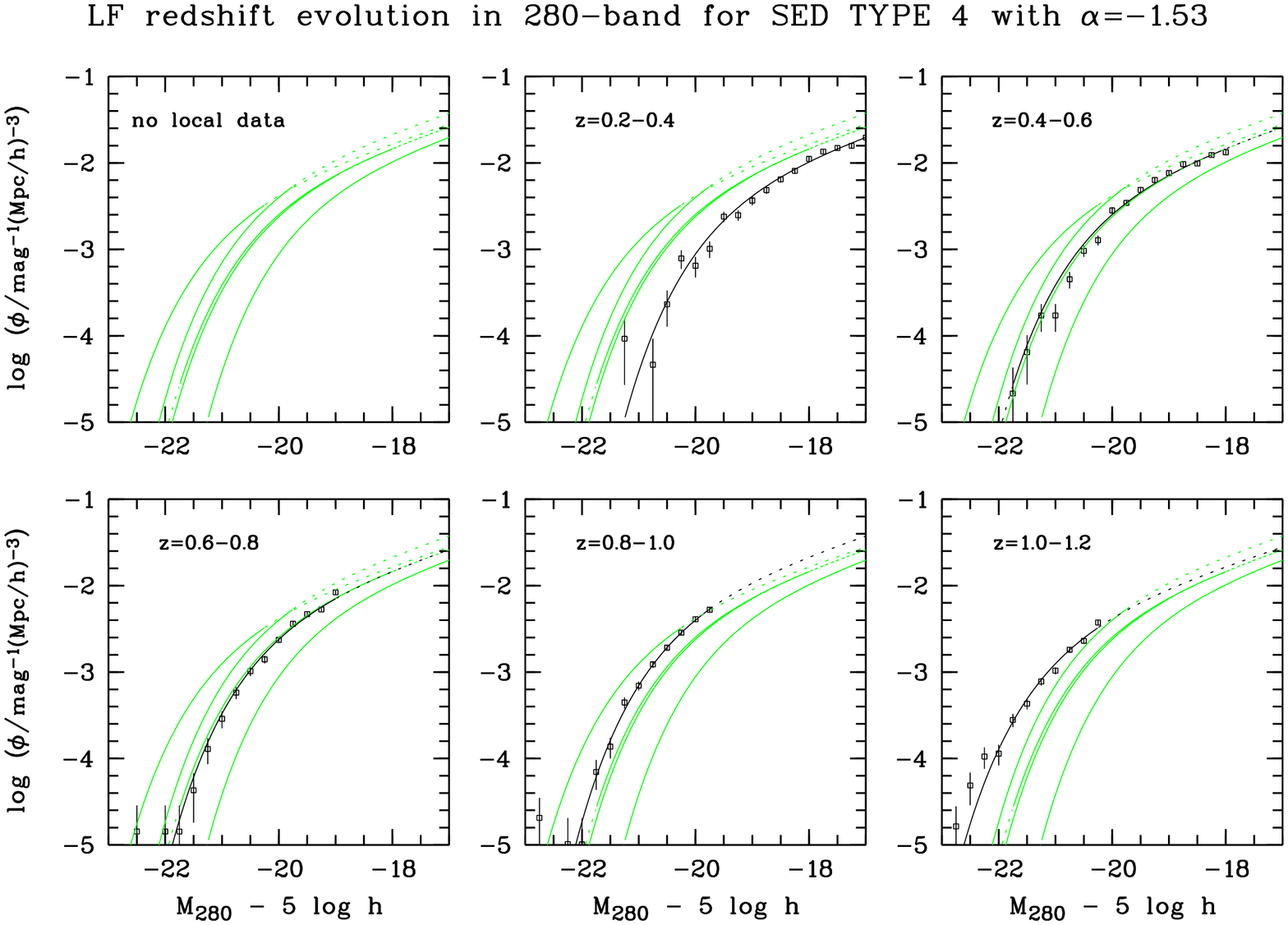,angle=270,clip=t,width=15.5cm}}}
\caption[ ]{Redshift evolution of $\phi(M_\mathrm{280})$ for type 4 galaxies:
$V_\mathrm{max}$ data points are shown with error bars for one redshift interval per panel.
The corresponding STY fit is plotted as a black line, while the fits for the other
redshifts are shown as grey lines. No local reference available.
\label{lf_evo4u}}
\end{figure*}

\begin{figure*}
\centerline{\hbox{
\psfig{figure=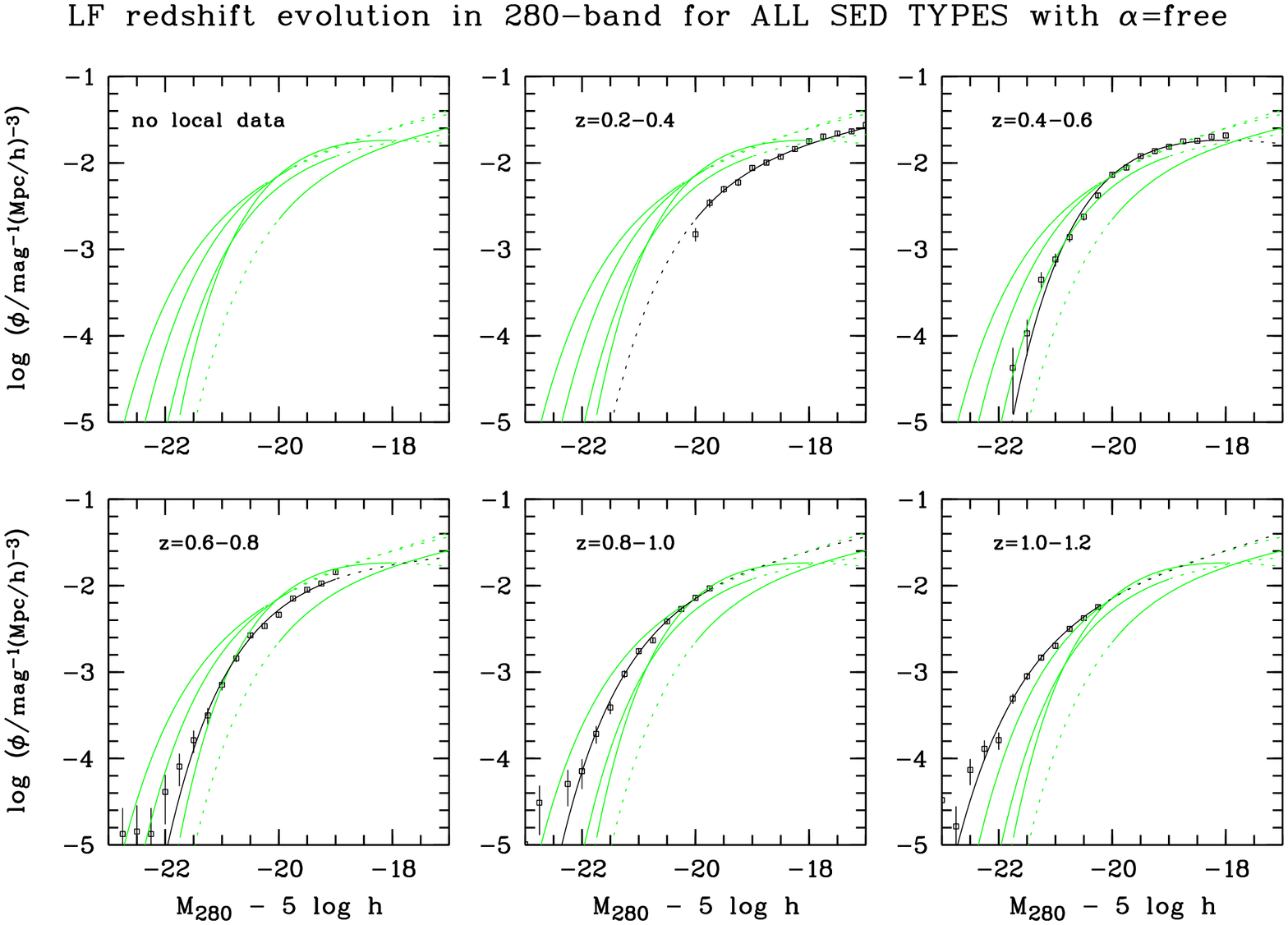,angle=270,clip=t,width=15.5cm}}}
\caption[ ]{Redshift evolution of $\phi(M_\mathrm{280})$ for all galaxies:
$V_\mathrm{max}$ data points are shown with error bars for one redshift interval per panel.
The corresponding STY fit is plotted as a black line, while the fits for the other
redshifts are shown as grey lines. No local reference available.
\label{lf_evo6u}}
\end{figure*}

\begin{table*} 
\caption{Redshift evolution of parameters for $\phi(M_\mathrm{280})$:
Results of STY fits in five redshift intervals centered at $z=[1.1,0.9,0.7,
0.5,0.3]$. The faint-end slope $\alpha$ was determined in the quasi-local 
sample and fixed for all redshifts. Listed are $M^*$, $\phi^*$ with its 
field-to-field variation, $j$ and the covariance between $\phi^*$ and $L^*$.
The $j$ values for the combined sample are actually sums over the $j_{type}$.
\label{lfpars_uc}}
\begin{center}
\scriptsize
\begin{tabular}{lcr@{$\pm$}rr@{$\pm$}rr@{$\pm$}rr@{$\pm$}rc}
parameter &$<z>$ & \multicolumn{2}{c}{$M^*-5$~log~$h$} & 
  \multicolumn{2}{c}{$\phi^* \times 10^{-4}$} & 
  \multicolumn{2}{c}{$\alpha$} & 
  \multicolumn{2}{c}{$j \times 10^{7} L_{\odot}$} & $c_{\phi^*,L^*}$ \\
 & & \multicolumn{2}{c}{(Vega mag)} & \multicolumn{2}{c}{$(h/$Mpc)$^{-3}$} &
  \multicolumn{2}{c}{} & \multicolumn{2}{c}{$(h/$Mpc$^3$)}  \\
\noalign{\smallskip} \hline \noalign{\smallskip}
        & 0.3 & $-17.81$ & $ 0.14$ & $ 39.49$ & $19.00$ & $ 0.56$ & $ 0.19$ & $  3.36$ & $ 1.62$ & $-0.413$ \\
        & 0.5 & $-18.49$ & $ 0.16$ & $ 23.19$ & $ 9.20$ &         &         & $  3.69$ & $ 1.47$ & $-0.176$ \\
type 1  & 0.7 & $-18.49$ & $ 0.14$ & $ 14.90$ & $ 5.15$ &         &         & $  2.38$ & $ 0.82$ & $-0.175$ \\
        & 0.9 & $-18.76$ & $ 0.19$ & $  4.09$ & $ 2.54$ &         &         & $  0.84$ & $ 0.52$ & $-0.599$ \\
        & 1.1 & $-19.34$ & $ 0.23$ & $  3.09$ & $ 2.05$ &         &         & $  1.08$ & $ 0.72$ & $-1.321$ \\
\noalign{\smallskip} \hline \noalign{\smallskip}
        & 0.3 & $-19.58$ & $ 0.20$ & $ 20.84$ & $11.66$ & $-1.20$ & $ 0.05$ & $  7.63$ & $ 4.27$ & $-0.687$ \\
        & 0.5 & $-20.03$ & $ 0.29$ & $ 29.04$ & $ 6.57$ &         &         & $ 16.09$ & $ 3.64$ & $-0.599$ \\
type 2  & 0.7 & $-19.70$ & $ 0.11$ & $ 34.23$ & $ 2.25$ &         &         & $ 14.00$ & $ 0.92$ & $-0.862$ \\
        & 0.9 & $-20.08$ & $ 0.13$ & $ 24.12$ & $ 7.81$ &         &         & $ 13.94$ & $ 4.51$ & $-1.130$ \\
        & 1.1 & $-20.51$ & $ 0.17$ & $ 16.73$ & $ 1.11$ &         &         & $ 14.43$ & $ 0.95$ & $-1.510$ \\
\noalign{\smallskip} \hline \noalign{\smallskip}
        & 0.3 & $-20.43$ & $ 0.25$ & $ 11.75$ & $ 5.51$ & $-1.48$ & $ 0.04$ & $ 13.64$ & $ 6.39$ & $-0.741$ \\
        & 0.5 & $-21.07$ & $ 0.44$ & $ 17.33$ & $ 4.04$ &         &         & $ 36.15$ & $ 8.44$ & $-0.695$ \\
type 3  & 0.7 & $-20.60$ & $ 0.16$ & $ 24.44$ & $ 3.24$ &         &         & $ 33.19$ & $ 4.39$ & $-0.949$ \\
        & 0.9 & $-20.83$ & $ 0.13$ & $ 23.29$ & $ 7.79$ &         &         & $ 39.29$ & $13.14$ & $-1.216$ \\
        & 1.1 & $-20.96$ & $ 0.14$ & $ 20.72$ & $ 4.18$ &         &         & $ 39.23$ & $ 7.92$ & $-1.412$ \\
\noalign{\smallskip} \hline \noalign{\smallskip}
        & 0.3 & $-19.39$ & $ 0.14$ & $ 74.43$ & $14.93$ & $-1.53$ & $ 0.05$ & $ 36.17$ & $ 7.26$ & $-0.977$ \\
        & 0.5 & $-20.09$ & $ 0.20$ & $ 65.77$ & $14.89$ &         &         & $ 61.21$ & $13.86$ & $-0.990$ \\
type 4  & 0.7 & $-20.01$ & $ 0.11$ & $ 68.99$ & $13.12$ &         &         & $ 59.90$ & $11.39$ & $-1.419$ \\
        & 0.9 & $-20.20$ & $ 0.09$ & $ 89.55$ & $24.69$ &         &         & $ 92.73$ & $25.57$ & $-1.823$ \\
        & 1.1 & $-20.81$ & $ 0.11$ & $ 48.60$ & $23.77$ &         &         & $ 88.54$ & $43.31$ & $-1.729$ \\
\noalign{\smallskip} \hline \noalign{\smallskip}
        & 0.3 & $-19.41$ & $ 0.08$ & $158.41$ & $50.65$ & $-1.30$ & $ 0.03$ & $ 60.80$ & $10.69$ & $-0.773$ \\
        & 0.5 & $-19.41$ & $ 0.05$ & $374.96$ & $71.17$ & $-0.72$ & $ 0.04$ & $117.16$ & $16.69$ & $-0.656$ \\
all     & 0.7 & $-19.80$ & $ 0.06$ & $193.29$ & $29.69$ & $-1.10$ & $ 0.05$ & $109.47$ & $12.27$ & $-1.221$ \\
        & 0.9 & $-20.31$ & $ 0.10$ & $148.33$ & $45.46$ & $-1.34$ & $ 0.09$ & $146.80$ & $29.10$ & $-1.580$ \\
        & 1.1 & $-20.80$ & $ 0.16$ & $ 88.56$ & $28.57$ & $-1.47$ & $ 0.17$ & $143.29$ & $44.05$ & $-1.694$ \\
\noalign{\smallskip} \hline
\end{tabular}
\end{center}
\end{table*}

\end{document}